\documentclass[useAMS,usenatbib,usegraphicx]{mn2e}
\usepackage{lscape}
\usepackage{multicol}
\usepackage{longtable}
\usepackage{pdfpages}
\usepackage{soul}

\title[Properties of massive star-forming clumps with infall motions]{Properties of massive star-forming clumps with infall motions}
\author[Yu-Xin~He et al.]{Yu-Xin~He$^{1,3}$\thanks{E-mail:heyuxin@xao.ac.cn}, Jian-Jun~Zhou$^{1,2}$\thanks{Email: zhoujj@xao.ac.cn},
Jarken~Esimbek$^{1,2}$, Wei-Guang~Ji$^{1,2}$, \newauthor Gang~Wu$^{1,2,3}$, Xin-Di~Tang$^{1,2,4}$, Toktarkhan~Komesh$^{5}$, Ye~Yuan$^{1,2}$,
\newauthor Da-Lei~Li$^{1,3}$ and W.A.~Baan$^{6}$\\
$^{1}$Xinjiang Astronomical Observatory, Chinese Academy of Sciences, Urumqi 830011, P. R. China\\
$^{2}$Key Laboratory of Radio Astronomy, Chinese Academy of Sciences, Urumqi 830011, P. R. China\\
$^{3}$University of the Chinese Academy of Sciences, Beijing 100080, P. R. China\\
$^{4}$Max-Planck-Institut f\"{u}r Radioastronomie, Bonn 53121, Germany\\
$^{5}$Department of Solid State Physics and Non-linear Physics, Faculty of Physics and Technology, Al-Farabi Kazakh National University, \\
Almaty 050040, Kazakhstan\\
$^{6}$Netherlands Institute for Radio Astronomy, 7991 PD Dwingeloo, The Netherlands}

\begin{document}

\pagerange{\pageref{firstpage}--\pageref{lastpage}} \pubyear{2002}

\maketitle

\label{firstpage}

\begin{abstract}
  In this work, we aim to characterise high-mass clumps with infall motions. We selected 327 clumps from the Millimetre Astronomy Legacy Team
  90-GHz (MALT90) survey, and identified 100 infall candidates. Combined with the results of \citet{2015MNRAS.450.1926H}, we obtained a sample of 732 high-mass clumps,
  including 231 massive infall candidates and 501 clumps where infall is not detected. Objects in our sample were classified as pre-stellar, proto-stellar, HII or photo-dissociation region (PDR). The detection rates of the infall candidates in the pre-stellar, proto-stellar, HII and PDR stages are 41.2\%, 36.6\%, 30.6\% and 12.7\%, respectively. The infall candidates have a higher H$_{2}$ column density and volume density compared with the clumps where infall is not detected at every stage. For the infall candidates, the median values of the infall rates at the pre-stellar, proto-stellar, HII and PDR stages
  are 2.6$\times$10$^{-3}$, 7.0$\times$10$^{-3}$, 6.5$\times$10$^{-3}$ and 5.5$\times$10$^{-3}$ M$_\odot$ yr$^{-1}$, respectively. These values indicate that infall candidates at later evolutionary stages are still accumulating material efficiently. It is interesting to find that both infall candidates and clumps where infall is not detected show a clear trend of increasing mass from the pre-stellar to proto-stellar, and to the HII stages. The power indices of the clump mass function (ClMF) are 2.04$\pm$0.16 and 2.17$\pm$0.31 for the infall candidates and clumps where infall is not detected, respectively, which agree well with the power index of the stellar initial mass function (2.35) and the cold \emph{Planck} cores (2.0).
\end{abstract}


\section{Introduction}
High-mass star formation is still poorly understood \citep{2014prpl.conf..149T}. Core accretion \citep{2002Natur.416...59M,2003ApJ...585..850M} and competitive accretion \citep{2001MNRAS.324..573B,2004MNRAS.349..735B} are two contending theories underlying current investigations.

The core accretion model essentially assumes that the evolution of the core is isolated from environmental effects outside the gravitational radius and that no material is added to the core after it forms. The final mass of a massive star is decided by the amount of mass that is necessary to be self-gravitating, which is thought to be supported either by thermal pressure, turbulence or magnetic fields \citep{2002Natur.416...59M,2003ApJ...585..850M}, and is not greatly influenced by a star's environment. The core accretion model imposes a direct link between core and star, and thus the core mass function (CMF) must have a similar shape to the stellar initial mass function (IMF) in that scenario. \citep[e.g.][]{1998A&A...336..150M}.

In the competitive accretion model, the mass of a star can be strongly influenced by its environment. Massive stars can reach the main sequence while still accreting in either scenario \citep{1996A&A...307..829B,2001A&A...373..190B,2010ApJ...721..478H}. Meanwhile, most higher mass cores start to fragment and do not continue to accrete significantly \citep{2006MNRAS.370..488B}. Finally, a few stars continue to accrete and become higher mass stars. In this case, competitive accretion can explain the full range and distribution of stellar masses \citep{2001MNRAS.324..573B}.

Thus, comprehensive knowledge about accretion is the key to understanding how massive stars form. Although accretion often cannot be observed directly, its presence can be inferred from the presence of large-scale inflows (infall) and outflows \citep{2007ApJ...663.1092K}. Therefore, high-mass infall candidates are important samples to study and understand high-mass star formation.

Evidence of infall motion is observable in self-absorbed, optically thick line profiles, which show a combination of double peaks with a brighter blue peak or a skewed single blue peak, or optically thin lines that peak at the self-absorption dip of the optically thick line \citep{1997ApJ...489..719M, 2000ApJ...533..440G, 2003ApJ...592L..79W}. Though rotation and outflows may also produce blue line asymmetries \citep{2009MNRAS.392..170S}, mapping observations with sufficient spatial resolution will make it possible to distinguish them from true infall motions.

In this paper, we will use the term ``clump" used by \citet{2000prpl.conf...97W} and \citet{2007ARA&A..45..339B} to refer to our sources. Clumps are defined as coherent regions in position-velocity space with typical size of 0.3-3 pc, while cores are defined as gravitationally bound regions, in which individual stars or stellar systems form, with typical size of 0.03-0.2 pc. We took advantage of five surveys to select a sample of 732 clumps, and in doing so we identified 231 infall candidates and derived their physical properties. It should be noted that 405 sources in our sample have been studied by \citet[][hereafter HYX15]{2015MNRAS.450.1926H}, of which 131 infall candidates were identified by them. The five surveys considered here are: the Millimetre Astronomy Legacy Team 90 GHz (MALT90) Survey \citep{2013PASA...30...57J}, the \emph{APEX} Telescope Large Survey of the Galaxy (ATLASGAL) 870 $\mu$m survey \citep{2009A&A...504..415S}, the GLIMPSE mid-infrared survey \citep{2003PASP..115..953B,2009PASP..121..213C}, the MIPSGAL survey \citep{2009PASP..121...76C}, and the \emph{Herschel} Infrared GALactic plane survey \citep{2010PASP..122..314M}. Infall candidates were identified by two optically thick lines, HCO$^{+}$(1-0) and HNC(1-0), and one optically thin line, N$_{2}$H$^{+}$(1-0). Mapping observations enabled us to identify the most probable infall candidates. A brief introduction to the surveys is given in Sections 2.1-2.4. The sample selection and classification in Section 3. In Section 4 the infall candidates are identified and investigated. Discussions are given in Section 5, followed by a summary of the important results in Section 6.

\section{DATA}
\subsection{The ATLASGAL survey}
The ATLASGAL Survey was the first systematic survey of the Inner Galactic plane in the submm band, tracing the thermal emission from dense clumps at 870 $\mu$m, It is complete to all massive clumps above 1000 M$_{\odot}$ to the far side of the Inner Galaxy ($\sim$20 kpc) \citep{2014MNRAS.443.1555U}. The survey was carried out with the Large \emph{APEX} Bolometer Camera \citep{2009A&A...497..945S}, which is an array of 295 bolometers observing at 870 $\mu$m (345 GHz). At this wavelength, the \emph{APEX} Telescope has a full width at halfmaximum (FWHM) beam size of 19.2 arcsec. The survey region covered the Galactic longitude region of $|\ell|<60^{\circ}$ and 280$^{\circ}$ $<$ $\ell$ $<$ 300$^{\circ}$, and Galactic latitude $|$b$|$ $<$ 1.5$^{\circ}$ and $-$2$^{\circ}$ $<$ $\ell$ $<$ 1$^{\circ}$, respectively. \citet{2014A&A...568A..41U} presented a compact source catalogue of this survey, which consists of $\sim$10163 sources and is 99 per cent complete at a $\sim$6$\sigma$ flux level, which corresponds to a flux sensitivity of $\sim$0.3-0.4 Jy beam$^{-1}$.

\subsection{The MALT90 survey}
The MALT90 Survey is a large international project that exploited the fast-mapping capability of the ATNF Mopra 22-m telescope to simultaneously image 16 molecular lines near 90 GHz (size of each map is 3$^{\prime}$$\times$3$^{\prime}$). These molecular lines characterize the physical and chemical conditions of high-mass star formation regions over a wide range of evolutionary states (from pre-stellar cores, to proto-stellar cores, and on to HII regions). The sample used in this study is a sub-sample of the ATLASGAL catalog which contains over 2,000 dense cores in the range of Galactic longitudes -60$^{\circ}$ to +20$^{\circ}$ \citep{2013PASA...30...57J}. The angular and spectral resolution of this survey are about 36$^{\prime\prime}$ and 0.11 km s$^{-1}$. The MALT90 data was obtained from the online archive\footnote{http://atoa.atnf.csiro.au/MALT90/}. N$_{2}$H$^{+}$(1-0), HNC(1-0), and HCO$^{+}$(1-0) emission lines were selected for the study in this research. At 15K, the corresponding critical densities are 1.5$\times$10$^{5}$, 2.9$\times$10$^{5}$ and 1.7$\times$10$^{5}$ cm$^{-3}$, respectively \citep{2014A&A...562A...3M}. The data were reduced using \caps{Gildas} (Grenoble Image and Line Data Analysis Software)\textbf{\footnote{https://www.iram.fr/IRAMFR/GILDAS/}}.

\subsection{The Spitzer surveys}
The Galactic Legacy Infrared Mid-Plane Survey Extraordinaire (GLIMPSE) survey is a mid-infrared survey (3.6, 4.5, 5.8, and 8.0 $\mu$m) of the Inner Galaxy performed with the \emph{Spitzer} Space Telescope. The angular resolution is better than 2$^{\prime\prime}$ at all wavelengths. GLIMPSE covers $5^{\circ}\leq|\ell|\leq65^{\circ}$ with $|b|\leq1^{\circ}$, $2\leq|\ell|<5^{\circ}$ with $|b|\leq1.5^{\circ}$, and $|\ell|<2^{\circ}$ with $|b|\leq2^{\circ}$. The MIPS/\emph{Spitzer} Survey of the Galactic Plane (MIPSGAL) is a survey of the same region as GLIMPSE at 24 and 70 $\mu$m, using the Multiband Imaging Photometer (MIPS) aboard the \emph{Spitzer} Space Telescope. The angular resolution at 24 and 70 $\mu$m is 6$^{\prime\prime}$ and 18$^{\prime\prime}$, respectively. The highly reliable point source catalogs \citep{2015AJ....149...64G} released from the MIPSGAL at 24 $\mu$m survey have been used in the following analysis.

\subsection{The Herschel Hi-GAL survey}
The \emph{Herschel} Infrared GALactic (Hi-GAL) plane survey is an Open Time Key Project on-board the ESA \emph{Herschel} Space Observatory \citep{2010A&A...518L...1P}, which mapped the Inner Galactic Plane at 70 and 160 $\mu$m with PACS \citep{2010A&A...518L...2P}, and 250, 350, and 500 $\mu$m with SPIRE \citep{2010A&A...518L...3G}. The angular resolutions are $\sim$6$^{\prime\prime}$, 12$^{\prime\prime}$, 18$^{\prime\prime}$, 25$^{\prime\prime}$, and 37$^{\prime\prime}$ for the five wavelength bands, respectively. The overall aim was to catalogue star-forming regions and study cold structures across the ISM. Using the broad spectral coverage available, the intention was to study the early phases of star formation, and particularly focus on providing an evolutionary sequence for the formation of massive stars within the Galactic Plane. The PACS data and SPIRE data used here are level 2.5 products produced by the \emph{Herschel} interactive processing environment (\caps{Hipe}) software, version 13.0 \citep{2010ASPC..434..139O}.

\section{Sample and classification}
We selected high-mass clumps with N$_{2}$H$^{+}$(1-0), HNC(1-0), and HCO$^{+}$(1-0) emission lines detected simultaneously from the MALT90 survey data, i.e.with a $S/N>3$, and searched for infall signatures using these three lines. To ensure the clumps in our sample are separated, and not contaminated by emission from an adjacent clump, each source with an angular separation of less than 36$^{\prime\prime}$ (the Mopra beam size at 90 GHz) from its nearest neighbour was excluded. HYX15 studied 405 high-mass clumps selected from the MALT90 survey (years 1 and 2) and identified 131 infall candidates; {in this work 327 high-mass clumps were selected from the remaining data from the MALT90 survey and an additional 100 infall candidates were identified} (see Section 4.1). In total we created a sample of 732 high-mass clumps, and identified 231 infall candidates. In Figure 1, we show the Galactic longitude, latitude and 870-$\mu$m integrated flux density distributions of the 732 clumps selected from the MALT90 Survey (blue histogram), and the corresponding distributions of all 8016 ATLASGAL sources located in the same region (filled grey histogram). Our sources show three peaks different from that of all ATLASGAL sources in longitude (10$^{\circ}$, -8$^{\circ}$, and -28$^{\circ}$), and three peaks different from that of all ATLASGAL sources in latitude (-0$^{\circ}$.3, 0$^{\circ}$.2 and 0$^{\circ}$.6), respectively. Moreover, our sample sources show obviously higher 870-$\mu$m integrated flux densities (see lower panel of Figure 1). \citet{2011ApJ...731...90D} suggests that the observed increase in flux density could be a reflection of an increasing dust temperature. These results suggest that selection bias exists in our sample. In fact, this is consistent with that we selected sources with N$_{2}$H$^{+}$(1-0), HNC(1-0), and HCO$^{+}$(1-0) emission lines (high-density tracers) detected simultaneously, i.e., we selected more dense and warm clumps.

\begin{figure}
  \begin{center}
  \includegraphics[width=0.45\textwidth]{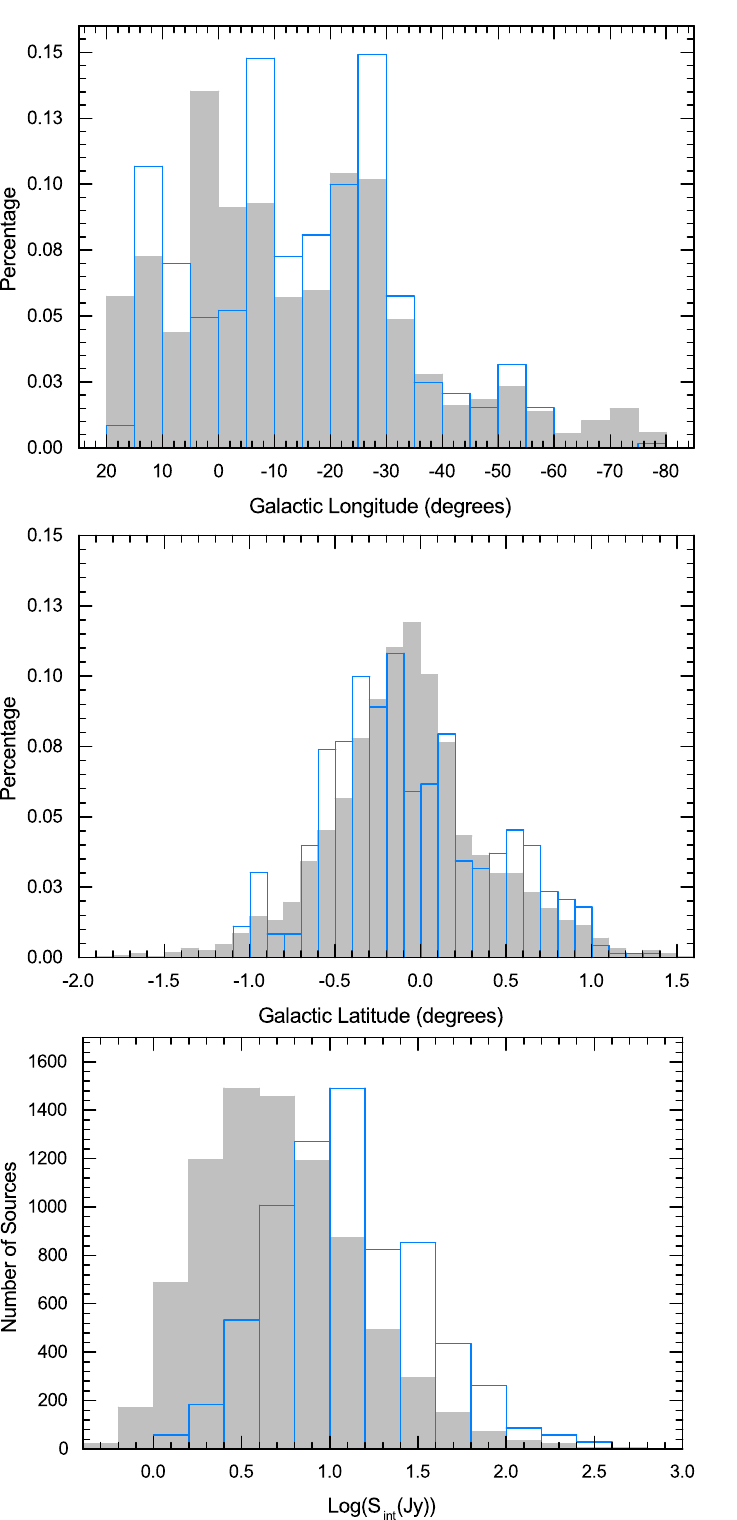}
  \end{center}
  \caption[dum]{Upper panel: Galactic longitude distribution of 8016 ATLASGAL sources (grey filled histogram) and 732 clumps in our sample (blue histogram). The bin size used is 5$^{\circ}$. Mid panel: Same as upper panel, but for Galactic latitude. The bin size used is 0$^{\circ}$.1. Lower panel: Same as upper panel, but for 870-$\mu$m integrated flux density distributions. The 732 clumps distribution has been scaled to the peak of the 8016 clumps distribution.}
\end{figure}

\citet[here after GAE15]{2015arXiv151100762G} classified the whole MALT90 survey clumps by visual inspection into four consecutive evolutionary stages: Quiescent (pre-stellar), proto-stellar, HII region, and photo-dissociation region (PDR). This evolutionary scheme is based on the \emph{Spitzer} 3.6, 4.5, 8.0 and 24$\mu$m images \citep[see][GAE15 for details]{2013ApJ...777..157H}. Figure 2 shows an example of a source at each evolutionary stage. Pre-stellar clumps appear dark at 3.6, 4.5, 8.0, 24 and 70 $\mu$m, and have no embedded high-mass young stellar objects. Clumps that have a 24 $\mu$m point source, 70 $\mu$m compact emission or are associated with extended 4.5 $\mu$m emission were categorized as proto-stellar. A clump was classified as HII if it was associated with a compact HII region(s) that emitted extensively in the 8.0 and 24 $\mu$m \emph{Spitzer} bands. Clumps classified as PDR are at a late evolutionary stage, where the ionizing radiation, winds and outflows provide feedback to the surrounding ISM and the expansion of the ionized gas finally disrupts the molecular envelope (show extended 8$\mu$m emission). This marks the transition to an observational stage characterized by an extended classical HII region and a photo-dissociation region. Taking into consideration that high-mass stars are usually formed in clusters \citep{2012ApJ...758L..28B}, and that the parental envelope of a clump in the PDR stage is finally disrupted, we cannot accurately discover the true parental clump for such evolved sources. Some nearby clumps, or those interacting with a classical HII region may be classified as PDR. In fact, it may be a pre-existing clump in an earlier evolutionary stage, which is common for bubble-like HII regions. We should keep in mind this caveat in the following analysis. Clumps named as ``uncertain" exhibit no clear mid-IR features that allow us to classify them unambiguously in above four categories. Our sample is a subsample of GAE15 except G303.268+01.247, G304.204+01.337 and G324.171+00.439.

\begin{figure}
  \begin{center}
  \includegraphics[width=0.48\textwidth]{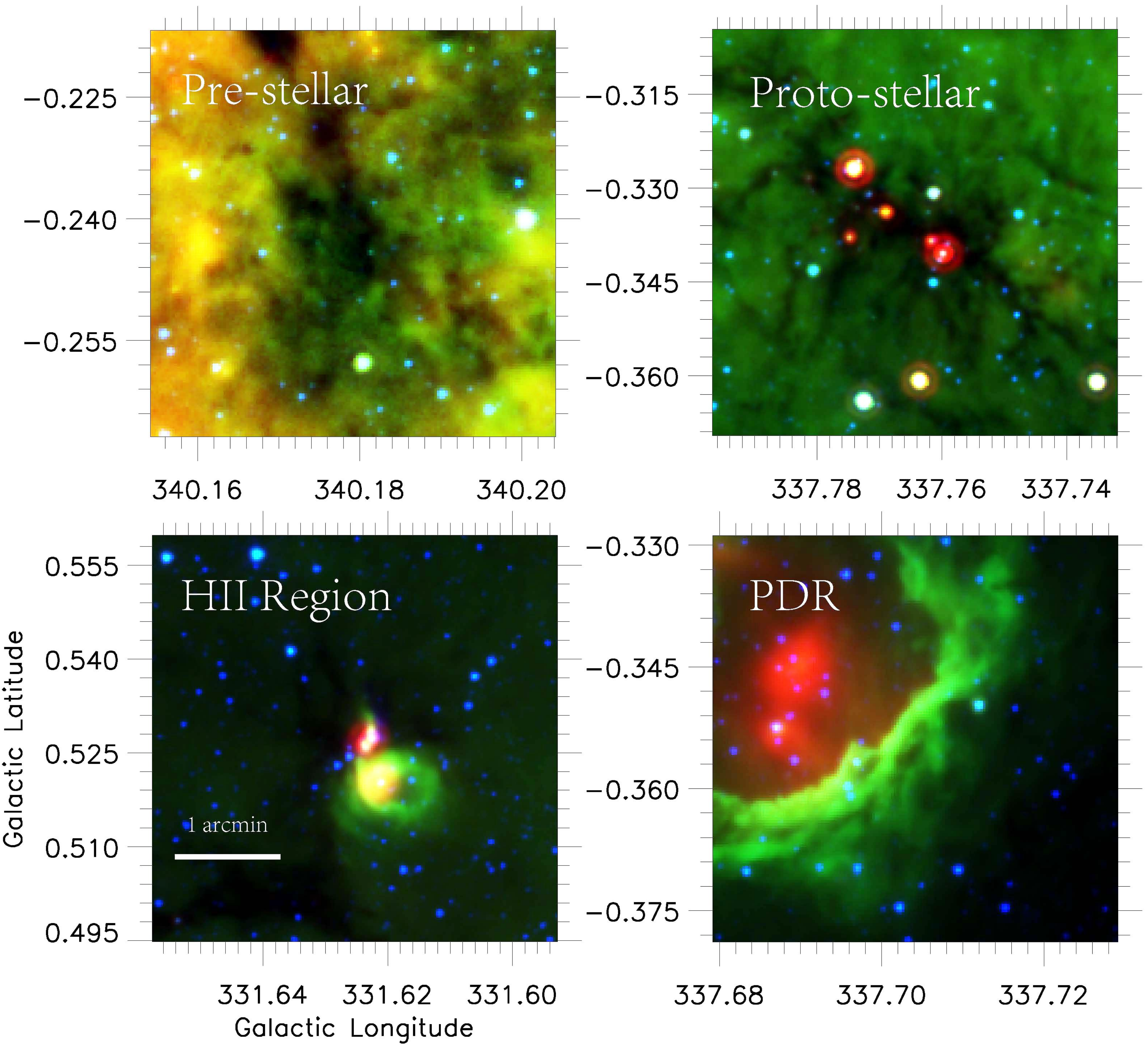}
  \end{center}
  \caption[dum]{Examples of IR clump classification. These three-color images show IRAC 3.6 $\mu$m as blue, IRAC 8.0 $\mu$m as green, and MIPS 24 $\mu$m as red.}
\end{figure}

For sources that were classified by visual inspection of \emph{Spitzer} image at 3.6, 4.5, 8.0, and 24$\mu$m (GAE15), this classification might not be completely reliable. Therefore, we checked the sources classified as "pre-stellar" and "uncertain" using the highly reliable 24 $\mu$m point source catalog \citep{2015AJ....149...64G}. A clump was reclassified as proto-stellar if it contained a 24 $\mu$m point source within a circular area of diameter 36$^{\prime\prime}$ centered on each clump. In this manner 15 pre-stellar and 14 uncertain clumps were classified as proto-stellar. We also checked the SIMBAD\footnote{http://simbad.u-strasbg.fr/simbad/} database and found that 36 sources with the pre-steller, PDR and uncertain classifications were associated with previously known HII regions. As such, 23 proto-stellar, 12 PDR, and one uncertain clumps were reclassified as HII. Note that G303.268+01.247, G304.204+01.337 and G324.171+00.439 were not in the sample of GAE15; here G303.268+01.247 and G304.204+01.337 are classified as uncertain, and G324.171+00.439 is classified as proto-stellar. The dust temperatures of these three clumps are derived in Section 4.2.

Finally, our sample includes 68 pre-stellar, 292 proto-stellar, 235 HII, 71 PDR, and 66 uncertain objects. Column 10 of Table 1 lists the evolutionary stages of all clumps in our sample.

\begin{table*}
\tiny
 \centering
  \begin{minipage}{177mm}
   \caption{Examples of the derived clump parameters. The columns are as follows: (1) Clump names; (2) peak submillimetre emission;
   (3) integrated submillimetre emission; (4) dust temperature; (5) heliocentric distance; (6) references and distance; (7) effective physical radius;
   (8) aspect ratios; (9) column density; (10) clump mass derived from the integrated 870$\,\umu$m emission; (11) virial mass; (12) volume density; and (13) \emph{Spitzer} classification.
   The full table is available online.}
    \begin{tabular}{rrrrrrrrrrrrr}
        \hline
        \multicolumn{1}{c}{Clump$^{\rm{a}}$} & \multicolumn{1}{c}{Peak flux} & \multicolumn{1}{c}{Int. flux} & \multicolumn{1} {c} {Temp.} & \multicolumn{2}{c}{\textbf{Distance}} & \multicolumn{1}{c}{Radius} &  \multicolumn{1}{c}{Aspect} & \multicolumn{1}{c}{Log(N(H$_2$))} & \multicolumn{1}{c}{Log(M$_{\texttt{clump}}$)} & \multicolumn{1}{c}{Log(M$_{\texttt{virial}}$)} & \multicolumn{1}{c}{Vol. density} & \multicolumn{1}{c}{\textbf{Mid-IR}}\\
        \multicolumn{1}{c}{name} & \multicolumn{1}{c}{(Jy beam$^{-1}$)} & \multicolumn{1}{c}{(Jy)} &  \multicolumn{1}{c}{(K)} & \multicolumn{1}{c}{(kpc)} & \multicolumn{1}{c}{Ref.} & \multicolumn{1}{c}{(pc)} & \multicolumn{1}{c}{Ratio} & \multicolumn{1}{c}{(cm$^{-2}$)} & \multicolumn{1}{c}{(M$_{\odot}$)} & \multicolumn{1}{c}{(M$_{\odot}$)}  & \multicolumn{1}{c}{$10^{4} cm^{-3}$} & \multicolumn{1}{c}{Classification}\\
        \multicolumn{1}{c}{(1)} & \multicolumn{1}{c}{(2)} & \multicolumn{1}{c}{(3)} & \multicolumn{1}{c}{(4)} & \multicolumn{1}{c}{(5)} & \multicolumn{1}{c}{(6)} & \multicolumn{1}{c}{(7)} & \multicolumn{1}{c}{(8)} & \multicolumn{1}{c}{(9)} & \multicolumn{1}{c}{(10)} & \multicolumn{1}{c}{(11)} & \multicolumn{1}{c}{(12)} & \multicolumn{1}{c}{(13)}\\
        \hline
      $*$G000.156$-$0.452    &    0.86   &    3.34  &  12.9 &  $-$      & $-$ &   $-$      &   4.20    &  22.64  &  $-$     &   $-$     &   $-$        &   Pre-stellar       \\
$*$G000.591$-$0.851    &    0.91   &    2.22  &  22.7 &  2.0      & 2   &   $-$      &   1.50    &  22.28  &  1.61    &   $-$     &   $-$        &   Pre-stellar       \\
$*$G010.199$-$0.316    &    1.59   &    3.59  &  24.1 &  6.0      & 2   &   $-$      &   2.17    &  22.48  &  2.74    &   $-$     &   $-$        &   Pre-stellar       \\
$*$G332.892$-$0.569    &    0.56   &    5.47  &  20.5 &  3.62     & 19  &   0.49     &   1.91    &  22.13  &  2.58    &   2.75    &  1.16        &   Pre-stellar       \\
$*$G340.179$-$0.242    &    0.71   &   15.46  &  15.4 &  12.01    & 19  &   3.26     &   1.89    &  22.43  &  4.27    &   3.84    &  0.19        &   Pre-stellar       \\
$*$G344.977$-$0.292    &    0.65   &    4.83  &  14.1 &  2.86     & 19  &   0.26     &   1.30    &  22.45  &  2.58    &   2.34    &  7.74        &   Pre-stellar       \\
$*$G348.383+0.537      &    0.61   &   10.59  &  11.4 &  1.29     & 19  &   0.18     &   1.73    &  22.59  &  2.39    &   1.92    &  15.06       &   Pre-stellar       \\
$*$G358.572$-$0.306    &    1.04   &   11.02  &  17.7 &  $-$      & $-$ &   $-$      &   1.74    &  22.50  &  $-$     &   $-$     &   $-$        &   Pre-stellar       \\
$*$G358.586$-$0.302    &    0.75   &    2.86  &  14.7 &  $-$      & $-$ &   $-$      &   1.00    &  22.48  &  $-$     &   $-$     &   $-$        &   Pre-stellar       \\
$*$G359.216$-$0.192    &    0.70   &    2.20  &  20.4 &  $-$      & $-$ &   $-$      &   1.08    &  22.23  &  $-$     &   $-$     &   $-$        &   Pre-stellar       \\
      \hline
     \end{tabular}
     \medskip
     $^{\rm{a}}$ Sources are named by galactic coordinates of ATLASGAL sources:  An $\ast$ indicates infall candidates.\\
  References --- Distance: (1) \citet{2004A&A...426...97F}, (2) \citet{2005MNRAS.363..405H}, (3) \citet{2012MNRAS.420.1656U}, (4) \citet{2013A&A...560A..76M}, (5) \citet{2013ApJS..206....9C},
  (6) \citet{2013MNRAS.431.1752U}, (7) \citet{2013ApJ...770...39E}, (8) \citet{2013A&A...559A..79R}, (9) \citet{2013A&A...550A..21S}, (10) \citet{2013ApJS..208...11L}, (11) \citet{2014ApJ...780...85V}, (12) \citet{2014MNRAS.443.1555U}, (13) \citet{2014MNRAS.437.1791U}, (14) \citet{2014ApJS..212....1A},
  (15) \citet{2014A&A...570A..65G}, (16) \citet{2015MNRAS.451.3089T}, (17) IRDC, (18) tangential point, (19) this paper, (20) $z>$120pc .\\

  \end{minipage}
\end{table*}
\normalsize

\section{Results}
\subsection{Infall candidates}
The asymmetry parameter, $\delta V =(V_{thick}-V_{thin})/dV_{thin}$ \citep[pioneered by][]{1997ApJ...489..719M}, has been used to quantify the asymmetry of an optical thick molecular line, where $V_{thick}$ is the peak velocities of an optically thick line, and $V_{thin}$ is an optically thin line velocity in units of the optically thin line full width at half-maximum $dV_{thin}$. A line was considered to be a blue skewed profile if $\delta$V $<$ -0.25, and to be red skewed profile if $\delta$V $>$ 0.25. The combination of the optically thick lines (HCO$^{+}$ $J$=1-0 and HNC $J$=1-0) and the optically thin line (N$_{2}$H$^{+}$ $J$=1-0\emph{}) can serve as good tracer of infall motions (e.g. HYX15). The optically thin line N$_{2}$H$^{+}$(1-0) shows hyperfine structure (hfs) and was fitted using the hfs method in CLASS\footnote{https://www.iram.fr/IRAMFR/GILDAS/} to obtain the peak velocity and full width at half maximum (FWHM). All of these parameters for each source are listed in Table 2. For 327 clumps, 68 red skewed profiles were identified using the HCO$^{+}$(1-0) lines, 39 red skewed profiles were identified using the HNC(1-0) lines, and 30 sources show red skewed profiles in both the HCO$^{+}$(1-0) and HNC(1-0) spectra.

Following the criterion used in HYX15, infall candidates must show a blue skewed profile at least in one optically thick line (HCO$^{+}$ $J$=1-0 or HNC $J$=1-0), no red skewed profile in the other optically thick line and no spatial difference in the mapping observation. Finally, we identified 100 infall candidates from 327 high-mass clumps. In Figure 3, we show a typical example of an infall candidate: G316.139-0.506. Infall candidates are marked with a asterisk in Table 2, and the corresponding profile asymmetries are also given in the table. Combined with the 131 infall candidates in HYX15, we now have 231 infall candidates in total. Among of them, 28 are pre-stellar, 107 are proto-stellar, 72 are HII, 9 are PDR, and 15 are uncertain. The detection rates of the infall candidates are 41.2\%, 36.6\%, 30.6\% and 12.7\%, respectively. This shows that the detection rates decrease as a function of evolutionary stage.

Furthermore, we used 1-dimensional spherically symmetric RATRAN models\footnote{http://home.strw.leidenuniv.nl/~michiel/ratran/} \citep{2000A&A...362..697H} of the HCO$^{+}$ and HNC lines to constrain the spatial scales on which infall takes place. Assuming a free-fall velocity profile \citep[e.g.][]{2013A&A...558A.126M}, a typical radial density and temperature profiles with power index 1.5, we vary the radii over which infall takes place. We assume abundances for HCO$^{+}$ and HNC of 2.51$\times$10$^{-8}$ and 3.73$\times$10$^{-8}$, which are typical for massive clumps \citep{2012ApJ...756...60S}. We found that our observations for the ground-state lines of HCO$^{+}$ and HNC are sensitive to radii $<$133000 and 5600 $-$ 28000 AU at a typical distance of 3.5kpc, respectively. The corresponding densities are 467 and 3063 cm$^{-3}$ and temperatures 4.5 and 7.3 K for HCO$^{+}$(1-0) and HNC(1-0), respectively. For the ground-state line of HCO$^{+}$, emissions from the outer regions of central infalling zone within one Mopra beam have little effects in the emission line profile. Therefore, the minimum radius was not given by the RATRAN model. Overall, these two lines are both good tracers of infall motion in envelope material.

\begin{figure}
  \begin{center}
  \includegraphics[width=0.45\textwidth]{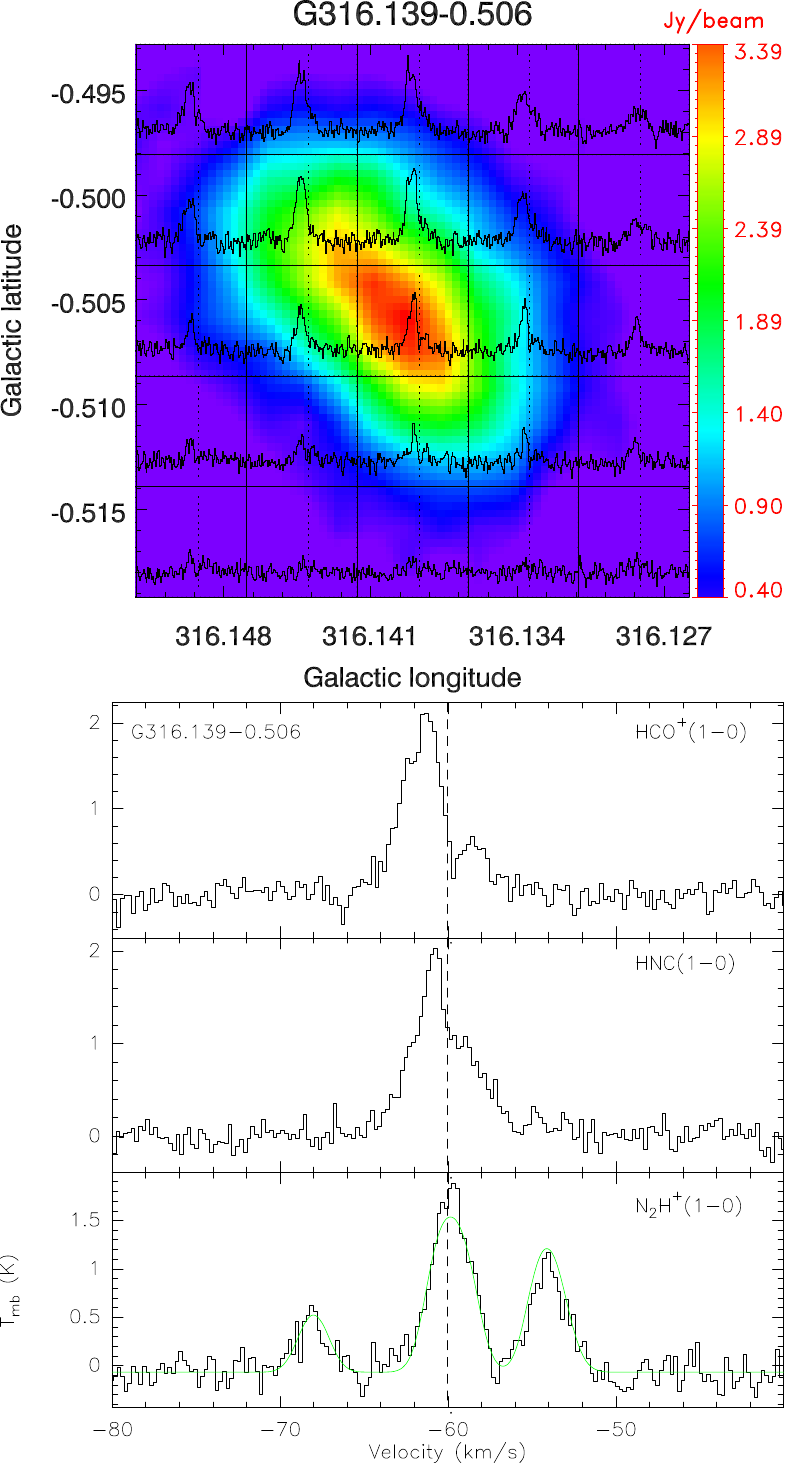}
  \end{center}
  \caption[dum]{Example of an infall source G316.139-0.506. Upper panel: the HCO$^{+}$(1-0) map grid (gridded to 1/2 beam size) superposed on the 870 $\mu$m
  continuum emission map (starting from a flux density of 0.4 Jy beam$^{-1}$, which corresponds to a peak flux above 6$\sigma$). Lower panel: the extracted
  spectra of HCO$^{+}$(1-0), HNC(1-0) and N$_{2}$H$^{+}$(1-0) from the central position of this clump. The dashed lines on the
  profiles indicate the velocity of N$_{2}$H$^{+}$(1-0).}
\end{figure}

The blue excess $E=(N_{blue}-N_{red})/N_{total}$ was used to quantify whether blue profiles dominate in a given sample, where $N_{blue}$ and $N_{red}$ are the number of sources that show a blue or red profile, respectively, and $N_{total}$ is the total number of sample sources \citep{1997ApJ...489..719M}. The blue excess $E$ values of the pre-stellar, proto-stellar, HII and PDR classifications are 0.29, 0.19, 0.09 and 0.06, respectively. These values suggest that red skewed profiles show an increased fraction for the later stages, suggesting that the star formation is contributing to removing the surrounding clump/cloud material. Also, the detection rate of infall candidates decrease as clumps evolve, which can be ascribed to increasingly stronger feedback from the central star. In order to show the blue excess more clearly, we plotted the $\delta$V values of the HNC and HCO$^{+}$ distributions in Figure 4.

\begin{figure}
  \begin{center}
  \includegraphics[width=0.45\textwidth]{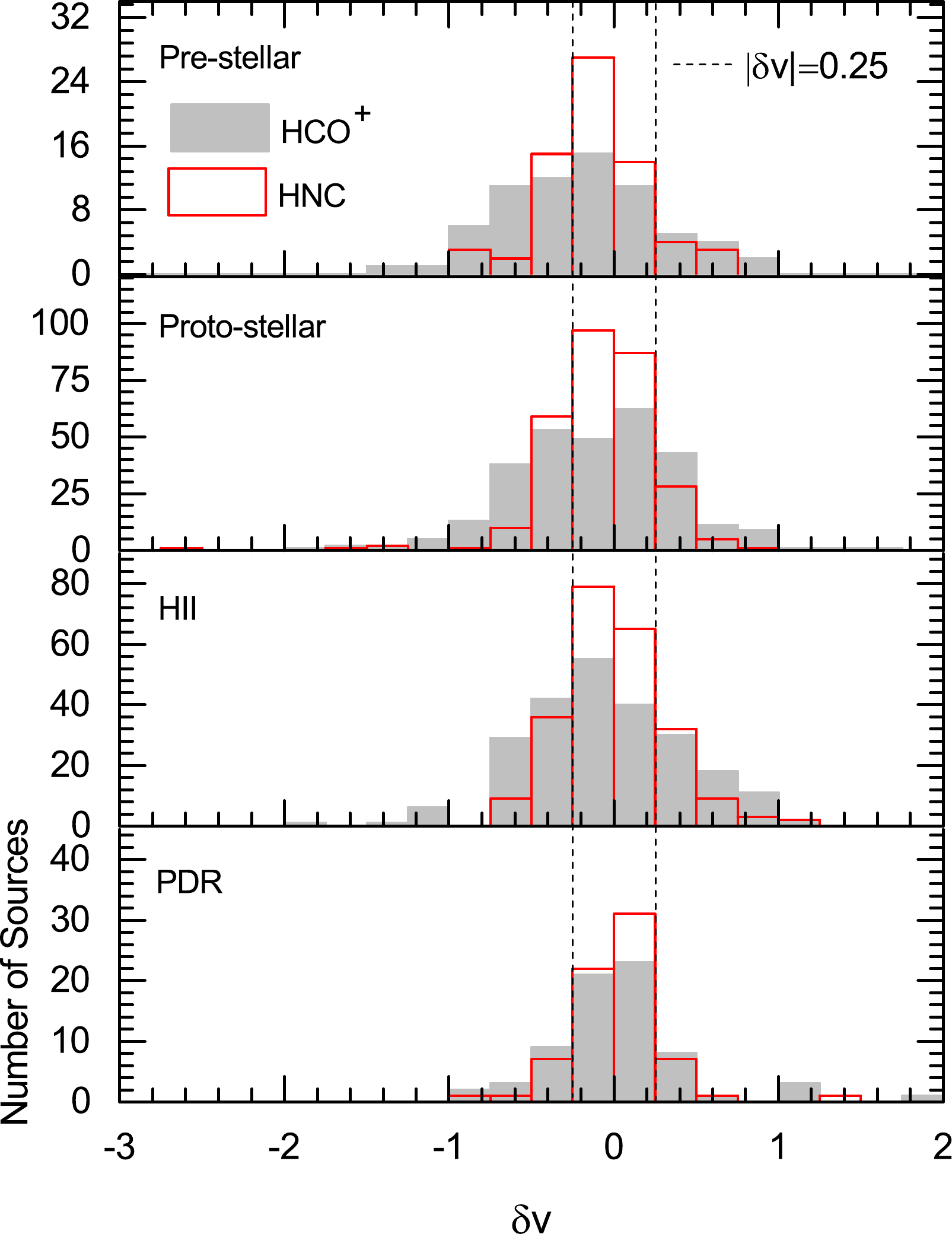}
  \end{center}
  \caption[dum]{The histogram of $\delta$$v$ for HNC(1-0) (solid red lines) overplotted on that of HCO$^{+}$(1-0) (shaded bars) for all classified sources separated by evolutionary stage. The corresponding stage are given on the top left of each panel. Dashed lines indicate $|$$\delta$$v$$|$=0.25.}
\end{figure}

\begin{table*}
\scriptsize
 \centering
  \begin{minipage}{155mm}
   \caption{Examples of the derived line parameters and profiles of the observed sources. Quantities in parentheses give the uncertainties in units of 0.01. The columns
   are as follows: (1) Clump names; (2) peak velocity of HCO$^{+}$(1-0); (3) peak velocity of HNC(1-0); (4) peak velocity of N$_{2}$H$^{+}$(1-0); (5) FWHM of N$_{2}$H$^{+}$(1-0);
   (6) asymmetry of HCO$^{+}$(1-0); (7) asymmetry of HNC(1-0); (8) profile of HCO$^{+}$(1-0) and HNC(1-0). The full table is available online. }
    \begin{tabular}{rrrrrrrrrrrrrrrrrrrrrrrr}
        \hline
        \multicolumn{1}{c}{Clump$^{\rm{a}}$} & \multicolumn{1}{c}{$V_{thick}$\,} & \multicolumn{1}{c}{$V_{thick}$\,} & \multicolumn{1}{c}{$V_{thin}$\,}
        & \multicolumn{1} {c} {$\Delta V$\,} & \multicolumn{1}{c}{$\delta v$\,} & \multicolumn{1}{c}{$\delta v$\,} & \multicolumn{1}{c}{Profile} \\
        \multicolumn{1}{c}{name} & \multicolumn{1}{c}{$HCO^{+}(1-0)$\,} & \multicolumn{1}{c}{$HNC(1-0)$\,)} & \multicolumn{1}{c}{$N_{2}H^{+}(1-0)$\,}
        & \multicolumn{1}{c}{$N_{2}H^{+}(1-0)$\,} & \multicolumn{1}{c}{$HCO^{+}(1-0)$\,} & \multicolumn{1}{c}{$HNC(1-0)$\,} & \\
        & \multicolumn{1}{c}{$km\ s^{-1}$\,} & \multicolumn{1}{c}{$km\ s^{-1}$\,} & \multicolumn{1}{c}{$km\ s^{-1}$\,} & \multicolumn{1}{c}{$km\ s^{-1}$\,} & & & \\
        \multicolumn{1}{c}{(1)} & \multicolumn{1}{c}{(2)} & \multicolumn{1}{c}{(3)} & \multicolumn{1}{c}{(4)} & \multicolumn{1}{c}{(5)} & \multicolumn{1}{c}{(6)}
        & \multicolumn{1}{c}{(7)}  & \multicolumn{1}{c}{(8)}\\
        \hline
      $*$G000.156$-$0.452  &   16.10(04)    &    16.19(09)      &    17.04(04)    &  2.08(09)  &  $-$0.45   &   $-$0.41   &  B,B   \\   
$*$G000.591$-$0.851  &   16.36(04)    &    16.83(05)      &    17.30(04)    &  2.58(07)  &  $-$0.36   &   $-$0.18   &  B,N   \\   
$*$G010.199$-$0.316  &    8.55(10)    &     9.15(08)      &    11.18(07)    &  4.13(11)  &  $-$0.64   &   $-$0.49   &  B,B   \\   
$*$G332.892$-$0.569  &  $-$57.96(05)  &   $-$57.44(09)    &   $-$57.03(07)  &  2.13(13)  &  $-$0.44   &   $-$0.19   &  B,N   \\   
$*$G340.179$-$0.242  &  $-$54.39(07)  &   $-$53.63(08)    &   $-$53.27(08)  &  3.38(14)  &  $-$0.33   &   $-$0.11   &  B,N   \\   
$*$G344.977$-$0.292  &  $-$29.70(08)  &   $-$28.50(08)    &   $-$28.02(05)  &  1.86(14)  &  $-$0.90   &   $-$0.26   &  B,B   \\   
$*$G348.383+0.537    &   $-$7.85(02)  &    $-$7.55(02)    &    $-$7.24(03)  &  1.27(08)  &  $-$0.48   &   $-$0.24   &  B,N   \\   
$*$G358.572$-$0.306  &   $-$5.41(07)  &    $-$4.87(05)    &    $-$4.26(04)  &  1.80(10)  &  $-$0.64   &   $-$0.34   &  B,B   \\   
$*$G358.586$-$0.302  &   $-$5.66(05)  &    $-$4.11(10)    &    $-$4.48(06)  &  2.74(11)  &  $-$0.43   &    0.14     &  B,N   \\   
$*$G359.216$-$0.192  &   15.05(05)    &    15.32(05)      &    15.55(04)    &  0.86(12)  &  $-$0.58   &   $-$0.27   &  B,B   \\   
      \hline
     \end{tabular}
     \medskip
      $^{\rm{a}}$ Sources are named by galactic coordinates of ATLASGAL sources:  An $\ast$ indicates infall candidates.\\
  NOTE. The HCO$^{+}$(1-0), and HNC(1-0) profiles are evaluated as follows: B denotes a blue profile, R denotes a red profile, and N denotes neither blue nor red.\\
   \end{minipage}
\end{table*}
\normalsize

\subsection{Dust temperatures}
The dust temperatures of the objects in our sample were all taken from GAE15, except for G303.268+1.247, G304.204+1.337 and G324.171+0.439. Since the sample of GAE15 do not include these three sources, we derived their temperatures ourself. Combining the \emph{Herschel} Hi-GAL observations at 70, 160, 250, 350, and 500 $\mu$m, we obtained the dust temperature of these three sources via spectral energy distribution (SED) fitting. Taking into consideration that dust emission is not optically thick at all wavelengths, and that the dust emissivity index is not a fixed value \citep{2012MNRAS.426..402F}, we adopt a modified blackbody model \citep{1990MNRAS.244..458W}
\begin{equation}
\rm F_{\nu}=\Omega\,B_{\nu}(T)(1-e^{-\tau})\,\,\rm{[Jy]},
\end{equation}
where $\Omega$ is the source solid angle, B$_{\nu}$$(T)$ is the Planck function at the dust temperature T, and $\tau$ is the optical depth. We calculated $\tau$ using the relation
\begin{equation}
\rm \tau$=$\left(\nu/\nu_{c}\right)^{\beta},
\end{equation}
where $\beta$ is the dust emissivity index, and $\nu_{c}$ is the critical frequency at which $\tau\,=\,1$. The free parameters in the fits were the dust temperature ,T, the dust emissivity index, $\beta$, and the critical frequency, $\nu_{c}$. We assigned an uncertainty of 20\% for each \emph{Herschel} flux, which was taken as the calibration error \citep{2012MNRAS.426..402F}. The quality of an SED fit was assessed using $\chi^{2}$ minimisation, by considering the observed flux at each of the five Hi-GAL wavebands available for every individual association.

Figure 5 shows the fitting results with the modified blackbody model for G303.268+1.247, G304.204+1.337 and G324.171+0.439, where the dust temperature, dust emissivity index and critical frequency of each source are given at the top right of each panel. The dust temperatures of these three sources are 19.0K, 19.5K and 22.0K, respectively. We also derived the dust temperature, dust emissivity index and critical frequency for all other clumps in our sample using the method. These results are listed in Table A1 in Appendix A in this paper. By comparing temperatures derived by us and GAE15, we found that our temperatures are systematically a few K higher (see Appendix A). The temperatures derived by GAE15 are more consistent with previous results \citep[e.g.][]{2013ApJ...777..157H}, so we prefer the temperatures of GAE15 for 729 sources in this paper except G303.268+1.247, G304.204+1.337 and G324.171+0.439.

\begin{figure}
  \begin{center}
  \includegraphics[width=0.4\textwidth]{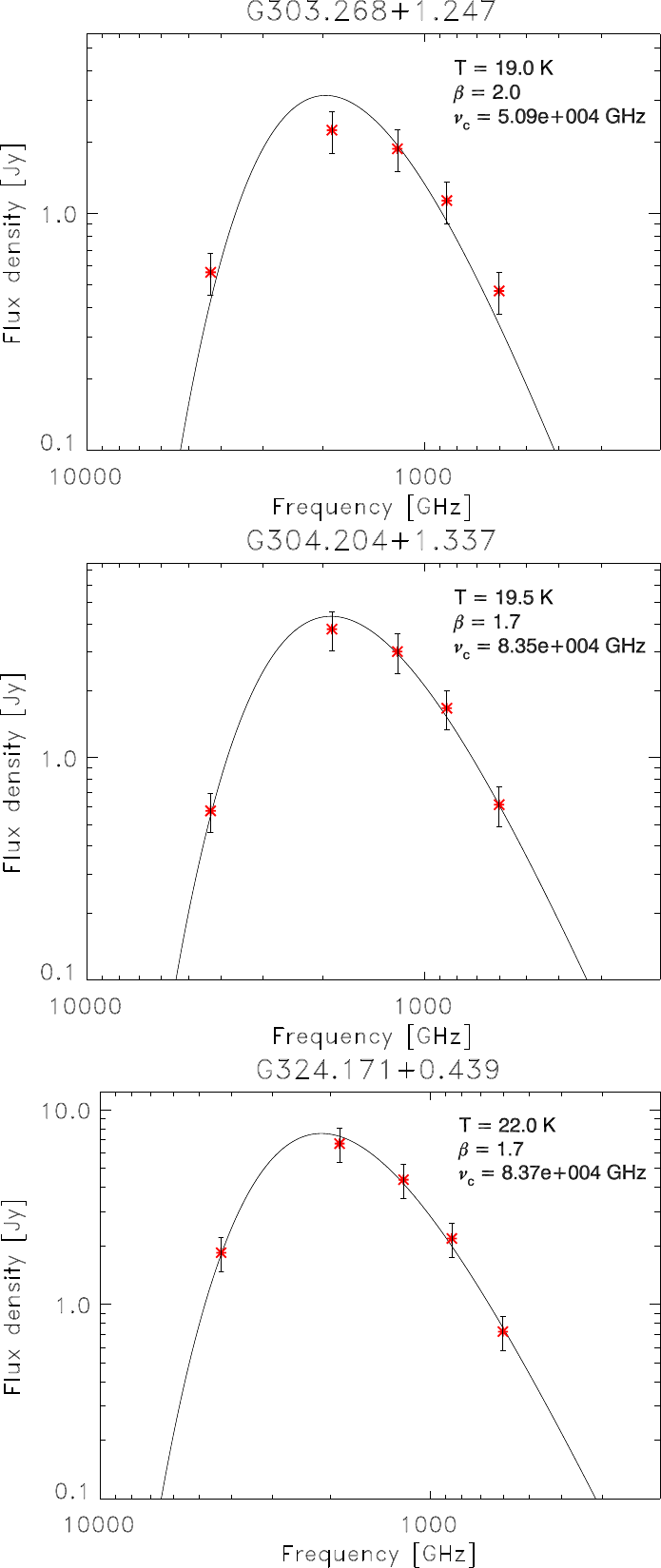}
  \end{center}
  \caption[dum]{Fitted results of G303.268+1.247, G304.204+1.337 and G324.171+0.439 using a modified blackbody. The black solid line represents the best fit to the data.
  The dust temperature, dust emissivity index and critical frequency of each sources are given on the top right of each panel.}
\end{figure}

The dust temperatures of the infall candidates and clumps where infall is not detected at different evolutionary stages are shown in Figure 6. It is clear that the median temperature increases monotonically with the evolutionary stage of the infall candidates (blue histograms) and clumps where infall is not detected (gray filled histograms), although there is overlap between the different categories. The median values of the dust temperature for the clumps where infall is not detected at pre-stellar, proto-stellar, HII and PDR stages are 17.4, 18.3, 22.8 and 25.7 K, respectively. The corresponding mean values are 18.6, 18.8, 23.6 and 26.3 K. The median value of the dust temperature for the pre-stellar, proto-stellar, HII and PDR clumps with infall motions are 17.0, 17.2, 22.1 and 26.1 K, respectively. The corresponding mean values are 17.2, 17.5, 22.2 and 28.3 K. The statistical parameters of these distributions are summarized in Table 4. The Kolmogorov-Smirnov (K-S) test is a robust method for measuring the similarity between two samples. Throughout this paper, we used the K-S test to quantify the similarity between two samples. The K-S test gives a probability of 49\% that the dust temperature distributions of infall candidates and clumps where infall is not detected originate from the same parent population for pre-stellar, the corresponding values are 0.03\%, $\ll$0.1\%, and 0.05\% for proto-stellar, HII, and PDR, respectively. There is no significant difference in temperature between the infall candidates and clumps where infall is not detected in pre-stellar stage. Since we are tracing infall on large scales and there is no detected protostar yet, the lack of difference may be genuine if infall is not yet translated into accretion onto a central object (which would result in shocks/radiation which heat the gas), or if this is too small an effect to be detected. Although K-S test suggests that the temperature distributions of infall candidates are significantly different from that of clumps where infall is not detected for proto-stellar and HII phases, the corresponding median and mean values show no obvious differences.

\begin{figure}
  \begin{center}
  \includegraphics[width=0.45\textwidth]{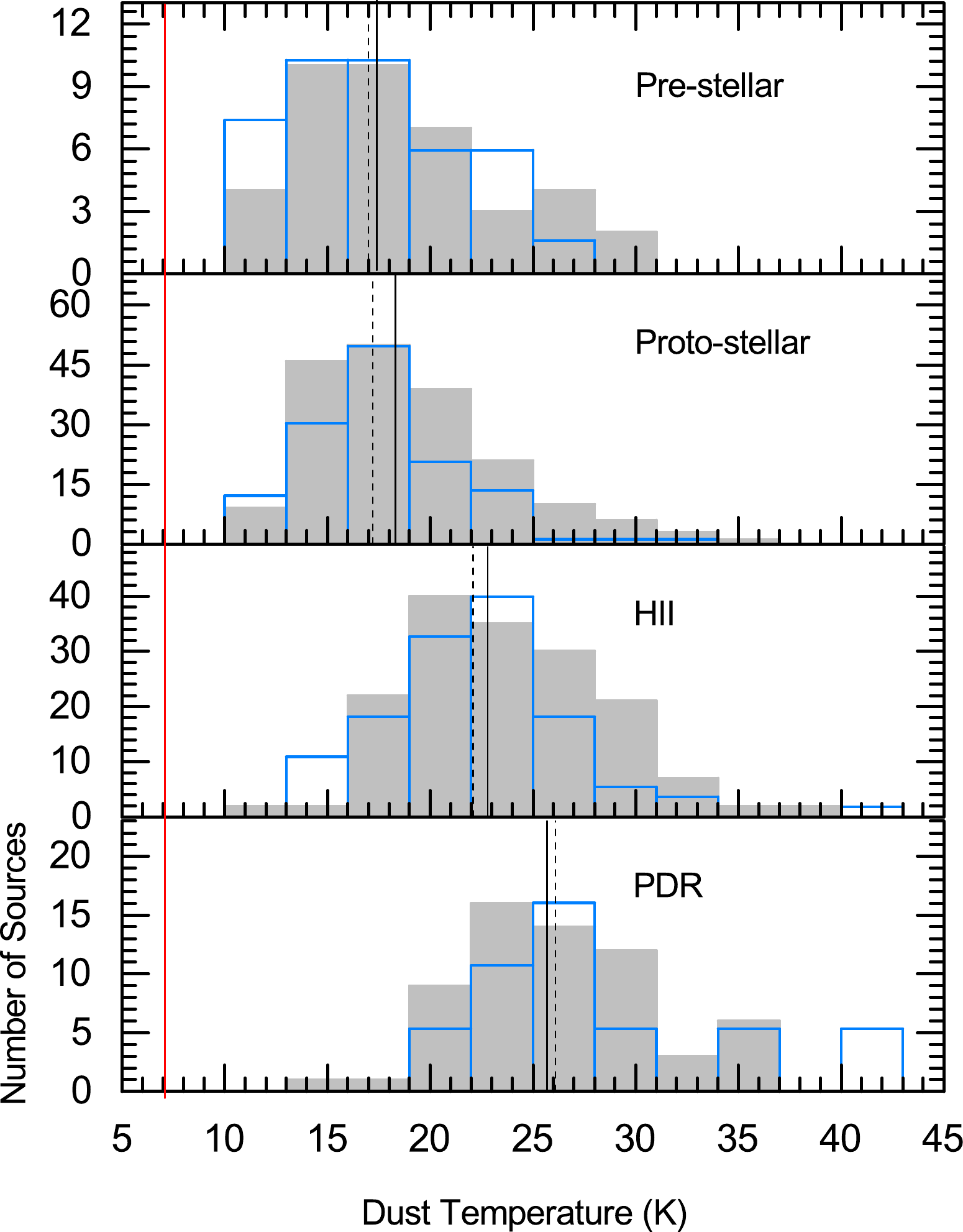}
  \end{center}
  \caption[dum]{Dust temperature distribution derived from \emph{Herschel}/Hi-GAL data separated by evolutionary stages. The median temperature for infall candidates (blue histogram)
  and clumps where infall is not detected (gray filled histogram) in each stage are indicated by the dashed vertical black line and the solid vertical black line, respectively.
  The corresponding stages are given on the top right of each panel. The infall candidates distribution has been scaled to the peak of the infall is not detected clump
  distribution. The solid red line indicates the dust temperature sensitivity limit.}
\end{figure}

\subsection{Kinematic distances}
We used the revised rotation curve of the Milky Way presented by \citet{2009ApJ...700..137R} to calculate the kinematic distances of 327 clumps selected in this work. The system velocity was determined via an hfs fit to the N$_{2}$H$^{+}$(1-0) line. Distances to 94 clumps have already been determined in the literature (see column 6 in Table 1). For the other 233 clumps, most of them lie inside the solar circle and so have both simultaneous near and far distances (so called kinematic distances ambiguity).

In order to resolve this ambiguity, we compared our sources with the IRDC catalogue \citep{2009A&A...505..405P}. IRDCs are seen as dark extinction feature against the bright mid-infrared background emission. We assumed that a clump is located at the near distance if it coincided with an IRDC along the same line of sight. Moreover, if a clump is located at the near distance, a cold and dense HI target clump will absorb warmer HI background continuum emission, and show self-absorption at the system velocity, whereas any clumps at the farther distance will not display this signature \citep{2006MNRAS.366.1096B}. The HI data cubes were obtained from the Southern Galactic Plane Survey (SGPS) archive\footnote{http://www.atnf.csiro.au/research/HI/sgps/queryForm.html} \citep{2005ApJS..158..178M}. When a clump is located near the tangent points, it will have similar near and far distances. Clumps are also considered at the near distance if a far allocation displacement from the Galactic mid-place, $z$, would result in a value larger than 120pc, which is inconsistent with star formation regions \citep{2014MNRAS.437.1791U}. Using these methods, we resolved the kinematic distances ambiguity and determined the kinematic distance to 193 clumps. Note that kinematic distances were not determined for 40 clumps towards the direction of the galactic center (within five degrees of the Galactic center direction).

Figure 7 shows two examples where we resolved the kinematic distance ambiguity using the HI self-absorption technique. The upper panel represents an unambiguous near-distance solution, where the N$_{2}$H$^{+}$ emission line coincides with the dip of HI absorption. The lower panel displays a far distance solution, where the N$_{2}$H$^{+}$ emission line does not coincide with the HI self-absorption dip. The N$_{2}$H$^{+}$ and overlaid HI spectral profiles of all 148 clumps are available online-only as supplementary material.

\begin{figure}
  \begin{center}
  \includegraphics[width=0.45\textwidth]{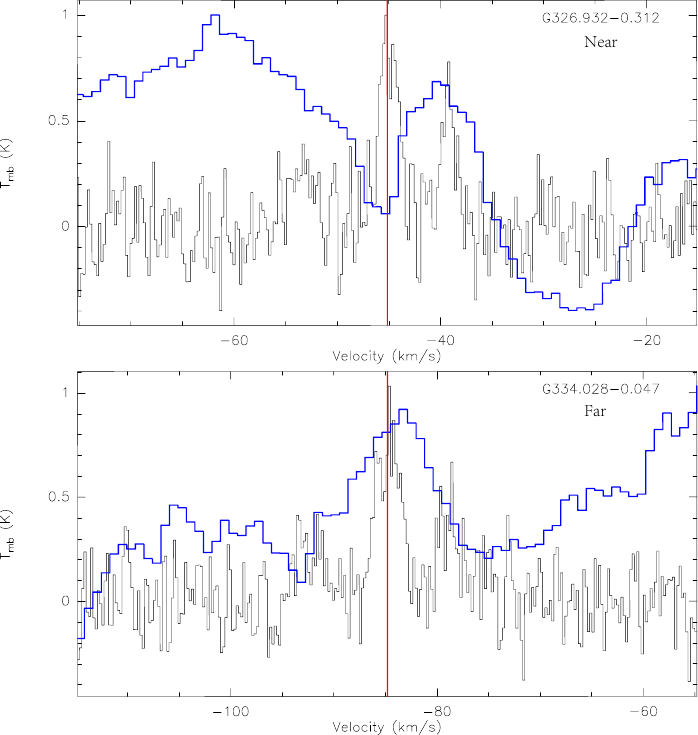}
  \end{center}
  \caption[dum]{Two sources located at the near and far distances, their kinematic distances ambiguity was resolved by HI self-absorption. The N$_{2}$H$^{+}$(1-0) spectra (solid black lines) overlaid with the HI 21-cm spectra (solid blue lines), with the HI data scaled to the peak of N$_{2}$H$^{+}$(1-0). The vertical solid red line indicates the velocity of the clump.
  HI spectra extracted from the SGPS archive.}
\end{figure}

Figure 8 shows the distribution of infall candidates (blue histogram) and clumps where infall is not detected (grey filled histogram) as a function of heliocentric distance. The K-S test gives a probability of 10\% that infall candidates and clumps where infall is not detected distributions originate from the same parent population. The two samples show a similar distribution. Figure 9 presents the distribution of infall candidates (blue pentacles) and clumps where infall is not detected (white squares) with known kinematic distances projected on an artist's impression of the Milky Way (R. Hurt: NASA/JPL-Caltech/SSC). It seems that infall candidates usually concentrate on the spiral arms of the Galaxy, while the distribution of clumps where infall is not detected is more scattered.

In Figure 10, we plotted the derived values of the effective radius, R = D$\theta$ for infall candidates (red cross) and clumps where infall is not detected (black point) as a function of distance. Here D is the heliocentric distance to the source and $\theta$ is diameter of the source in radian. We assumed a $\sim$10 percent error in the distance. In this plot, the dotted line indicates roughly the limiting size (19.2$^{\prime\prime}$ = 1 FWHM is chosen as a guide) to which the survey is sensitive. Using a K-S test, we calculated a probability of 10\% that the kinetic distance distributions of the infalling clumps and those where infall is not detected, and 0.36\% that the effective radius distributions of infall candidates and clumps where infall is not detected originate from the same parent populations.

\begin{figure}
  \begin{center}
  \includegraphics[width=0.45\textwidth]{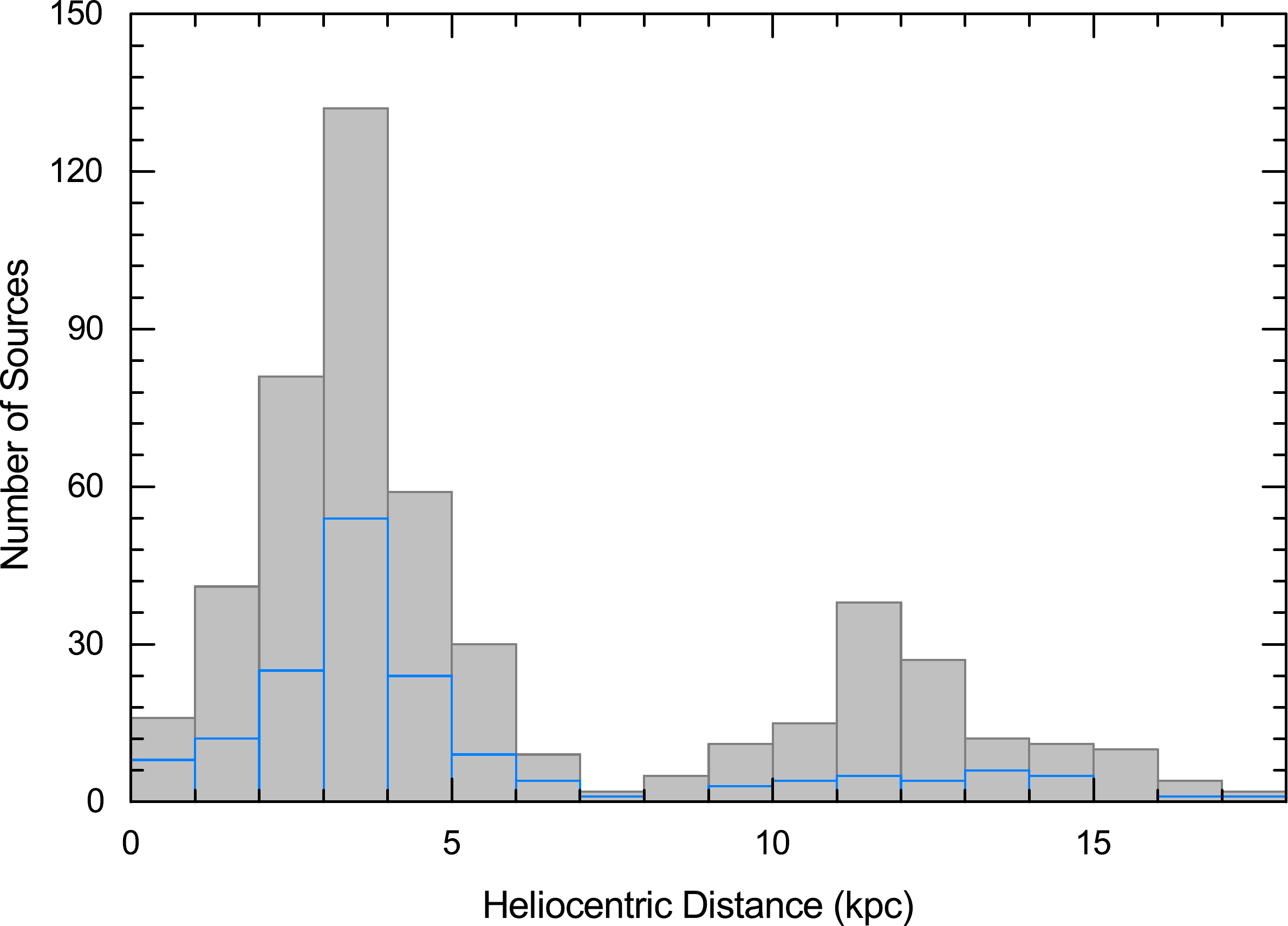}
  \end{center}
  \caption[dum]{Heliocentric distance distribution for the infall candidates (blue histogram) and the clumps where infall is not detected (grey filled histogram). The bin size is 1 kpc.}
\end{figure}

\begin{figure}
  \begin{center}
  \includegraphics[width=0.45\textwidth]{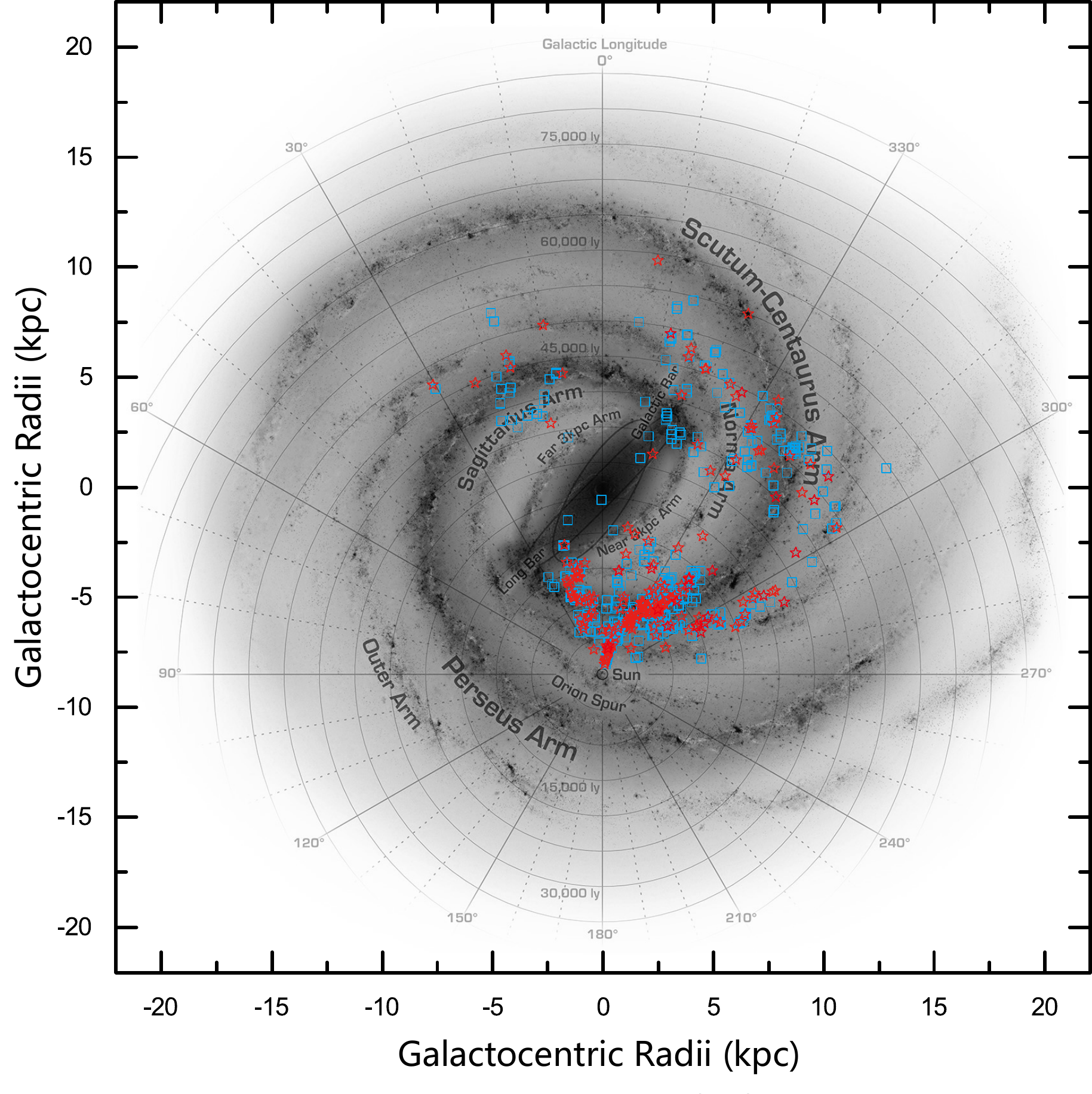}
  \end{center}
  \caption[dum]{Galactic distribution of the infall candidates and clumps where infall is not detected with known distances. We show the kinematic position of infall candidates as red pentacles and clumps where infall is not detected as cyan squares. Our source distribution overlaid on the central part of the informed artist impression of the Milky Way (R. Hurt: NASA/JPL-Caltech/SSC).}
\end{figure}

\begin{figure}
  \begin{center}
  \includegraphics[width=0.45\textwidth]{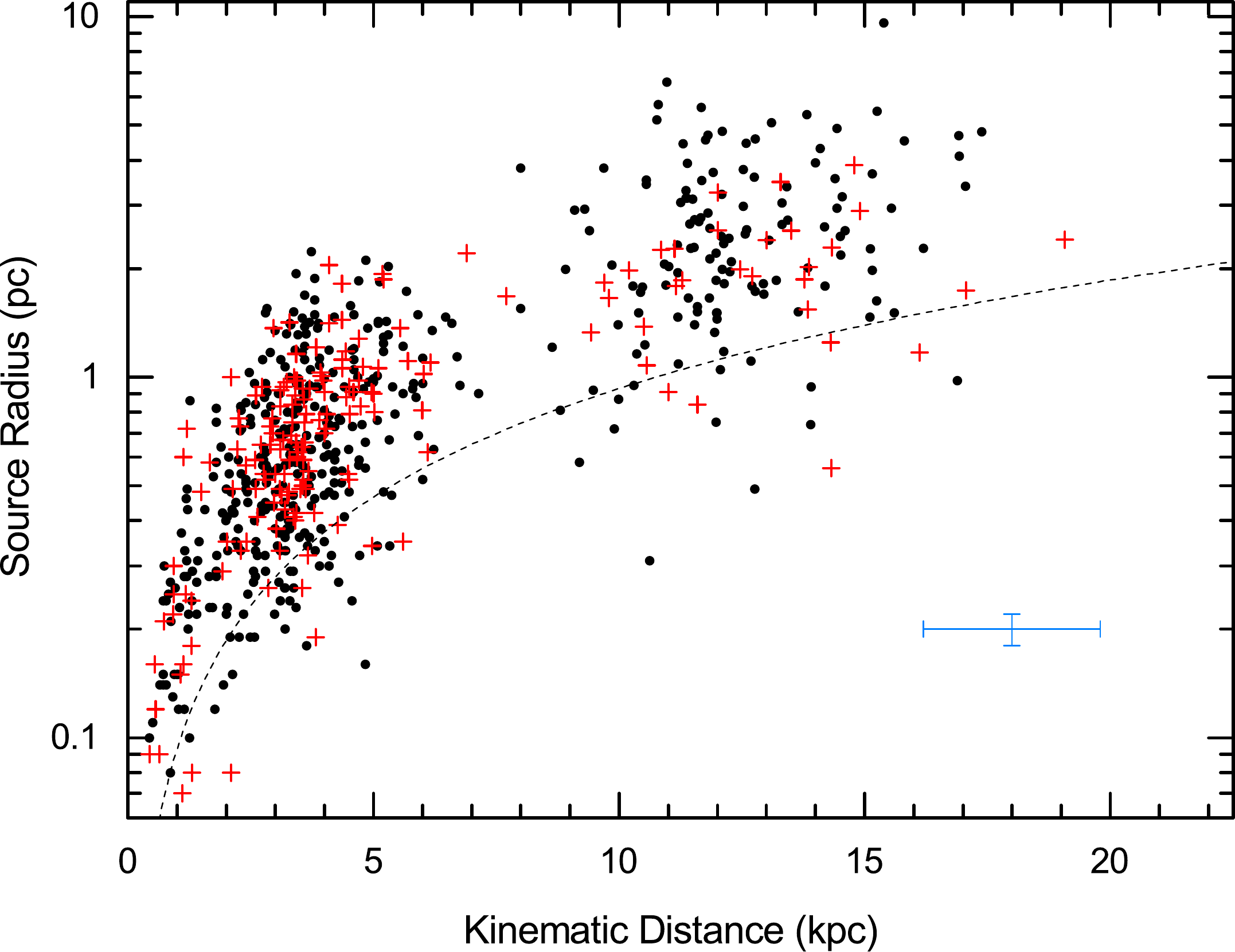}
  \end{center}
  \caption[dum]{The sources radius distribution as a function of heliocentric distance for the infall candidates (red cross) and the clumps where infall is not detected (black point). The dashed curved line indicates the physical resolution of the survey based on the APEX 19.2 arcsec beam at 870 $\mu$m Characteristic error bars for these parameters are shown in the lower-right corner of the plot.}
\end{figure}

\subsection{Aspect ratio}
In Figure 11, we plotted the derived values of the aspect ratios as functions of heliocentric distance. We found that, for sources with distances larger than 8.5 kpc, the infall candidates tend to have relatively smaller aspect ratios. Figure 12 shows the distribution of the aspect ratio for all infall candidates (blue histogram) and all clumps where infall is not detected (gray filled histogram). The median aspect ratio of the infall candidates (1.50) is smaller than that of the clumps where infall is not detected (1.57). The K-S test gives a probability of $\ll$0.01\% that the aspect ratio distributions of the infall candidates and clumps where infall is not detected originate from the same parent population, suggesting that the aspect ratios of infall candidates are significantly different from that of clumps where infall is not detected. The median aspect ratio values of infall candidates and clumps where infall is not detected that we obtained are larger than the values of 1.33 in HII-region associated clumps and 1.4 in methanol-maser associated clumps reported by \citet{2013MNRAS.435..400U}.

The aspect ratios of the infall and infall is not detected sources separated by evolutionary stage are shown in upper panel of Figure 13. The median values of the aspect ratios of the pre-stellar, proto-stellar, HII and PDR clumps with infall are 1.65, 1.53, 1.38 and 1.58, respectively. The corresponding mean values are 1.76, 1.59, 1.51 and 1.77. While the median values of the aspect ratios for the clumps where infall is not detected at the pre-stellar, proto-stellar, HII and PDR stages are 1.72, 1.53, 1.52 and 1.70, respectively. The corresponding mean values are 1.85, 1.64, 1.60 and 1.85. These results show that there are no obvious differences in the aspect ratios of the infall candidates and the clumps where infall is not detected in all evolutionary stages, except for HII. The aspect ratio distribution of clumps in the HII stage reveals significant difference between the infall candidates and clumps where infall is not detected, which indicates that infall clumps in the HII stage have a more spherical structure. The aspect ratios of both infall candidates and clumps where infall is not detected display a marginal tendency of decrease from pre-stellar to proto-stellar, and to HII stages. In lower panel of Figure 11, we plotted the distribution of the aspect ratio for all infall candidates expect HII regions (blue histogram) and all clumps where infall is not detected expect HII regions (gray filled histogram). The K-S test gives a probability of 46\% that these two distributions originate from the same parent population. It is prove that the difference found in the global distribution (see Figure 12) due to the HII regions alone. The statistical parameters of each of these distributions are summarized in Table 4.

\begin{figure}
  \begin{center}
  \includegraphics[width=0.45\textwidth]{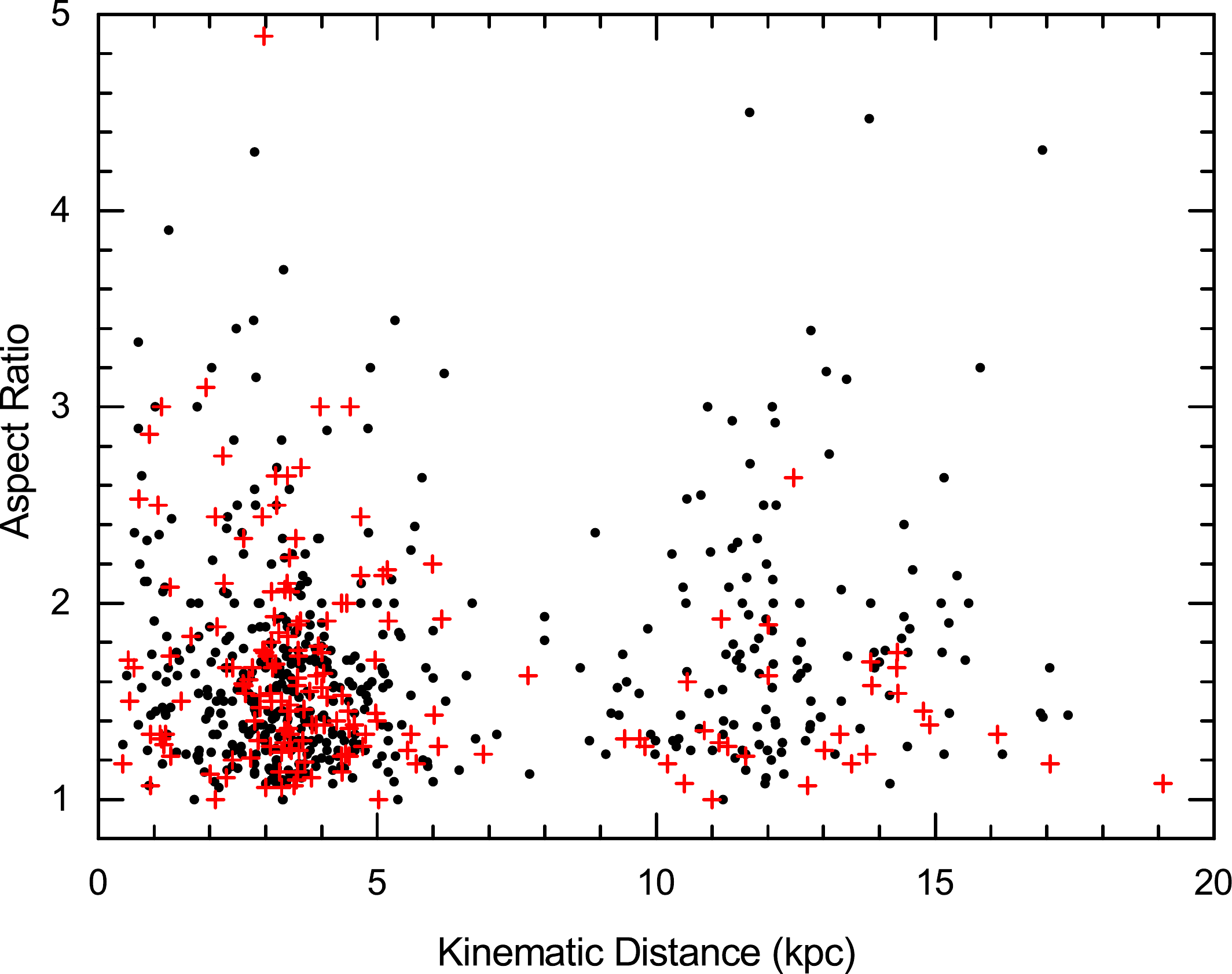}
  \end{center}
  \caption[dum]{Same as Figure 10, but showing aspect ratio distribution as a function of heliocentric distance.}
\end{figure}

\begin{figure}
  \begin{center}
  \includegraphics[width=0.45\textwidth]{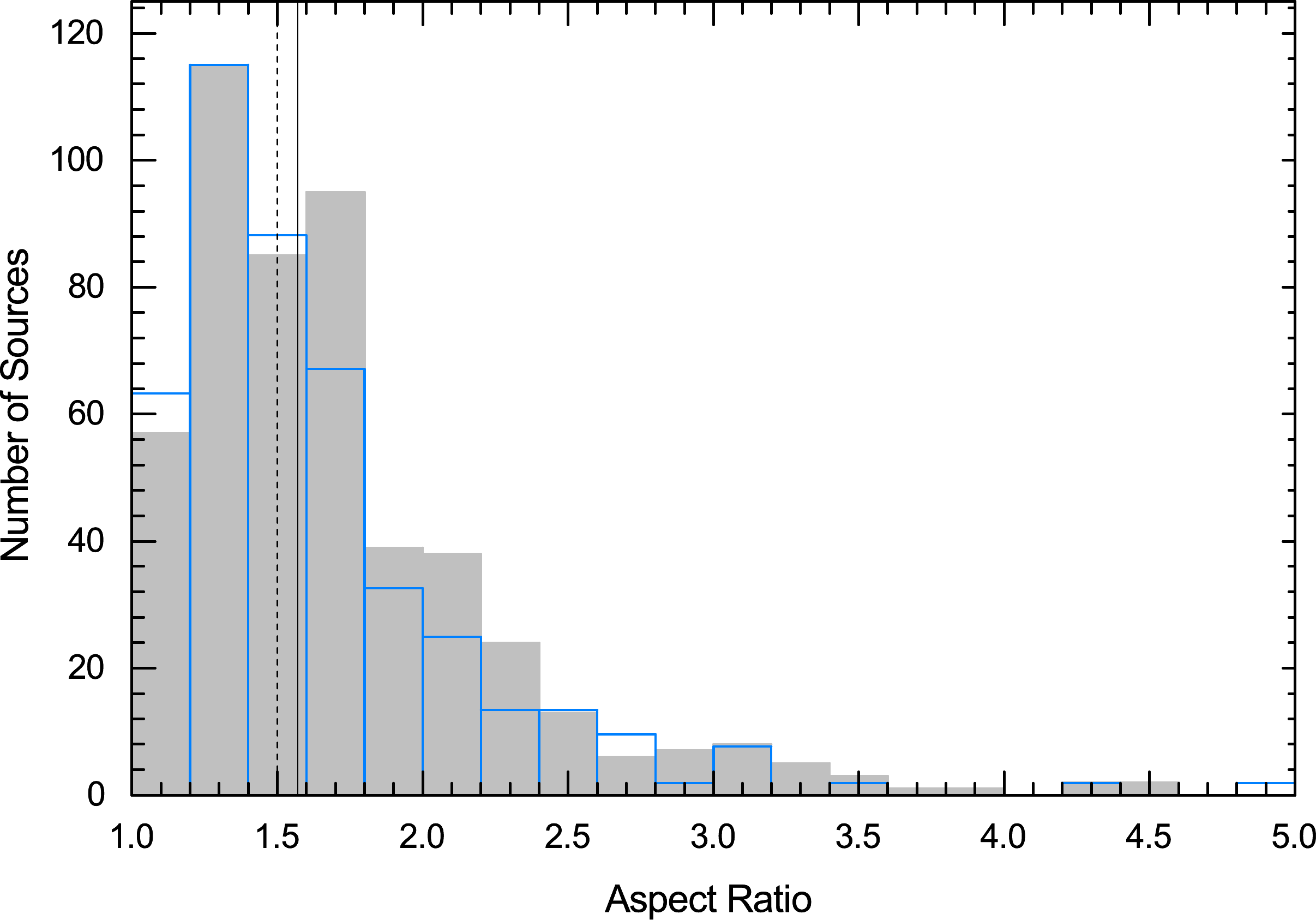}
  \end{center}
  \caption[dum]{The aspect ratio for infall candidates and clumps where infall is not detected are shown in blue histogram and gray filled histogram, respectively. The infall candidates distribution has been scaled to the peak of the infall is not detected clump distribution. Median values are indicated by the dashed and solid black vertical lines for the infall candidates and clumps where infall is not detected, respectively.}
\end{figure}

\begin{figure}
  \begin{center}
  \includegraphics[width=0.45\textwidth]{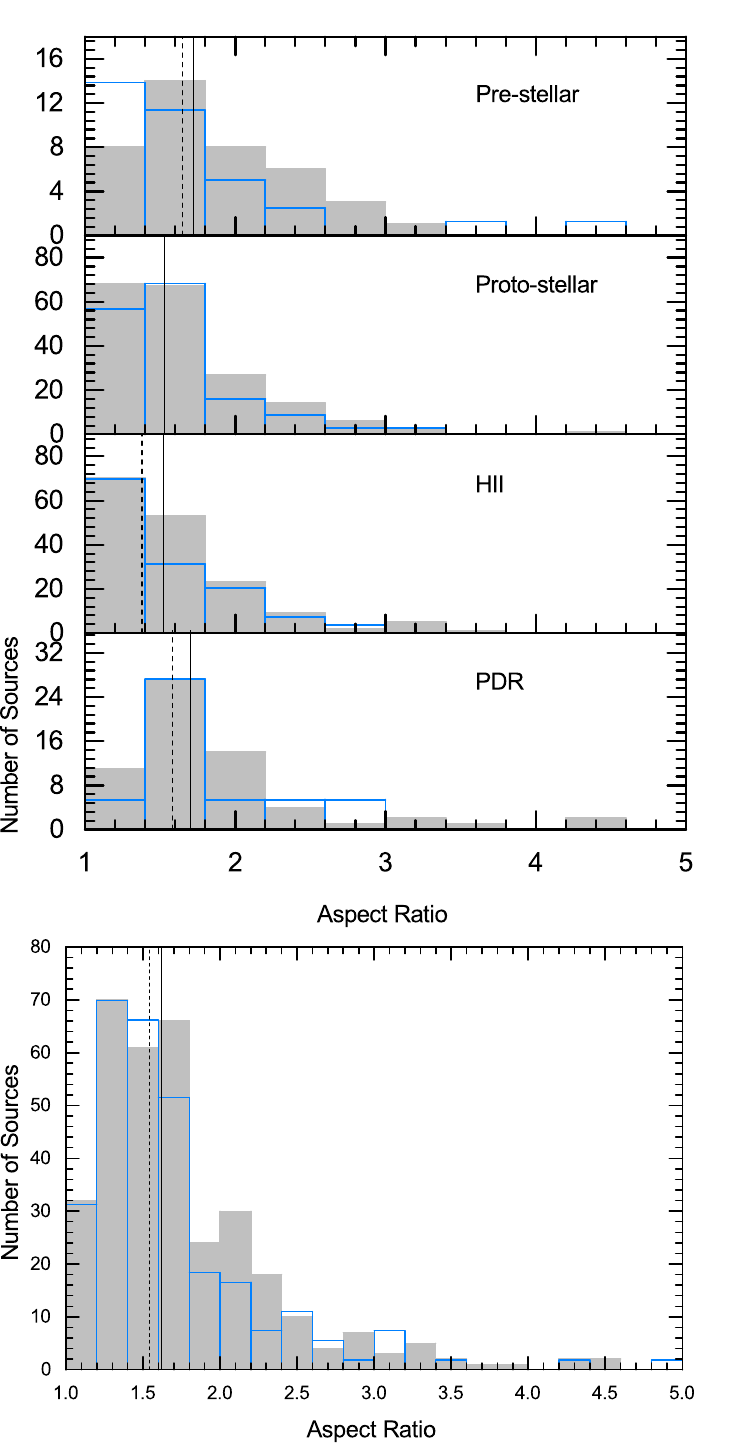}
  \end{center}
  \caption[dum]{Upper panel: Same as Figure 12, but showing aspect ratio for sources separated by their \emph{Spitzer} IR classification. Lower panel: Same as Figure 12, but showing a distribution of everything except HII regions.}
\end{figure}

\subsection{Clump masses and densities}
The beam-averaged H$_{2}$ column densities were derived by using the peak flux density at 870-$\mu$m via the formula
\begin{equation}
\rm N_{H_2}={{S_{peak}}R\over{B_\nu}(T_D){\Omega}{\kappa_\nu}{\mu}{m_H}},
\end{equation}
where S$_{peak}$ is the peak 870 $\mu$m flux density, R is the gas-to-dust mass ratio (assumed to be 100), B$_{\nu}$ is the Planck function for a given dust temperature T$_{D}$ (from GAE15), and D is the heliocentric distance to the source. The masses were derived by using the integrated 870-$\mu$m flux of the whole source via the formula
\begin{equation}
\rm M={{D^2}{S_{int}}{R}\over{B_\nu}(T_D){\kappa_\nu}},
\end{equation}
where $\Omega$ is the beam solid angle, $\mu$ is the mean molecular weight of the interstellar medium (assumed to be 2.8), m$_{H}$ is the mass of a hydrogen atom, S$_{int}$ is the integrated 870-$\mu$m flux, and $\kappa$$_{\nu}$ is the dust-absorption coefficient \citep[taken as 1.85 cm$^{2}$ g$^{-1}$, interpolated to 870 $\mu$m from Col. 5 of Table 1 for thin ice mantles and 10$^{5}$ yrs coagulation at 10$^{6}$ cm$^{-3}$ from][]{1994A&A...291..943O}. As in \citet{2014MNRAS.443.1555U}, the uncertainty in the derived clump mass and column density is $\sim$50\%.

In Figure 14, we present the distribution of clump mass as a function of heliocentric distance. It is clear that we are sensitive to all clumps with mass above 1000 M$_{\odot}$ within 20 kpc, and our statistics should be complete above this level. For a distance-limited sub-sample (between 2.5 and 3.5 kpc), the median values are 776 and 588 M$_{\odot}$ for the infall candidates and clumps where infall is not detected, respectively. We performed a K-S test on them, and got a probability of 4.4\% that the clump mass distributions of the infall candidates and clumps where infall is not detected originate from the same parent population. This reveals that the mass difference is not caused by distance/selection bias.

Figure 15 shows the distribution of 870-$\mu$m peak flux densities (Figure 15a) and H$_{2}$ column densities (Figure 15b) of the infall candidates (blue histogram) and clumps where infall is not detected (gray filled histogram) separated by their evolutionary stages. The median values of the peak 870-$\mu$m flux densities of the pre-stellar, proto-stellar, HII and PDR objects with infall are 0.83, 1.44, 2.56 and 2.40 Jy beam$^{-1}$, respectively. The corresponding median values for the H$_{2}$ column densities are 3.09$\times$10$^{22}$, 4.68$\times$10$^{22}$, 5.50$\times$10$^{22}$ and 3.63$\times$10$^{22}$ cm$^{-2}$. The K-S test gives a probability of 0.1\% for pre-stellar and proto-stellar, $\ll$0.1\% for proto-stellar and HII, and 7.8\% for HII and PDR distributions, respectively. The median values of the peak 870-$\mu$m flux densities of the clumps where infall is not detected at the pre-stellar, proto-stellar, HII and PDR stages are 0.92, 1.40, 2.57 and 1.87 Jy beam$^{-1}$, respectively. The corresponding median values of H$_{2}$ column densities are 2.82$\times$10$^{22}$, 4.37$\times$10$^{22}$, 4.90$\times$10$^{22}$ and 3.16$\times$10$^{22}$ cm$^{-2}$. The K-S test gives probabilities of $\ll$0.1\%, $\ll$0.1\%, and 1.5\% for pre-stellar and proto-stellar, proto-stellar and HII, and HII and PDR clump distributions originate from the same parent populations, respectively. The above results suggest that the peak column densities of the infall candidates and clumps where infall is not detected really increase from pre-stellar to proto-stellar to HII. Note that the median values of the peak column densities of the infall candidates at every stage are greater than the corresponding values of the clumps where infall is not detected. The K-S test gives a probability of 0.8\%, $\ll$0.1\%, $\ll$0.1\% and 0.3\% for infall candidates and clumps where infall is not detected distributions in pre-stellar, proto-stellar, HII and PDR originate from the same parent populations, respectively. This suggests that clumps with infall are evidently accumulating more materials.

The H$_{2}$ column density distribution of all infall candidates (blue histogram) and clumps where infall is not detected (gray filled histogram) are displayed in Figure 15(c), where the corresponding median values are 4.47$\times$10$^{22}$ and 3.89$\times$10$^{22}$ cm$^{-2}$. The K-S test gives a probability of $\ll$0.1\% that the H$_{2}$ column density distribution of all infall candidates and clumps where infall is not detected originate from the same parent population, and so the column density difference between infall candidates and clumps where infall is not detected is probably true. The statistical parameters of these distributions are summarized in Table 4.

Figure 16 presents the same distributions as Figure 15, but for integrated 870-$\mu$m flux densities (Figure 16a) and masses (Figure 16b and 16c). The median values of the integrated 870$\mu$m flux densities of the infall candidates and clumps where infall is not detected increase slightly as a function of evolutionary stage. The median values of the masses of the infall candidates at the pre-stellar, proto-stellar, HII and PDR stages are 501, 1023, 1778 and 1071 M$_{\odot}$ (Figure 10b), respectively. The corresponding median values of the clumps where infall is not detected are 589, 871, 2290 and 1413 M$_{\odot}$. The median values of masses for both infall candidates and the clumps where infall is not detected increase from pre-stellar to proto-stellar, and to HII stages, then decrease from HII to PDR stages. The K-S test gives probabilities of $\ll$0.1\% for pre-stellar and proto-stellar, $\ll$0.1\% for proto-stellar and HII, and 0.05\% for HII and PDR distributions for infall candidates originate from the same parent populations. The corresponding probabilities are $\ll$0.1\%, $\ll$0.1\% and $\ll$0.1\% for clumps where infall is not detected. The K-S test supports that clumps accumulate material continuously and efficiently as they evolve, which is consistent with the fact that the beam-averaged H$_{2}$ column densities of the clumps increase as they evolve (see above paragraph).

This result also supports the idea that molecular clouds gather material from large scales down to the centre of their gravitational well, and thus increase their masses \citep{2013A&A...555A.112P}. Of course, it is necessary to check this notion on a larger sample. Both infall candidates and clumps where infall is not detected show a decreasing trend of mass from the HII to the PDR stage, which is easy to understand in the sense that star-formation feedback finally disrupts the parent clump in the PDR stage.

The median values of the masses for all infall candidates and clumps where infall is not detected are 1318 and 1349 M$_{\odot}$, respectively, which are larger than the completeness limit (Figure 16c). \citet{2013MNRAS.435..400U} reported that clumps associated with compact ($\sim$ 10000 M$_{\odot}$) and UC HII ($\sim$ 5000 M$_{\odot}$) both have significantly larger median mass values than the median masss values of infall candidates and clumps where infall is not detected in this paper. The difference may be due to the simplifying assumption that all of the clumps have the same temperature 20 K by \citet{2013MNRAS.435..400U}. The statistical parameters of these distributions are summarized in Table 4.

HII regions are generally further away than low-mass pre-stellar and proto-stellar sources, and have higher masses. This may be the cause of the difference in mass distribution between the HII region and younger sources (Figure 16b). So it is necessary to determine whether the masses difference is caused by distance bias between HII region and younger clumps (pre-stellar and proto-stellar). We performed a K-S test on a distance-limited sample (between 3 and 5 kpc), the median mass values for the HII regions and young clumps are 1445 and 750 M$_{\odot}$, respectively. The K-S test gives a probability of $\ll$0.1\% that these two distributions originate from the same parent population, and suggests that the mass difference between the HII region and younger clumps (pre-stellar and proto-stellar) is not caused by distance bias. This is consistent with the following result that most of these clumps satisfy the criteria to form massive stars (see section 5.1).

The average H$_{2}$ volume density was calculated from the relation
\begin{equation}
\rm \bar{n}=3M/(4{\pi}{R^3}{\mu}{m_H})
\end{equation}
using the clump mass, M, as derived from the dust continuum emission at 870 $\mu$m, and the effective radius, R, as given in \citet{2014A&A...568A..41U}. We have taken the mean particle mass to be $\mu$ = 2.72m$_{H}$ \citep{1973asqu.book.....A}. The upper panel of Figure 17 shows the number distribution of the H$_{2}$ volume density of the infall candidates (blue histogram) and the clumps where infall is not detected (gray filled histogram) of each evolutionary stage. The median values of each distribution are 1.83$\times$10$^{4}$, 1.70$\times$10$^{4}$, 8.20$\times$10$^{3}$ and 5.80$\times$10$^{3}$ cm$^{-3}$ for the pre-stellar, proto-stellar, HII and PDR clumps with infall motions, respectively. As for clumps where infall is not detected at the different stages, the corresponding values
are 1.16$\times$10$^{4}$, 1.17$\times$10$^{4}$, 6.50$\times$10$^{3}$ and 5.00$\times$10$^{3}$ cm$^{-3}$, respectively. In general, both infall candidates and clumps where infall is not detected display a trend of decreasing volume density with the evolution. The upper panel of Figure 17 shows the average H$_{2}$ volume density distributions for all infall candidates and clumps where infall is not detected, the corresponding median values are 1.26$\times$10$^{4}$ and 8.32$\times$10$^{3}$ cm$^{-3}$.

On the other hand, Figure 18 shows the distribution of effective radius separated by their evolutionary stages (upper panel) and clump mass vs. volume density (lower panel). The median values of the effective radius at the pre-stellar, proto-stellar, HII and PDR stages are 0.47, 0.63, 0.97 and 0.94 pc, respectively. In addition, there is an anti-correlation between clump mass and volume density. So the decreasing volume density could be ascribed to an increasing effective radius with the evolution of the clumps. All infall candidates exhibit volume densities between 600 cm$^{-3}$ and 2.98$\times$10$^{5}$ cm$^{-3}$, with a median of 1.26$\times$10$^{4}$ cm$^{-3}$, while all clumps where infall is not detected show volume densities between 300 cm$^{-3}$ and 5.51$\times$10$^{5}$ cm$^{-3}$, with a median of 8.3$\times$10$^{3}$ cm$^{-3}$ (see lower panel of Figure 17). The K-S test gives a probability of $\ll$0.1\% that these two distributions originate from the same parent population, this suggests that infall candidates tend to be more compact.

\begin{figure}
  \begin{center}
  \includegraphics[width=0.45\textwidth]{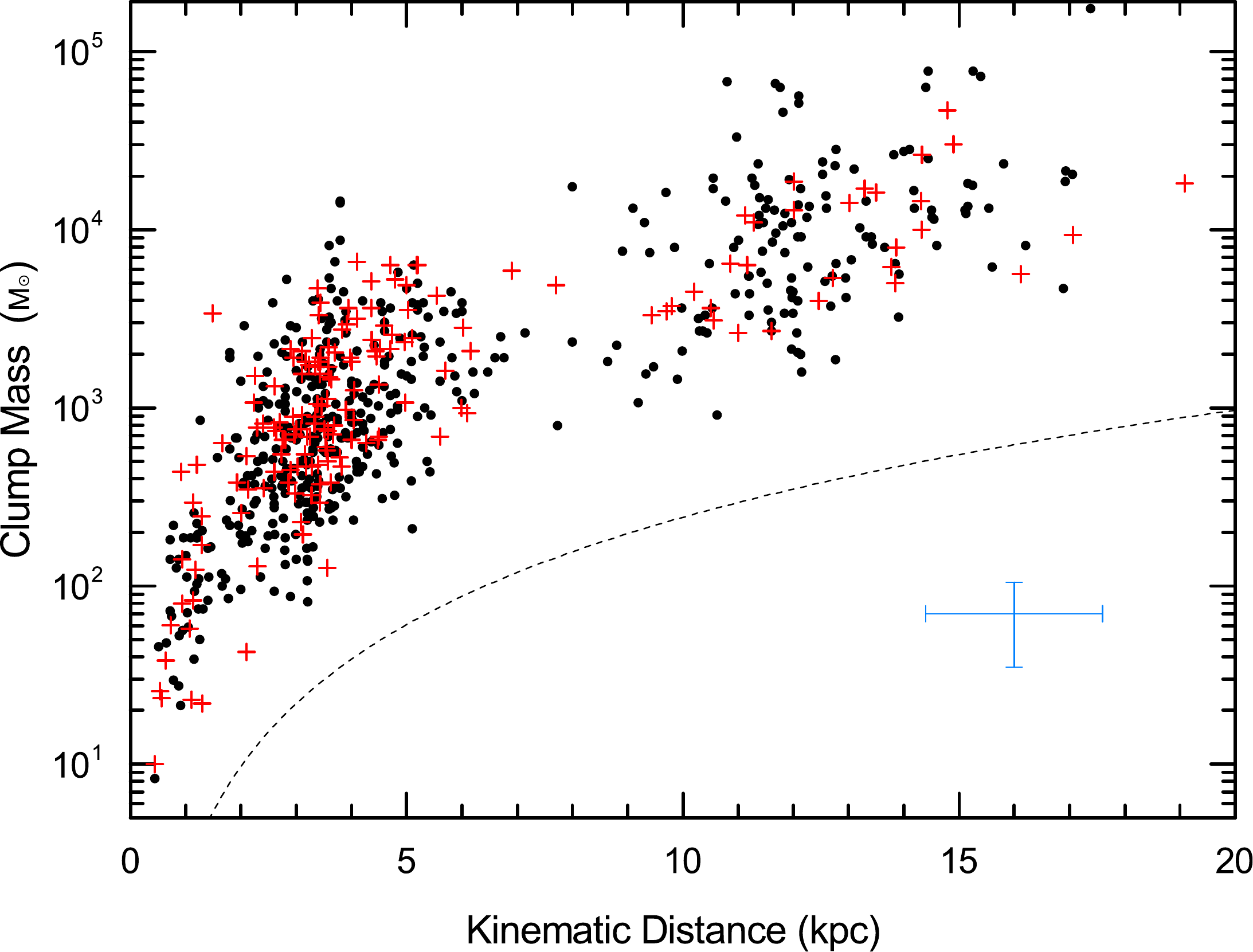}
  \end{center}
  \caption[dum]{Same as Figure 10, but showing dust mass distribution as a function of heliocentric distance. The dashed black line indicates the mass sensitivity limit of the survey with a dust temperature of 20K.}
\end{figure}

\begin{figure}
  \begin{center}
  \includegraphics[width=0.45\textwidth]{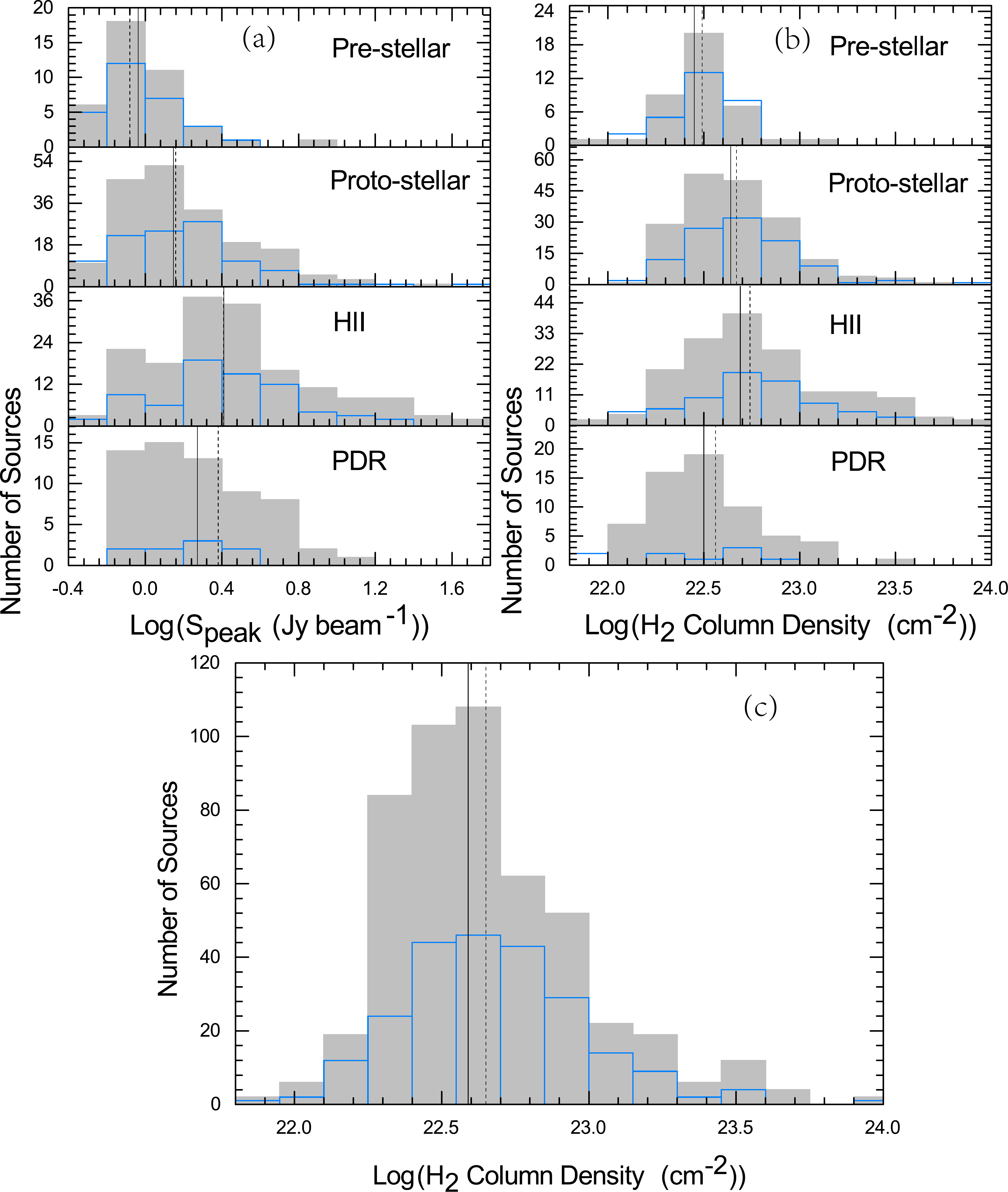}
  \end{center}
  \caption[dum]{The 870-$\mu$m fluxes and H$_{2}$ column densities for infall candidates and clumps where infall is not detected. (a). The 870-$\mu$m peak flux density distributions for
  infall candidates (blue histogram) and clumps where infall is not detected (gray filled histogram) separated by evolutionary stage. The median values of each stage are indicated by
  dashed and solid black vertical lines. (b). same as (a), but showing H$_{2}$ beam-averaged column densities. (c). The H$_{2}$ beam-averaged column density distributions for all infall candidates and clumps where infall is not detected, the corresponding median values are indicated by dashed and solid black vertical lines.}
\end{figure}

\begin{figure}
  \begin{center}
  \includegraphics[width=0.45\textwidth]{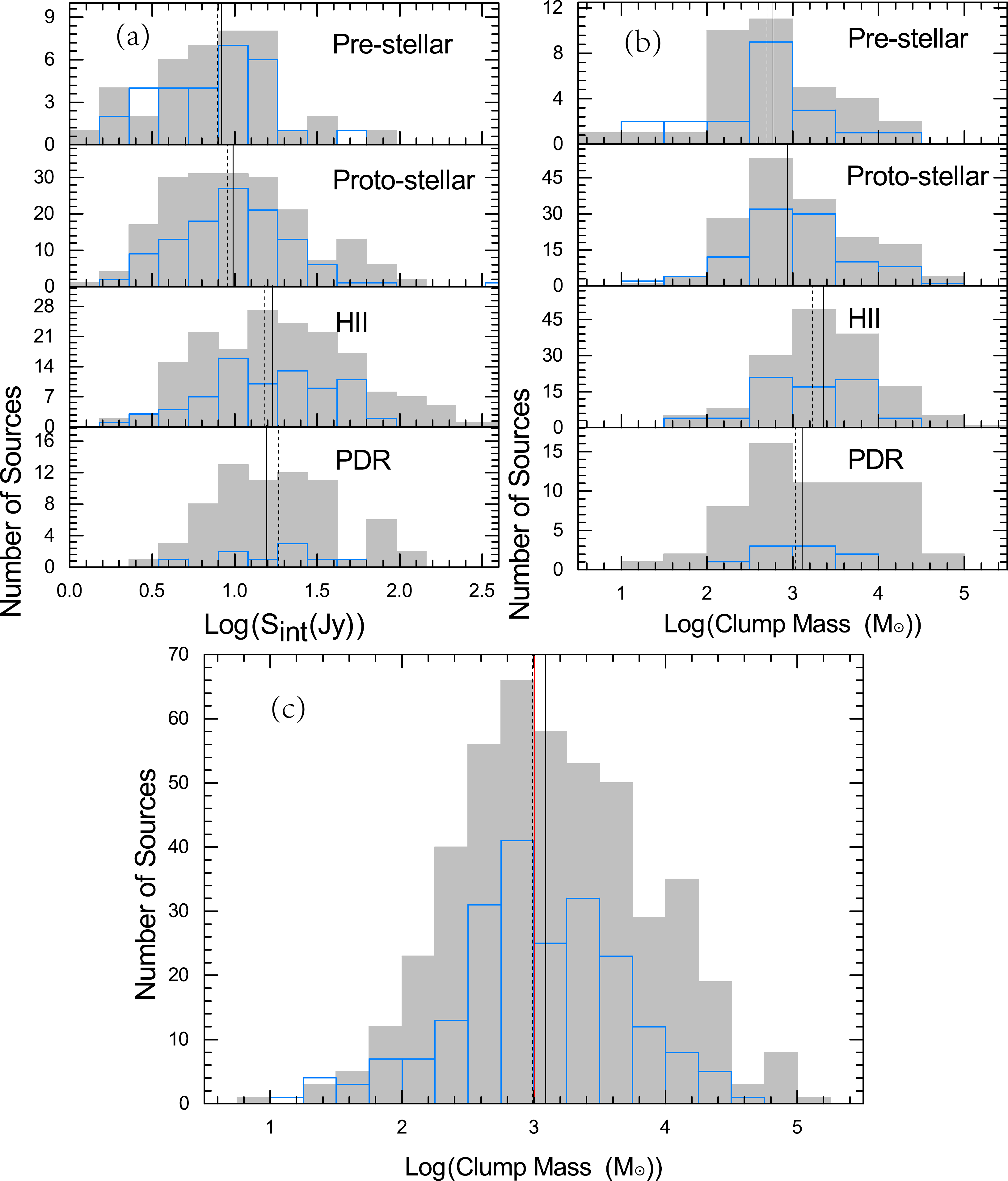}
  \end{center}
  \caption[dum]{Same as Figure 15, but showing 870-$\mu$m integrated flux density and clump mass. The vertical red line indicates the completeness limit of 1000 M$_{\odot}$.}
\end{figure}

\begin{figure}
  \begin{center}
  \includegraphics[width=0.45\textwidth]{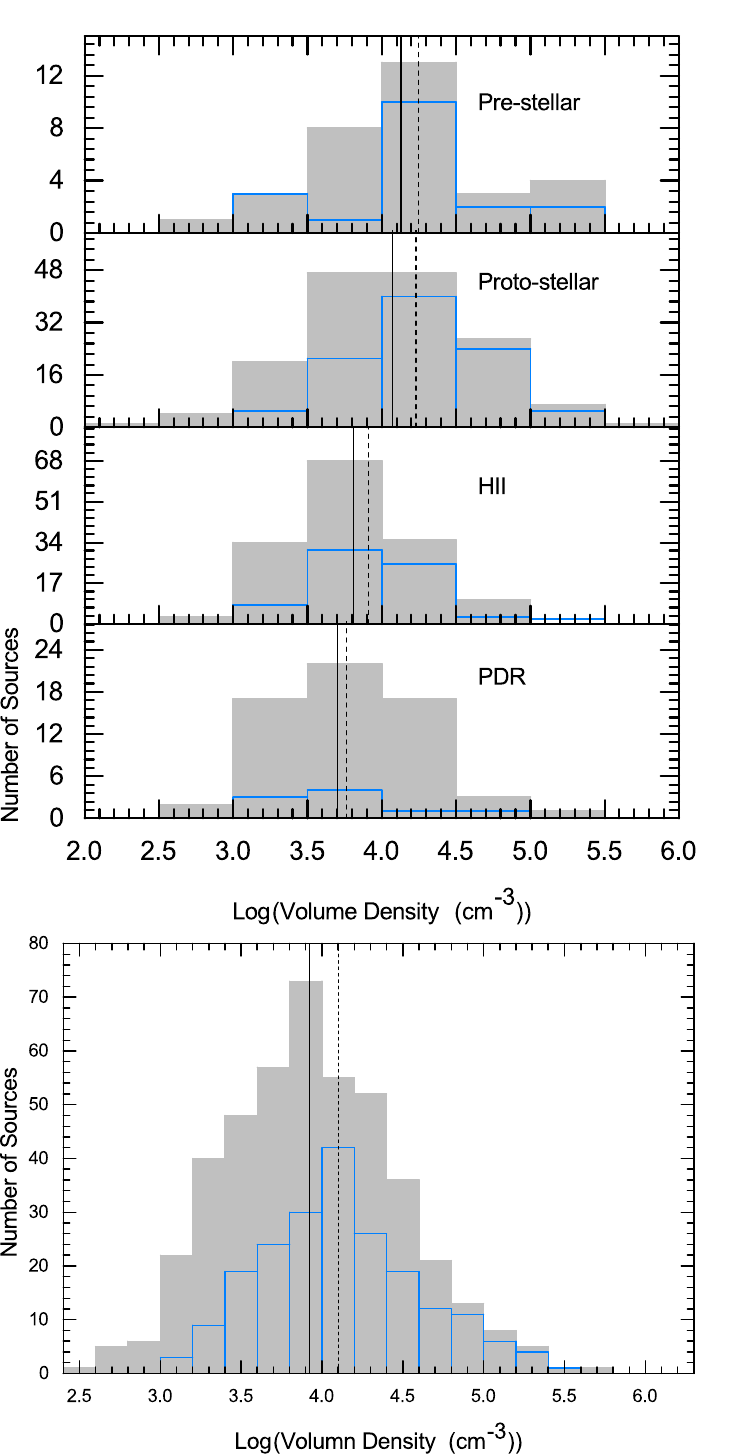}
  \end{center}
  \caption[dum]{Upper panel: Histograms of the average H$_{2}$ $\bar{n}$ of the infall candidates (blue histogram) and clumps where infall is not detected (gray filled histogram) separated by evolutionary stage. The median values of each stage are indicated by dashed and solid black vertical lines. Lower panel: The average H$_{2}$ volume density distributions for all infall candidates and clumps where infall is not detected, the corresponding median values are indicated by dashed and solid black vertical lines.}
\end{figure}

\begin{figure}
  \begin{center}
  \includegraphics[width=0.45\textwidth]{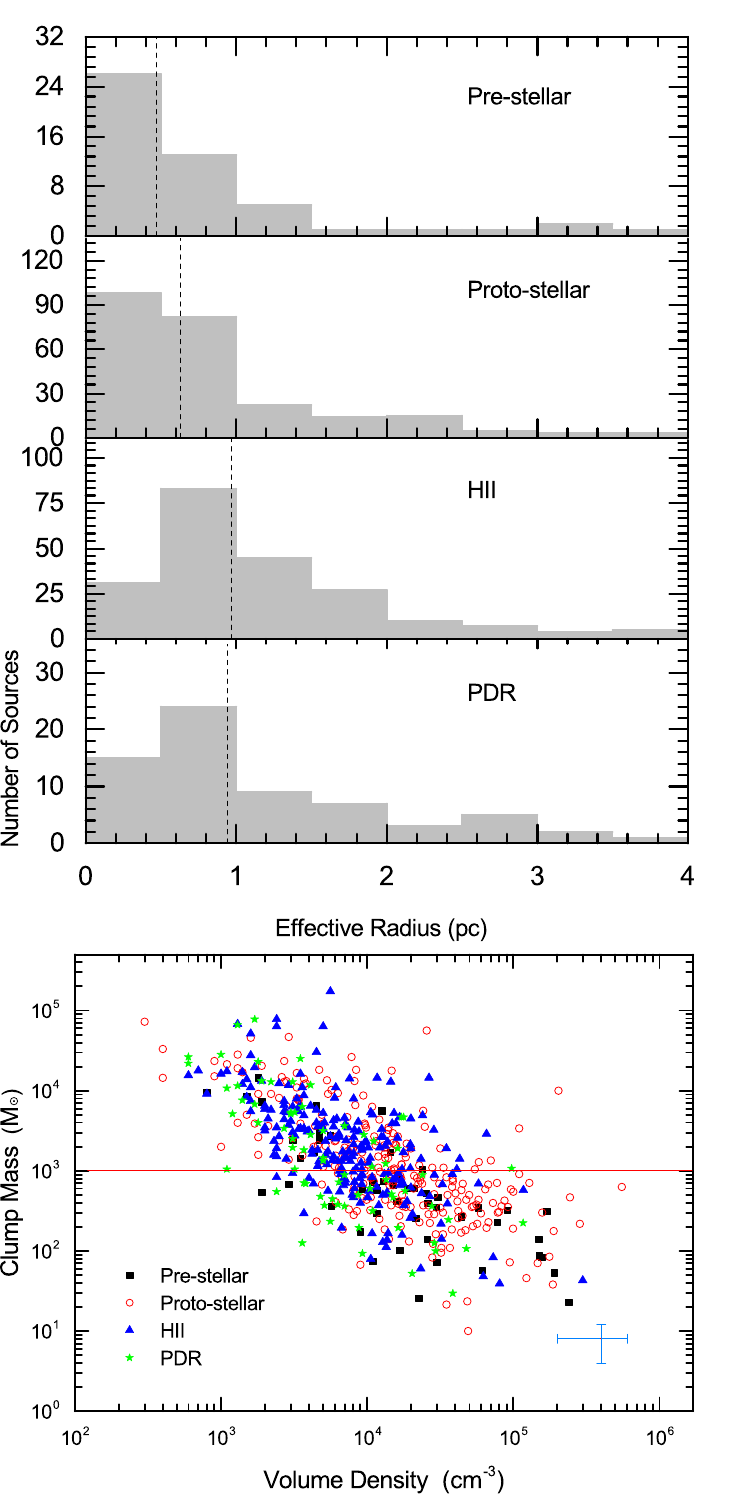}
  \end{center}
  \caption[dum]{Histograms of the source radius and clump mass vs. volume density relation for clumps separated by evolutionary stage in upper and lower panels, respectively. The horizontal red line displays the completeness limit of 1000 M$_{\odot}$. Characteristic error bars for these parameters are shown in the lower-right corner of the plot.}
\end{figure}

\section{Discussion}
\subsection{Empirical criterion for massive-star formation}
\citet{2010ApJ...716..433K} investigated cloud fragments in several molecular clouds that are forming (Orion A, G10.15-0.34, G11.11-0.12) or not forming ($\leq$10M$_{\odot}$, Pipe Nebula, Taurus, Perseus, and Ophiuchus) massive stars, and derived a threshold for the formation of massive stars. High-mass star-forming regions obey the empirical relationship m(r)$\geq$580M$_{\odot}$(R$_{eff}$ pc$^{-1}$)$^{1.33}$, which was confirmed by \citet{2013MNRAS.431.1752U,2013MNRAS.435..400U}. In Figure 19, we present the mass--size relationship for 206 infall candidates and 445 clumps where infall is not detected with determined distances. The sample shows a fairly continuous distribution over about four orders of magnitude in mass and two orders of magnitude in radius. The dashed blue and solid black lines indicate the least-squares fit to the infall candidates and the clumps where infall is not detected expressed as the empirical relation Log(M$_{clump}$)=3.12$\pm$1.85+(1.82$\pm$0.06)$\times$Log(R$_{eff}$), with a correlation coefficient of 0.81, and Log(M$_{clump}$)=3.14$\pm$1.54+(1.84$\pm$0.02)$\times$Log(R$_{eff}$), with a correlation coefficient of 0.87.  The difference in the fits is within the uncertainties in the parameters. The yellow shaded region in Figure 19 shows the parameter space that is devoid of massive-star formation. We find that 193 infall candidates (94\%) and 428 clumps where infall is not detected (96\%) have masses greater than the limiting mass for their size, as determined by \citet{2010ApJ...716..433K} for massive-star formation. Based on Figure 19, most of our sample of clumps should be capable of forming high-mass stars.

Sources located at the green shaded region (``Massive Proto-cluster Candidates", MPCs) in the upper portion of the Mass--radius diagram of Figure 19 have the potential to form a massive proto-cluster \citep{2012ApJ...758L..28B,2013MNRAS.431.1752U}. Using more accurate dust temperatures from GAE15, we identified two new MPCs (G333.018+0.766 and G348.531-0.972) in this work, four MPCs identified in HYX15 (G008.691-0.401, G348.183+0.482, G348.759-0.946 and G351.774-0.537) were confirmed again. However, the other three MPC candidates identified in HYX15 (G333.299-0.351, G338.459+0.024 and G345.504+0.347) were not confirmed here because of their masses changed lower.

\begin{figure}
  \begin{center}
  \includegraphics[width=0.45\textwidth]{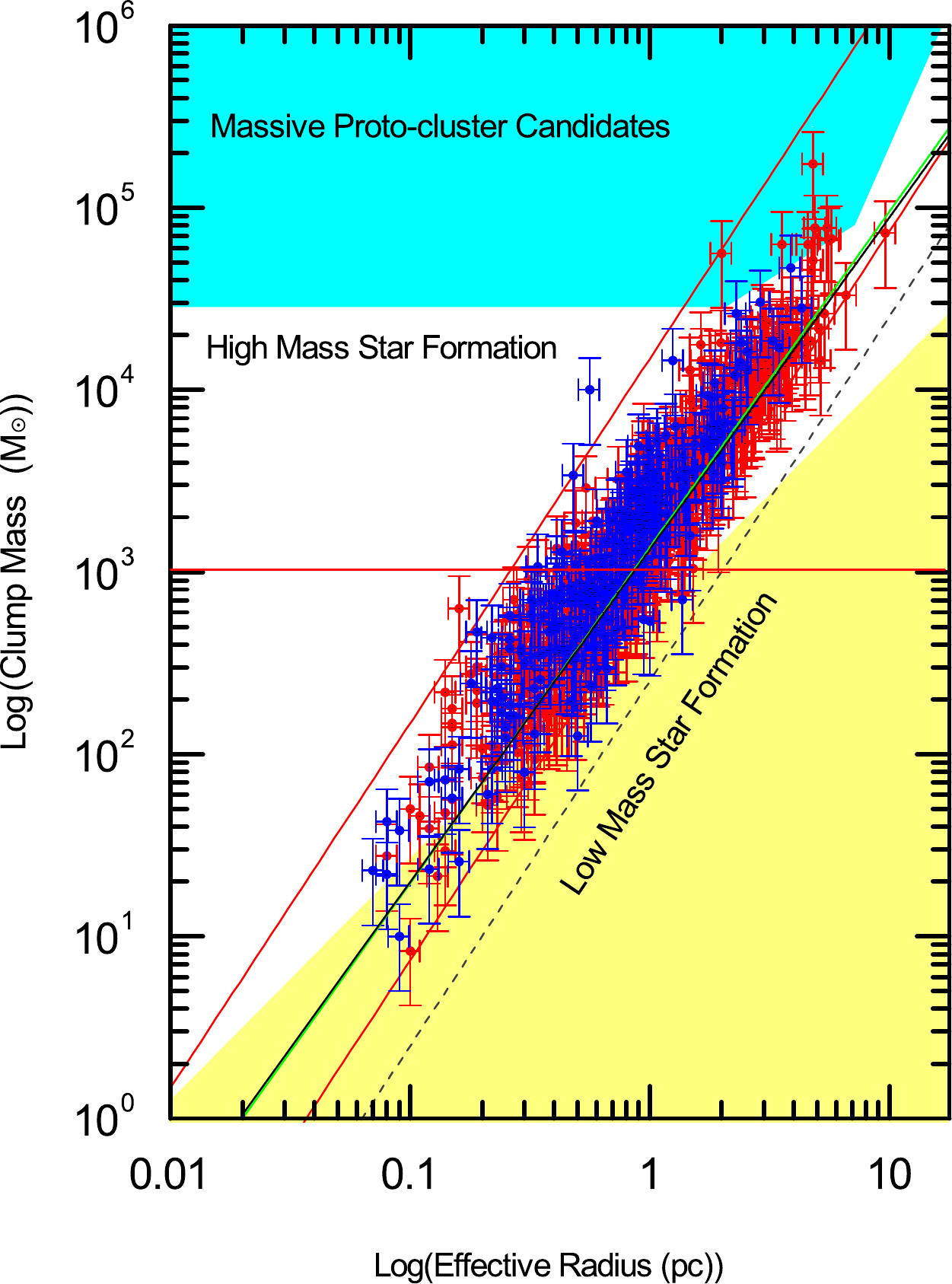}
  \end{center}
  \caption[dum]{The mass-size relationship of 212 infall candidates (blue dots) and 443 clumps where infall is not detected (red dots) with mass determined. The yellow shaded region is the parameter space to be devoid of massive star formation, where m(r)$\leq$580M$_{\odot}$(R$_{eff}$ pc$^{-1}$)$^{1.33}$) \citep[cf.][]{2010ApJ...723L...7K}. The cyan shaded region indicates the region in which young massive cluster progenitors are expected to be found \citep[i.e.][]{2012ApJ...758L..28B}. The solid black and solid green lines show the results of linear least-squares fits to the infall candidates and clumps where infall is not detected. The grey dashed line shows the sensitivity of the ATLASGAL survey, the upper and lower solid red line shows the surface densities of 1 and 0.05 g cm$^{-2}$. The horizontal red line displays the completeness limit of 1000 M$_{\odot}$.}
\end{figure}

\subsection{Virial mass and clump mass function}
To obtain the virial mass, we assumed a clump density profile of $\rho \propto r^{-1.5}$ for the 688 resolved clumps that have kinetic distances and effective physical radii. These were determined using the following formulae \citep{2013MNRAS.431.1752U}:
\begin{equation}
\rm ({{M_{vir}}\over{M_{\odot}}})\,=\,{{783}\over{7ln2}}({{R_{eff}}\over{pc}})({{\Delta v_{avg}}\over{km\,s^{-1}}})^2,
\end{equation}
where $R_{eff}$ is the effective radius of the clump (listed in Table 1). The average velocity dispersion of the total column of gas $\Delta v_{avg}$ can be calculated via
\begin{equation}
\rm ({\Delta v_{avg}})^2\,=\,({\Delta v_{corr}})^2+8ln2 \times {{k_{b}T_{kin}}\over{m_{H}}}({{1\over\mu_{p}}-{1\over\mu_{N_{2}H^{+}}}}),
\end{equation}
where $\Delta v_{corr}$ is the observed N$_{2}$H$^{+}$(1-0) line width corrected for the resolution of the spectrometer (0.21 km s$^{-1}$). $k_{b}$ is the Boltzmann constant, and $T_{kin}$ is the kinetic temperature of the gas (approximated as the dust temperature). $\mu_{p}$ and $\mu_{N_{2}H^{+}}$ are the mean molecular masses of molecular hydrogen (2.33) and N$_{2}$H$^{+}$ (29), respectively. To calculate the virial mass, we assume that the observed N$_{2}$H$^{+}$(1-0) line width is representative of the whole clump. However, we not that due to its resistance to depletion at low temperatures and high densities, N$_{2}$H$^{+}$(1-0) traces the cold, outer-parts of the clump which may not be as stirred by outflows or as closely related to the ongoing star formation. As such, we may underestimate $\Delta v_{avg}$ and thus $M_{vir}$. The uncertainties in the virial masses come from measurement errors of the line widths and kinetic temperatures. The error in the virial mass is of the order of 20\% allowing for an $\sim$10\% error in the distance \citep{2013MNRAS.435..400U}. Note that this estimate considers the simplest case of virial equilibrium where only gravity and the velocity dispersion of the gas are taken into account, thus neglecting the effects of e.g. external pressure and magnetic fields. The virial mass of the 688 sources are listed in Table 1.

In Figure 20, we plot the ratio ($M_{clump}/M_{vir}$) vs. the clump mass for the infall candidates (upper panel) and clumps where infall is not detected (lower panel). The dashed horizontal line marks the line of gravitational stability. The ratio $M_{clump}/M_{vir}$ describes the competition between the internal supporting energy against the gravitational energy. Clumps with ratios less than unity are likely to be unbound, while those with ratios higher than unity may be unstable and collapsing. The percentages of sources with ratios ($M_{clump}/M_{vir}$) greater than unity and also showing infall are 61\%, 75\%, 74\%, and 56\% of the pre-stellar, proto-stellar, HII and PDR sources (upper panel in Figure 20). The corresponding values of the clumps where infall is not detected are 56\%, 69\%, 70\% and 50\%, respectively (lower panel in Figure 20). For those infalling clumps which are not gravitationally bound, this probably suggests that there is a significant contribution from additional support mechanisms (e.g. magnetic fields, turbulence, external pressure). Thus, not all clumps showing infall motions are gravitationally bound or collapsing.

The clump mass function (ClMF) describes the mass distribution of clumps, and is related to the stellar initial mass function (IMF) \citep{2001ApJ...559..307J}. To obtain the parameter index of the ClMF, we fixed the bin widths and counted the number of clumps per bin. The power-law correlations of the ClMF are $dN/dM=10^{5.60\pm0.65}M^{-2.04\pm0.16}$ and $dN/dM=10^{6.19\pm1.04}M^{-2.37\pm0.29}$ for the clump mass and virial mass of the infall candidates (upper panel of Figure 21), respectively. For clumps where infall is not detected, the corresponding ClMFs are $dN/dM=10^{6.60\pm1.29}M^{-2.17\pm0.31}$ and $dN/dM=10^{5.55\pm0.75}M^{-2.02\pm0.20}$ (lower panel of Figure 21), respectively. It is evident that the ClMF of the infall candidates is similar to that of the clumps where infall is not detected. Our power indices for clump mass (2.04 for infall candidates and 2.02 for clumps where infall is not detected; see Figure 21) are between the IMF of 2.35 \citep{1955ApJ...121..161S} and CMF of 1.85-1.91 presented by \citet{2014MNRAS.443.1555U}. The ClMF is shallower than the IMF, suggesting that the star formation efficiency is lower in more massive clumps. However, they are close to the value of cold \emph{Planck} sources (2.0) reported by \citet{2015A&A...584A..92M}.

\begin{figure}
  \begin{center}
  \includegraphics[width=0.45\textwidth]{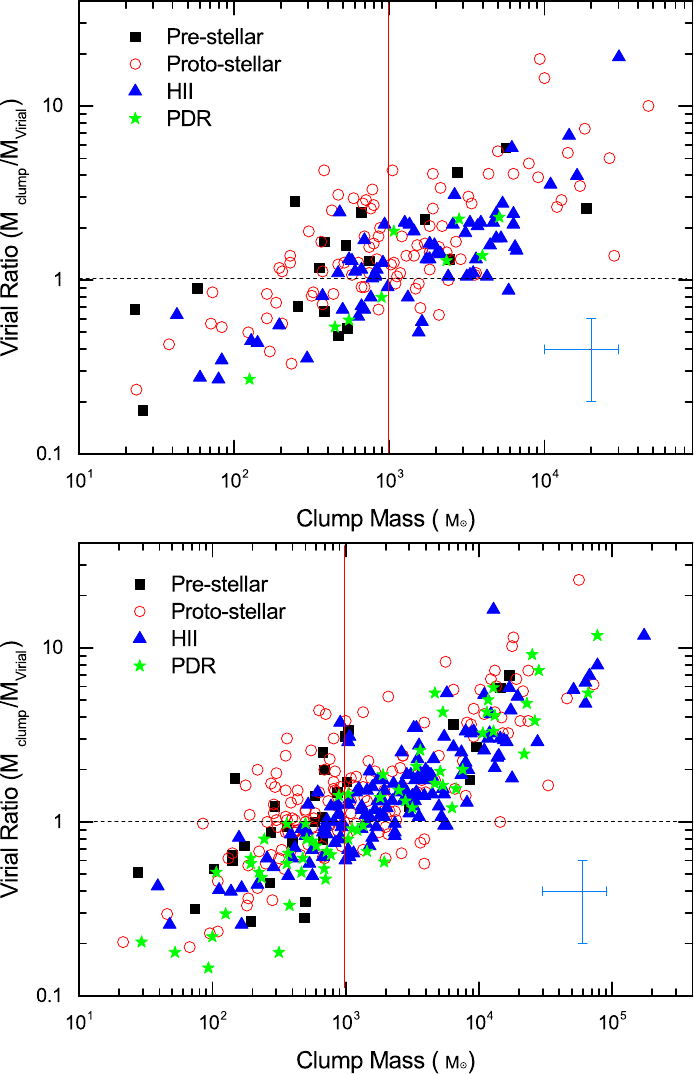}
  \end{center}
  \caption[dum]{Virial ratio ($M_{clump}/M_{vir}$) as a function of clump mass for infall candidates (upper panel) and clumps where infall is not detected (lower panel) separated by evolutionary stage. The black horizontal line indicates the locus of gravitational equilibrium for thermal and kinematic energies. The vertical red lines display the completeness limit of 1000 M$_{\odot}$. Characteristic error bars for these parameters are shown in the lower-right corner of the plot.}
\end{figure}

\begin{figure}
  \begin{center}
  \includegraphics[width=0.45\textwidth]{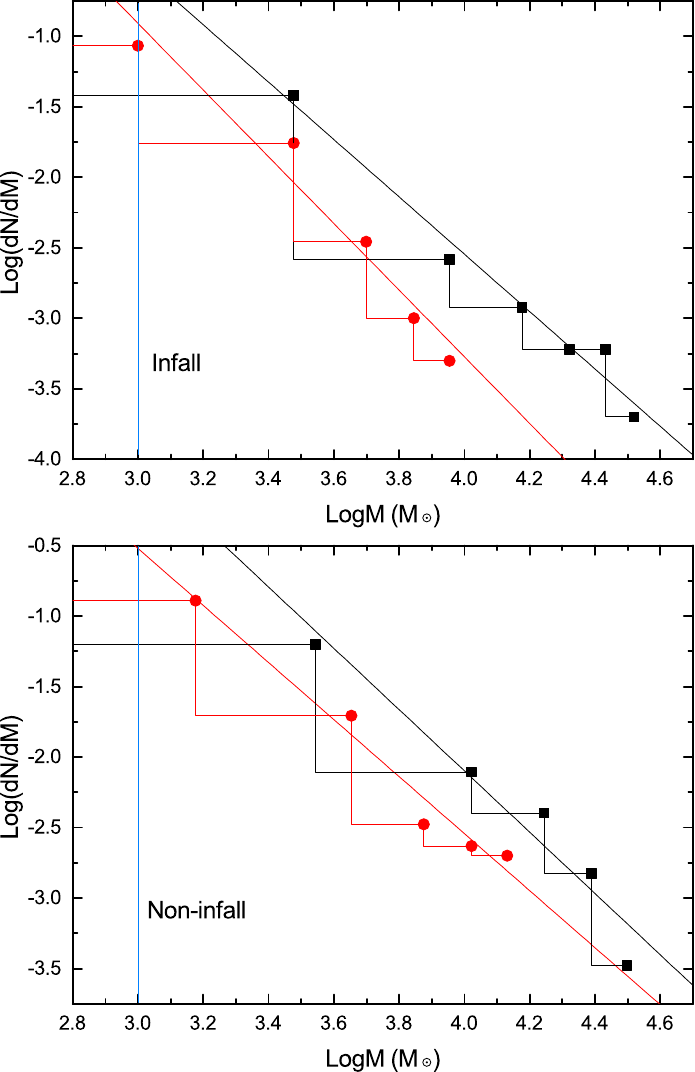}
  \end{center}
  \caption[dum]{Upper panel: Clump mass function (ClMF) of infall candidates. The black squares and red points are logarithmic values of the number of clumps per unit mass against logarithmic clump mass and logarithmic virial mass, respectively. The corresponding ClMFs are shown by the black and red solid lines. Lower panel: same as upper panel, but for clumps where infall is not detected. The horizontal blue lines display the completeness limit of 1000 M$_{\odot}$.}
\end{figure}

\subsection{The evolution of infall}
For the 231 sources identified as infall candidates, the infall rate can be roughly estimated from formulae \citep[Eq. 5;][]{2010A&A...517A..66L}:
\begin{equation}
\dot{M}_{inf} =4{\pi}{R^2}{V_{inf}}{n}
\end{equation}
where V$_{inf}$ = V$_{N_{2}H^{+}}$ $-$ V$_{HCO^{+}}$ is a rough estimate of the infall velocity, and $n$ is the volume density described in Section 4.5. The obtained mass infall rates are listed in Table 3. Figure 22 shows the distribution of the infall rate for the infall candidates separated into evolutionary stage. The median values of the distributions are indicated by black dashed lines for pre-stellar, proto-stellar, HII and PDR, which correspond to 2.6$\times$10$^{-3}$, 7.0$\times$10$^{-3}$, 6.5$\times$10$^{-3}$ and 5.5$\times$10$^{-3}$ M$_\odot$ yr$^{-1}$. They display an increasing trend with evolutionary stage from pre-stellar to proto-stellar, which then decreases from proto-stellar to HII, and then to PDR. Note that the infall rate reaches its maximum value at the proto-stellar stage, and then start to decrease. The K-S test gives probabilities of 0.1\% for pre-stellar and proto-stellar, 0.5\% for proto-stellar and HII, and 1.9\% for HII and PDR distributions originate from the same parent population. The statistical parameters of these distributions are summarized in Table 4.

In Figure 23, we show a relationship between the infall velocity and the total clump mass ($M_{clump}>100M_{\odot}$). The power-law correlation is $V_{in}=(0.62\pm0.12)M_{clump}^{0.03\pm0.03}$ (red dashed line). The uncertainty on the power-law fit result is basically flat, i.e. no relationship between V$_{in}$ and M$_{clump}$. Our power index (0.03) for clumps is significantly less than the power index of high mass cores of 0.36 presented by \citet{2013ApJ...776...29L}. This suggests that infall rate of high-mass cores is significantly greater than that of high-mass clumps.

Assuming a constant star formation efficiency, we derived depletion times (M$_{clump}$/$\dot{M}$$_{inf}$) for all infall candidates. Figure 24 displays the distributions of depletion times for each evolutionary stage. The median values of infall times for pre-stellar, proto-stellar, HII and PDR are 1.60$\times$10$^{5}$, 1.67$\times$10$^{5}$, 2.70$\times$10$^{5}$ and 3.33$\times$10$^{5}$ yr, respectively. These results show that the infall times increase as a function of evolution. Since the mass infall rate is similar for proto-stellar sources and HII regions, the change in depletion time is due to HII clumps typically having larger mass, and so can sustain star formation for longer. The K-S test (probability of $\ll$0.01\%) supports that infall rate increase from proto-stellar to HII stage. As the small number of members of pre-stellar and PDR, we did not do K-S test to them.

\begin{table*}
  \tiny
  \centering
  \begin{minipage}{175mm}
   \caption{Mass infall rates of infall candidates.}
    \begin{tabular}{lrrlrrlrr}
      \hline
      \multicolumn{1}{c}{Clump}              &  \multicolumn{1}{c}{$\dot{M}$}                         & \multicolumn{1}{c}{$Spitzer$}       &
      \multicolumn{1}{c}{Clump}              &  \multicolumn{1}{c}{$\dot{M}$}                         & \multicolumn{1}{c}{$Spitzer$}       &
      \multicolumn{1}{c}{Clump}              &  \multicolumn{1}{c}{$\dot{M}$}                         & \multicolumn{1}{c}{$Spitzer$}       \\
      \multicolumn{1}{c}{name}               &  \multicolumn{1}{c}{0.001 M$_{\odot}$ yr$^{-1}$}       & \multicolumn{1}{c}{classification}  &
      \multicolumn{1}{c}{name}               &  \multicolumn{1}{c}{0.001 M$_{\odot}$ yr$^{-1}$}       & \multicolumn{1}{c}{classification}  &
      \multicolumn{1}{c}{name}               &  \multicolumn{1}{c}{0.001 M$_{\odot}$ yr$^{-1}$}       & \multicolumn{1}{c}{classification}  \\
      \hline
      G000.156$-$0.452   &   $-$     &   Pre-stellar        &       G331.709+0.602     &   13.51   &   Proto-stellar      &     G308.791+0.171     &   4.15    &   HII                \\ 
G000.591$-$0.851   &   $-$     &   Pre-stellar        &       G332.241$-$0.044   &   10.92   &   Proto-stellar      &     G308.917+0.122     &   4.21    &   HII                \\ 
G003.416$-$0.354   &   $-$     &   Pre-stellar        &       G332.254$-$0.056   &   4.39    &   Proto-stellar      &     G309.421$-$0.621   &   5.27    &   HII                \\ 
G003.434$-$0.572   &   $-$     &   Pre-stellar        &       G332.276$-$0.071   &   3.77    &   Proto-stellar      &     G309.981+0.546     &   4.88    &   HII                \\ 
G005.917$-$0.311   &   11.35   &   Pre-stellar        &       G332.559$-$0.147   &   1.95    &   Proto-stellar      &     G310.014+0.387     &   2.75    &   HII                \\ 
G007.234$-$0.101   &   6.39    &   Pre-stellar        &       G332.604$-$0.167   &   7.29    &   Proto-stellar      &     G311.492+0.404     &   6.12    &   HII                \\ 
G010.199$-$0.316   &   $-$     &   Pre-stellar        &       G332.999$-$0.639   &   2.09    &   Proto-stellar      &     G312.108+0.309     &   9.58    &   HII                \\ 
G012.776$-$0.211   &   2.54    &   Pre-stellar        &       G333.711$-$0.217   &   2.16    &   Proto-stellar      &     G314.264+0.091     &   6.46    &   HII                \\ 
G013.276$-$0.334   &   9.21    &   Pre-stellar        &       G334.651+0.469     &   3.91    &   Proto-stellar      &     G316.139$-$0.506   &   10.91   &   HII                \\ 
G013.299$-$0.326   &   5.56    &   Pre-stellar        &       G335.262$-$0.114   &   8.50    &   Proto-stellar      &     G320.201+0.859     &   4.17    &   HII                \\ 
G014.626$-$0.562   &   2.06    &   Pre-stellar        &       G335.349+0.412     &   6.16    &   Proto-stellar      &     G320.247+0.404     &   1.66    &   HII                \\ 
G332.892$-$0.569   &   2.21    &   Pre-stellar        &       G337.098$-$0.929   &   3.24    &   Proto-stellar      &     G320.596$-$0.196   &   2.47    &   HII                \\ 
G333.449$-$0.182   &   6.23    &   Pre-stellar        &       G337.176$-$0.032   &   10.98   &   Proto-stellar      &     G325.514+0.414     &   8.15    &   HII                \\ 
G333.524$-$0.269   &   5.83    &   Pre-stellar        &       G337.258$-$0.101   &   22.41   &   Proto-stellar      &     G329.524+0.084     &   7.10    &   HII                \\ 
G340.179$-$0.242   &   19.61   &   Pre-stellar        &       G337.761$-$0.339   &   8.34    &   Proto-stellar      &     G331.134$-$0.484   &   7.91    &   HII                \\ 
G342.114+0.504     &   2.67    &   Pre-stellar        &       G337.916$-$0.477   &   10.77   &   Proto-stellar      &     G331.512$-$0.102   &   36.71   &   HII                \\ 
G343.489$-$0.416   &   2.14    &   Pre-stellar        &       G338.327$-$0.409   &   2.33    &   Proto-stellar      &     G331.626+0.526     &   2.53    &   HII                \\ 
G344.977$-$0.292   &   7.53    &   Pre-stellar        &       G338.926+0.634     &   17.37   &   Proto-stellar      &     G332.296$-$0.094   &   8.62    &   HII                \\ 
G348.383+0.537     &   2.55    &   Pre-stellar        &       G339.584$-$0.127   &   12.20   &   Proto-stellar      &     G332.469$-$0.131   &   1.31    &   HII                \\ 
G350.772+0.796     &   0.52    &   Pre-stellar        &       G339.924$-$0.084   &   18.36   &   Proto-stellar      &     G332.677$-$0.614   &   19.34   &   HII                \\ 
G350.784+0.801     &   0.39    &   Pre-stellar        &       G340.232$-$0.146   &   23.20   &   Proto-stellar      &     G333.089$-$0.352   &   0.89    &   HII                \\ 
G352.511+0.776     &   0.63    &   Pre-stellar        &       G340.901$-$0.346   &   11.63   &   Proto-stellar      &     G333.386+0.032     &   11.14   &   HII                \\ 
G354.944$-$0.537   &   2.39    &   Pre-stellar        &       G340.969$-$1.021   &   9.46    &   Proto-stellar      &     G334.838$-$0.201   &   0.83    &   HII                \\ 
G358.572$-$0.306   &   $-$     &   Pre-stellar        &       G341.181$-$0.277   &   5.86    &   Proto-stellar      &     G335.177$-$0.242   &   2.31    &   HII                \\ 
G358.586$-$0.302   &   $-$     &   Pre-stellar        &       G341.217$-$0.212   &   9.39    &   Proto-stellar      &     G335.586$-$0.291   &   7.56    &   HII                \\ 
G359.216$-$0.192   &   $-$     &   Pre-stellar        &       G341.219$-$0.259   &   7.19    &   Proto-stellar      &     G336.411$-$0.256   &   9.56    &   HII                \\ 
G359.216$-$0.206   &   $-$     &   Pre-stellar        &       G341.236$-$0.271   &   13.55   &   Proto-stellar      &     G337.671$-$0.192   &   2.51    &   HII                \\ 
G359.929+0.196     &   $-$     &   Pre-stellar        &       G341.287+0.179     &   4.59    &   Proto-stellar      &     G337.844$-$0.376   &   4.31    &   HII                \\ 
G000.484$-$0.702   &   $-$     &   Proto-stellar      &       G343.738$-$0.112   &   3.80    &   Proto-stellar      &     G338.436+0.057     &   12.65   &   HII                \\ 
G000.534$-$0.644   &   $-$     &   Proto-stellar      &       G343.902$-$0.672   &   10.47   &   Proto-stellar      &     G338.916+0.382     &   23.79   &   HII                \\ 
G002.534+0.199     &   $-$     &   Proto-stellar      &       G344.004$-$0.579   &   15.44   &   Proto-stellar      &     G339.476+0.184     &   17.34   &   HII                \\ 
G004.627$-$0.666   &   19.05   &   Proto-stellar      &       G344.036$-$0.519   &   2.78    &   Proto-stellar      &     G339.589+0.082     &   8.71    &   HII                \\ 
G005.359+0.014     &   1.66    &   Proto-stellar      &       G346.369$-$0.647   &   8.06    &   Proto-stellar      &     G340.054$-$0.244   &   9.49    &   HII                \\ 
G005.617$-$0.082   &   12.14   &   Proto-stellar      &       G348.376+0.524     &   1.90    &   Proto-stellar      &     G341.266$-$0.302   &   10.59   &   HII                \\ 
G006.216$-$0.609   &   7.60    &   Proto-stellar      &       G348.419+0.106     &   5.32    &   Proto-stellar      &     G341.282+0.166     &   1.22    &   HII                \\ 
G007.166+0.131     &   11.49   &   Proto-stellar      &       G348.579$-$0.919   &   5.89    &   Proto-stellar      &     G341.932$-$0.174   &   10.47   &   HII                \\ 
G008.407$-$0.299   &   $-$     &   Proto-stellar      &       G348.599$-$0.912   &   48.73   &   Proto-stellar      &     G341.942$-$0.166   &   17.35   &   HII                \\ 
G008.459$-$0.222   &   6.45    &   Proto-stellar      &       G350.411$-$0.064   &   16.24   &   Proto-stellar      &     G342.323+0.294     &   4.05    &   HII                \\ 
G008.486$-$0.979   &   4.75    &   Proto-stellar      &       G350.521$-$0.349   &   3.18    &   Proto-stellar      &     G343.128$-$0.062   &   11.67   &   HII                \\ 
G008.504$-$0.982   &   2.71    &   Proto-stellar      &       G350.719+0.876     &   0.66    &   Proto-stellar      &     G344.102$-$0.662   &   9.90    &   HII                \\ 
G008.642$-$0.397   &   $-$     &   Proto-stellar      &       G350.757+0.942     &   35.94   &   Proto-stellar      &     G344.424+0.046     &   15.44   &   HII                \\ 
G008.684$-$0.367   &   35.76   &   Proto-stellar      &       G350.939+0.757     &   2.39    &   Proto-stellar      &     G345.196$-$0.744   &   6.07    &   HII                \\ 
G008.706$-$0.414   &   10.88   &   Proto-stellar      &       G350.941+0.746     &   3.32    &   Proto-stellar      &     G346.076$-$0.056   &   12.09   &   HII                \\ 
G009.284$-$0.147   &   9.54    &   Proto-stellar      &       G351.444+0.659     &   23.38   &   Proto-stellar      &     G348.533$-$0.982   &   30.14   &   HII                \\ 
G010.284$-$0.114   &   41.91   &   Proto-stellar      &       G351.809+0.644     &   1.24    &   Proto-stellar      &     G348.549$-$0.979   &   2.90    &   HII                \\ 
G011.004$-$0.071   &   13.86   &   Proto-stellar      &       G352.238$-$0.161   &   3.24    &   Proto-stellar      &     G351.041$-$0.336   &   4.52    &   HII                \\ 
G011.082$-$0.534   &   7.69    &   Proto-stellar      &       G352.502+0.784     &   2.45    &   Proto-stellar      &     G352.226$-$0.161   &   $-$     &   HII                \\ 
G013.169+0.077     &   7.19    &   Proto-stellar      &       G353.012+0.557     &   0.58    &   Proto-stellar      &     G352.246$-$0.064   &   4.57    &   HII                \\ 
G013.178+0.059     &   8.93    &   Proto-stellar      &       G353.579+0.659     &   0.65    &   Proto-stellar      &     G352.492+0.796     &   1.28    &   HII                \\ 
G013.234$-$0.071   &   8.89    &   Proto-stellar      &       G354.139+0.089     &   4.06    &   Proto-stellar      &     G353.089+0.444     &   0.90    &   HII                \\ 
G013.243$-$0.086   &   7.96    &   Proto-stellar      &       G354.629$-$0.611   &   1.18    &   Proto-stellar      &     G353.204+0.889     &   0.26    &   HII                \\ 
G013.281$-$0.321   &   16.08   &   Proto-stellar      &       G355.196$-$0.486   &   2.29    &   Proto-stellar      &     G353.396$-$0.071   &   11.55   &   HII                \\ 
G013.902$-$0.516   &   4.03    &   Proto-stellar      &       G355.264$-$0.269   &   3.21    &   Proto-stellar      &     G354.616+0.472     &   7.41    &   HII                \\ 
G014.227$-$0.511   &   6.99    &   Proto-stellar      &       G355.412+0.102     &   18.75   &   Proto-stellar      &     G013.213+0.039     &   5.53    &   PDR                \\ 
G014.777$-$0.487   &   5.19    &   Proto-stellar      &       G355.638$-$0.057   &   $-$     &   Proto-stellar      &     G308.744+0.532     &   4.57    &   PDR                \\ 
G019.472+0.171     &   51.73   &   Proto-stellar      &       G356.482+0.189     &   $-$     &   Proto-stellar      &     G319.899+0.791     &   1.85    &   PDR                \\ 
G300.218$-$0.111   &   1.87    &   Proto-stellar      &       G357.609$-$0.617   &   $-$     &   Proto-stellar      &     G331.496$-$0.079   &   10.44   &   PDR                \\ 
G304.672+0.257     &   $-$     &   Proto-stellar      &       G358.979+0.084     &   $-$     &   Proto-stellar      &     G331.498$-$0.102   &   11.61   &   PDR                \\ 
G304.886+0.636     &   1.70    &   Proto-stellar      &       G000.166$-$0.446   &   $-$     &   HII                &     G332.990$-$0.441   &   0.26    &   PDR                \\ 
G305.822$-$0.114   &   7.05    &   Proto-stellar      &       G003.998+0.327     &   $-$     &   HII                &     G337.704$-$0.354   &   5.77    &   PDR                \\ 
G309.154$-$0.349   &   6.08    &   Proto-stellar      &       G004.003+0.339     &   2.45    &   HII                &     G337.934$-$0.507   &   7.45    &   PDR                \\ 
G309.911+0.324     &   8.87    &   Proto-stellar      &       G005.834$-$0.514   &   4.14    &   HII                &     G343.489$-$0.064   &   1.65    &   PDR                \\ 
G309.991+0.507     &   3.02    &   Proto-stellar      &       G005.897$-$0.444   &   8.28    &   HII                &     G303.268+1.247     &   1.93    &   Unknown            \\ 
G310.374$-$0.304   &   4.33    &   Proto-stellar      &       G008.401$-$0.289   &   3.46    &   HII                &     G331.029$-$0.431   &   12.74   &   Unknown            \\ 
G316.084$-$0.674   &   $-$     &   Proto-stellar      &       G008.411$-$0.347   &   6.64    &   HII                &     G331.034$-$0.419   &   10.69   &   Unknown            \\ 
G323.799+0.017     &   9.48    &   Proto-stellar      &       G008.671$-$0.356   &   21.35   &   HII                &     G331.051$-$0.419   &   7.75    &   Unknown            \\ 
G326.626+0.834     &   2.44    &   Proto-stellar      &       G008.736$-$0.362   &   9.50    &   HII                &     G333.076$-$0.417   &   2.47    &   Unknown            \\ 
G326.796+0.382     &   1.55    &   Proto-stellar      &       G010.288$-$0.124   &   5.50    &   HII                &     G336.958$-$0.224   &   15.56   &   Unknown            \\ 
G327.393+0.199     &   6.31    &   Proto-stellar      &       G010.404$-$0.201   &   49.02   &   HII                &     G337.927$-$0.432   &   2.12    &   Unknown            \\ 
G327.918$-$0.612   &   2.61    &   Proto-stellar      &       G011.109$-$0.397   &   19.97   &   HII                &     G338.384+0.011     &   2.91    &   Unknown            \\ 
G328.549+0.271     &   3.04    &   Proto-stellar      &       G011.902$-$0.141   &   12.07   &   HII                &     G340.206$-$0.049   &   8.85    &   Unknown            \\ 
G329.164$-$0.286   &   12.77   &   Proto-stellar      &       G012.418+0.506     &   7.17    &   HII                &     G340.261$-$0.196   &   4.42    &   Unknown            \\ 
G329.312$-$0.302   &   3.35    &   Proto-stellar      &       G012.878$-$0.287   &   2.04    &   HII                &     G348.554$-$0.941   &   2.39    &   Unknown            \\ 
G329.721+1.165     &   6.13    &   Proto-stellar      &       G013.131$-$0.152   &   4.93    &   HII                &     G351.173+0.632     &   1.59    &   Unknown            \\ 
G331.374$-$0.314   &   $-$     &   Proto-stellar      &       G302.021+0.251     &   5.02    &   HII                &     G351.434+0.676     &   4.44    &   Unknown            \\ 
G331.571$-$0.229   &   1.05    &   Proto-stellar      &       G304.586+0.582     &   0.67    &   HII                &     G353.022+0.502     &   1.67    &   Unknown            \\ 
G331.709+0.582     &   19.93   &   Proto-stellar      &       G308.646+0.647     &   4.17    &   HII                &     G353.262+0.626     &   0.60    &   Unknown            \\ 
      \hline
     \end{tabular}
     \medskip
     NOTE. Columns are (from left to right)  Clump names, mass infall rates, \emph{Spitzer} classification, Clump names, mass infall rates, \emph{Spitzer} classification,
     Clump names, mass infall rates, \emph{Spitzer} classification. The units are unit of 0.001 M$_{\odot}$ yr$^{-1}$. \\
  \end{minipage}
\end{table*}
\normalsize

\begin{figure}
  \begin{center}
  \includegraphics[width=0.45\textwidth]{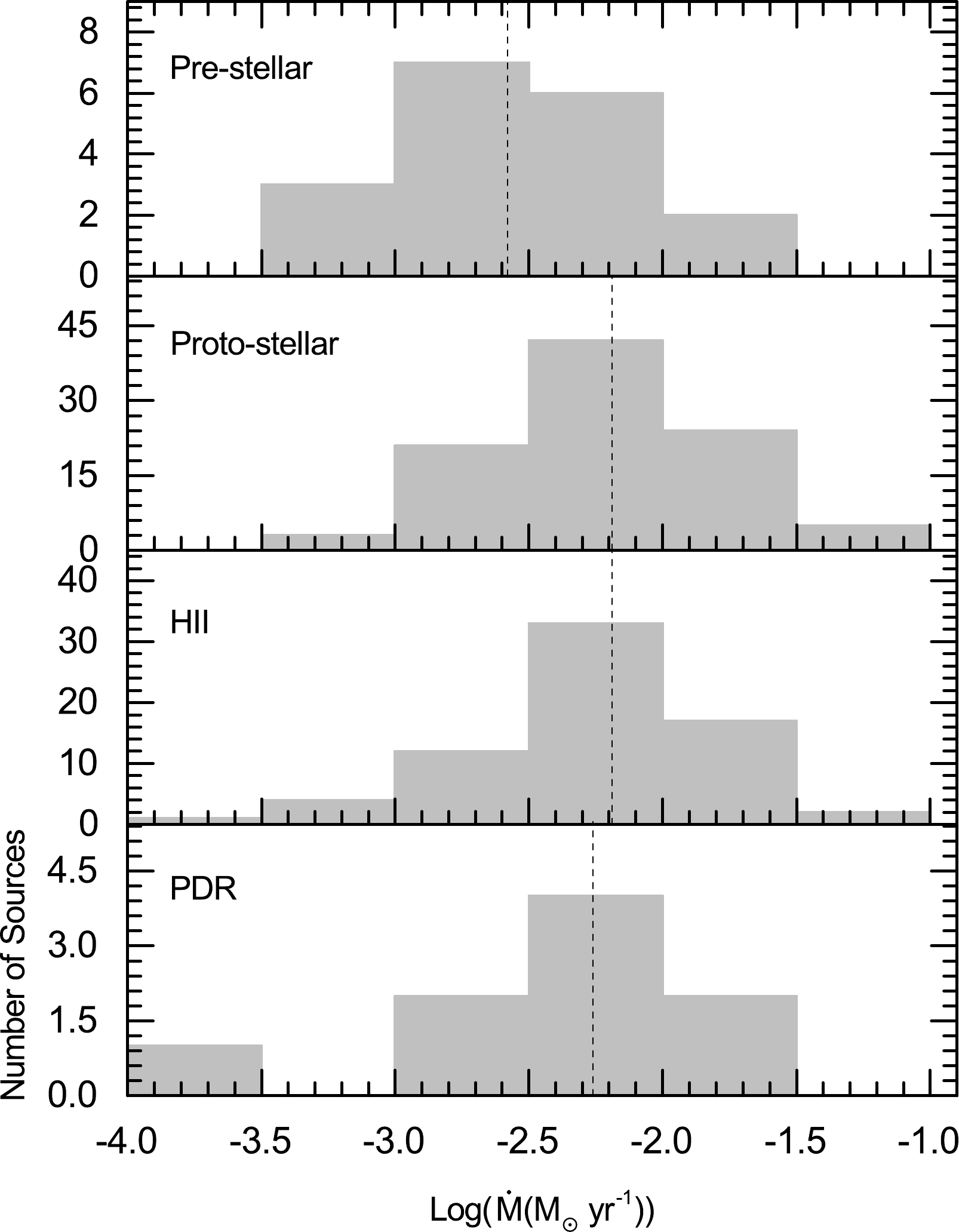}
  \end{center}
  \caption[dum]{Mass infall rate distribution for different evolutionary stages. Median values are indicated by the dashed black vertical lines.}
\end{figure}

\begin{figure}
  \begin{center}
  \includegraphics[width=0.45\textwidth]{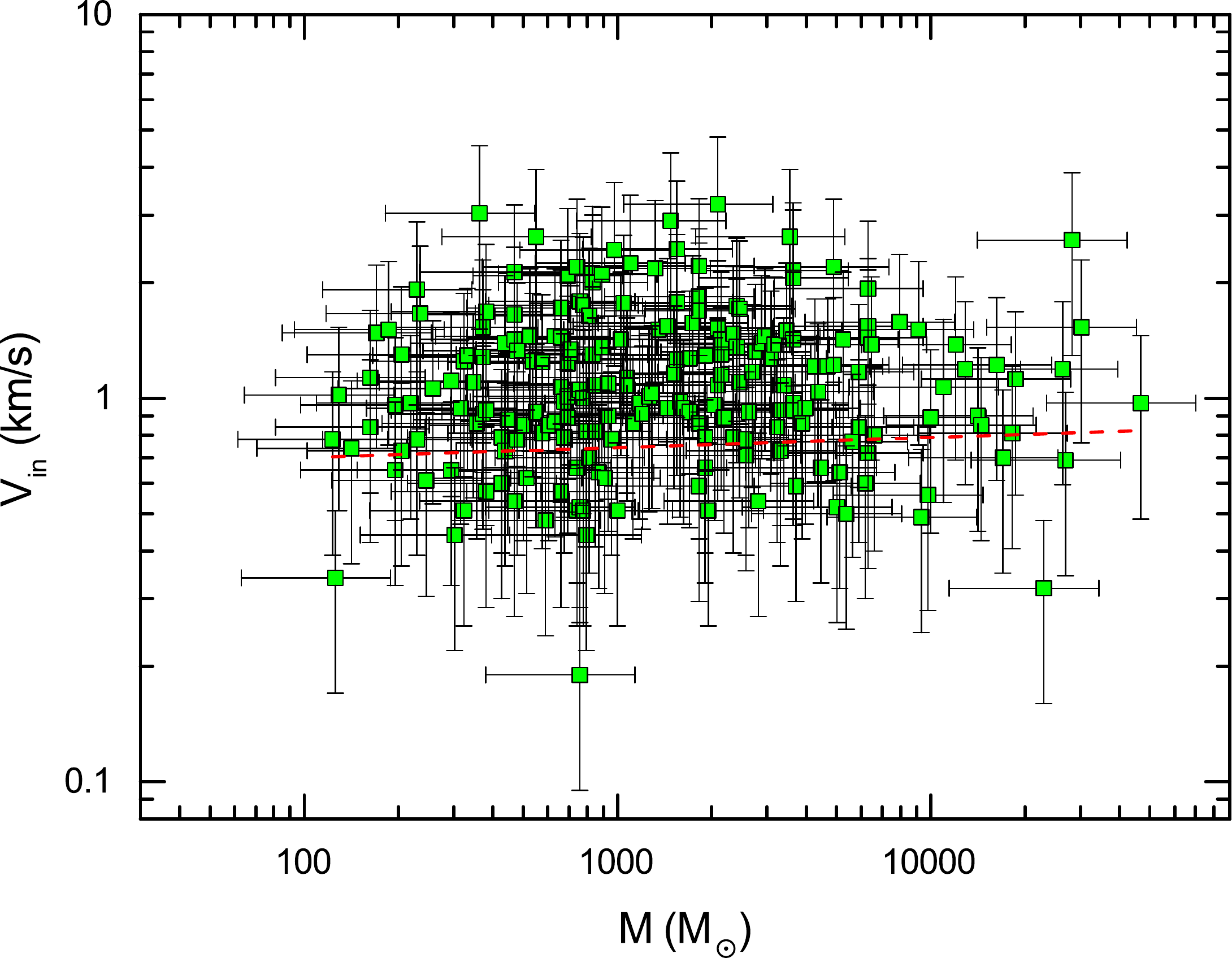}
  \end{center}
  \caption[dum]{Relationship between infall velocity and the total dust mass. The red dashed line represents the power-law fit to all infall candidates whose mass have been determined. We assume a 50\% error in the measurements of both infall velocity and mass.}
\end{figure}

\begin{figure}
  \begin{center}
  \includegraphics[width=0.45\textwidth]{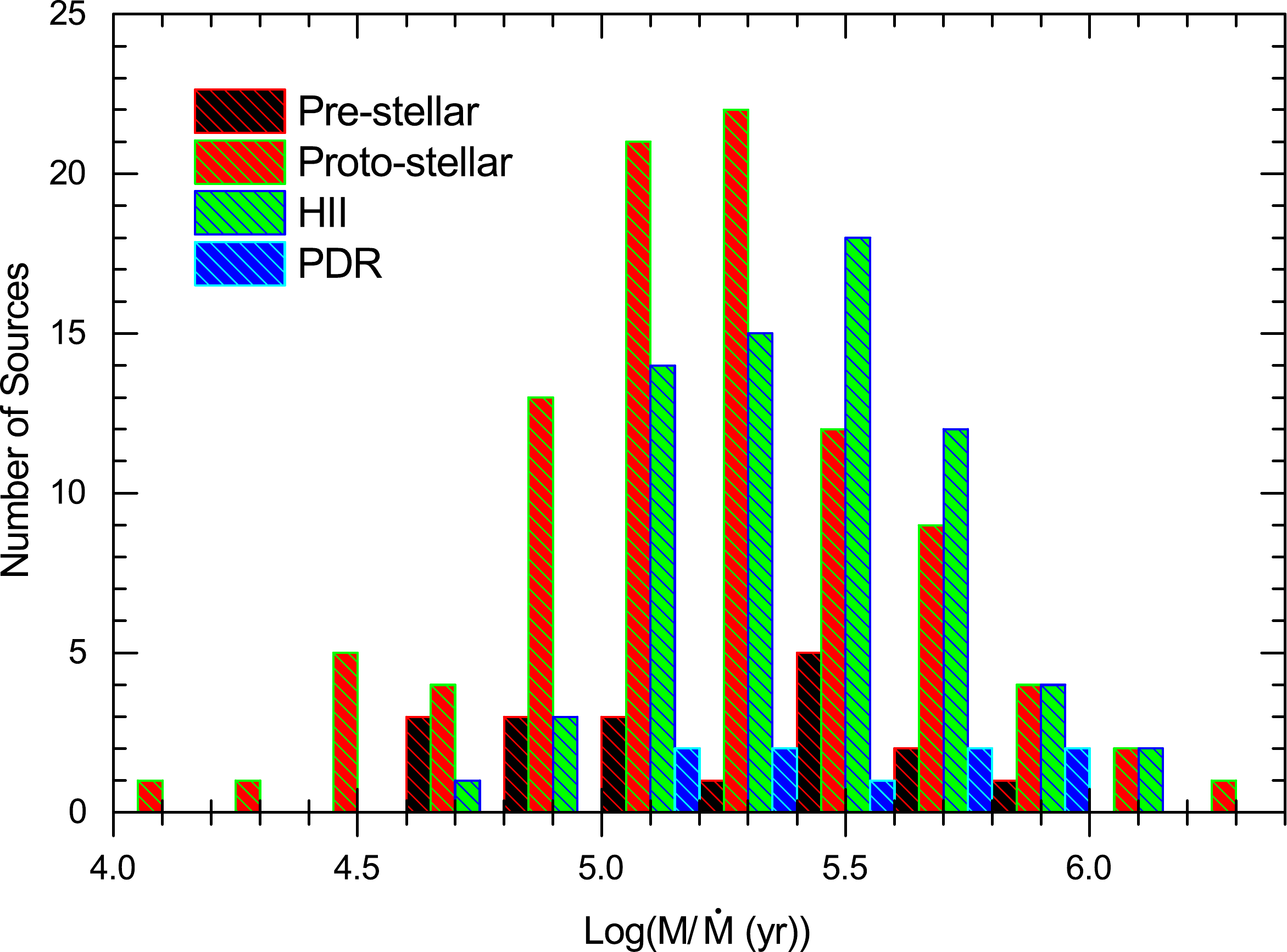}
  \end{center}
  \caption[dum]{Histograms of the depletion timescales of the infall candidates separated by evolutionary stage.}
\end{figure}

\begin{table*}
\tiny
 \centering
  \begin{minipage}{165mm}
   \caption{Summary of the derived infall candidates and clumps where infall is not detected properties.}
    \begin{tabular}{lrrrrrrrrr}
      \hline
      \multicolumn{1}{c}{Property}             &  \multicolumn{1}{c}{Group}     & \multicolumn{1}{c}{Notes}     &  \multicolumn{1}{c}{Counts} & \multicolumn{1}{c}{Mean}  &
      \multicolumn{1}{c}{Standard Error} & \multicolumn{1}{c}{Standard Deviation}   &  \multicolumn{1}{c}{Median}   &  \multicolumn{1}{c}{Minimum}  &
      \multicolumn{1}{c}{Maximum} \\
      \hline
       $\Delta V$(HCO$^{+}$ $J$=1-0)&    All sources     &   1     &      100      &     2.08   &   0.07       &       0.71        &         2.03     &  0.70  &  5.04       \\ 
($km\ s^{-1}$)               &                    &   2     &      227      &     2.06   &   0.04       &       0.68        &         1.99     &  0.64  &  4.18       \\
                             &    Pre-stellar     &   1     &      12       &     2.15   &   0.27       &       0.94        &         2.11     &  0.74  &  4.13       \\
                             &                    &   2     &      22       &     2.06   &   0.13       &       0.62        &         2.04     &  1.16  &  3.30       \\
                             &    Proto-stellar   &   1     &      49       &     2.10   &   0.10       &       0.69        &         1.99     &  1.11  &  5.04       \\
                             &                    &   2     &      100      &     1.97   &   0.07       &       0.67        &         1.82     &  0.70  &  4.14       \\
                             &    HII             &   1     &      30       &     2.00   &   0.12       &       0.65        &         2.00     &  0.70  &  3.35       \\
                             &                    &   2     &      63       &     2.20   &   0.08       &       0.67        &         2.15     &  0.64  &  4.18       \\
                             &    PDR             &   1     &      5        &     2.33   &   0.15       &       0.35        &         2.13     &  2.08  &  2.99       \\
                             &                    &   2     &      19       &     1.92   &   0.12       &       0.54        &         1.82     &  0.90  &  3.02       \\
$\delta v$(HCO$^{+}$ $J$=1-0)&    All sources     &   1     &      100      &    -0.54   &   0.03       &       0.28        &        -0.49     & -1.92  & -0.10       \\
                             &                    &   2     &      227      &     0.13   &   0.02       &       0.32        &         0.10     & -0.88  &  0.92       \\
                             &    Pre-stellar     &   1     &      12       &    -0.58   &   0.06       &       0.21        &        -0.61     & -1.05  & -0.33       \\
                             &                    &   2     &      22       &     0.03   &   0.08       &       0.38        &        -0.05     & -0.88  &  0.84       \\
                             &    Proto-stellar   &   1     &      49       &    -0.54   &   0.04       &       0.27        &        -0.51     & -1.92  & -0.10       \\
                             &                    &   2     &      100      &     0.14   &   0.03       &       0.32        &         0.11     & -0.86  &  0.84       \\
                             &    HII             &   1     &      30       &    -0.53   &   0.06       &       0.33        &        -0.47     & -1.90  & -0.21       \\
                             &                    &   2     &      63       &     0.12   &   0.04       &       0.32        &         0.08     & -0.55  &  0.92       \\
                             &    PDR             &   1     &      5        &    -0.50   &   0.10       &       0.21        &        -0.45     & -0.90  & -0.26       \\
                             &                    &   2     &      19       &     0.09   &   0.04       &       0.17        &         0.12     & -0.20  &  0.43       \\
$\delta v$(HNC $J$=1-0)      &    All sources     &   1     &      100      &    -0.26   &   0.02       &       0.21        &        -0.25     & -1.75  &  0.14       \\
                             &                    &   2     &      227      &     0.07   &   0.01       &       0.21        &         0.06     & -0.80  &  0.72       \\
                             &    Pre-stellar     &   1     &      12       &    -0.27   &   0.06       &       0.20        &        -0.25     & -0.69  &  0.14       \\
                             &                    &   2     &      22       &     0.03   &   0.06       &       0.29        &         0.02     & -0.80  &  0.72       \\
                             &    Proto-stellar   &   1     &      49       &    -0.27   &   0.04       &       0.25        &        -0.25     & -1.75  &  0.06       \\
                             &                    &   2     &      100      &     0.07   &   0.02       &       0.21        &         0.08     & -0.62  &  0.51       \\
                             &    HII             &   1     &      30       &    -0.26   &   0.03       &       0.15        &        -0.25     & -0.57  &  0.13       \\
                             &                    &   2     &      63       &     0.07   &   0.03       &       0.21        &         0.04     & -0.42  &  0.67       \\
                             &    PDR             &   1     &      5        &    -0.22   &   0.06       &       0.13        &        -0.23     & -0.41  & -0.03       \\
                             &                    &   2     &      19       &     0.08   &   0.03       &       0.14        &         0.07     & -0.19  &  0.33       \\
T                            &    All sources     &   1     &      231      &     19.6   &   0.3        &       5.0         &         18.9     &  10.6  &  42.6       \\
($K$)                        &                    &   2     &      501      &     21.6   &   0.2        &       5.5         &         21.0     &  10.5  &  38.7       \\
                             &    Pre-stellar     &   1     &      28       &     17.2   &   0.8        &       4.1         &         17.0     &  10.6  &  25.2       \\
                             &                    &   2     &      40       &     18.6   &   0.8        &       5.0         &         17.4     &  10.5  &  30.7       \\
                             &    Proto-stellar   &   1     &      107      &     17.5   &   0.4        &       3.7         &         17.2     &  11.0  &  32.0       \\
                             &                    &   2     &      185      &     18.8   &   0.3        &       4.6         &         18.3     &  10.6  &  36.8       \\
                             &    HII             &   1     &      72       &     22.2   &   0.5        &       4.5         &         22.1     &  13.0  &  41.5       \\
                             &                    &   2     &      163      &     23.6   &   0.4        &       4.9         &         22.8     &  12.0  &  38.7       \\
                             &    PDR             &   1     &      9        &     28.3   &   2.1        &       6.3         &         26.1     &  20.9  &  42.6       \\
                             &                    &   2     &      62       &     26.3   &   0.6        &       4.8         &         25.7     &  15.8  &  36.9       \\
Aspect ratio                 &    All sources     &   1     &      231      &     1.63   &   0.03       &       0.53        &         1.50     &  1.00  &  4.89       \\
                             &                    &   2     &      501      &     1.70   &   0.02       &       0.56        &         1.57     &  1.00  &  4.50       \\
                             &    Pre-stellar     &   1     &      28       &     1.76   &   0.13       &       0.69        &         1.65     &  1.00  &  4.20       \\
                             &                    &   2     &      40       &     1.85   &   0.09       &       0.54        &         1.72     &  1.11  &  3.33       \\
                             &    Proto-stellar   &   1     &      107      &     1.59   &   0.04       &       0.42        &         1.53     &  1.00  &  3.10       \\
                             &                    &   2     &      185      &     1.64   &   0.04       &       0.48        &         1.53     &  1.00  &  4.30       \\
                             &    HII             &   1     &      72       &     1.51   &   0.05       &       0.41        &         1.38     &  1.00  &  2.69       \\
                             &                    &   2     &      163      &     1.60   &   0.04       &       0.47        &         1.52     &  1.00  &  3.40       \\
                             &    PDR             &   1     &      9        &     1.77   &   0.15       &       0.46        &         1.58     &  1.21  &  2.64       \\
                             &                    &   2     &      62       &     1.85   &   0.09       &       0.69        &         1.70     &  1.00  &  4.50       \\
Effective radius             &    Pre-stellar     &   3     &      50       &     0.77   &   0.12       &       0.83        &         0.47     &  0.07  &  3.15       \\
($pc$)                       &    Proto-stellar   &   3     &      249      &     1.00   &   0.07       &       1.12        &         0.63     &  0.09  &  9.63       \\
                             &    HII             &   3     &      219      &     1.29   &   0.07       &       1.00        &         0.97     &  0.08  &  5.71       \\
                             &    PDR             &   3     &      71       &     1.44   &   0.16       &       1.34        &         0.94     &  0.14  &  5.60       \\
$S_{peak}$                   &    Pre-stellar     &   1     &      28       &     0.99   &   0.10       &       0.52        &         0.83     &  0.52  &  3.08       \\
($Jy\ beam^{-1}$)            &                    &   2     &      40       &     1.17   &   0.16       &       1.01        &         0.92     &  0.50  &  6.73       \\
                             &    Proto-stellar   &   1     &      107      &     2.42   &   0.44       &       4.59        &         1.44     &  0.48  &  40.80      \\
                             &                    &   2     &      185      &     2.30   &   0.22       &       3.04        &         1.40     &  0.49  &  32.86      \\
                             &    HII             &   1     &      72       &     3.66   &   0.43       &       3.62        &         2.56     &  0.62  &  18.93      \\
                             &                    &   2     &      163      &     5.28   &   0.65       &       8.29        &         2.57     &  0.44  &  59.26      \\
                             &    PDR             &   1     &      9        &     2.02   &   0.34       &       1.01        &         2.40     &  0.79  &  3.75       \\
                             &                    &   2     &      62       &     2.48   &   0.27       &       2.12        &         1.87     &  0.67  &  13.10      \\
Log($N_{H_{2}}$)             &    All sources     &   1     &      231      &     22.67  &   0.02       &       0.31        &         22.65    &  21.91 &  23.92      \\
($cm^{-2}$)                  &                    &   2     &      501      &     22.65  &   0.01       &       0.33        &         22.59    &  21.90 &  23.93      \\
                             &    Pre-stellar     &   1     &      28       &     22.48  &   0.03       &       0.17        &         22.49    &  22.08 &  22.78      \\
                             &                    &   2     &      40       &     22.47  &   0.03       &       0.20        &         22.45    &  21.97 &  23.09      \\
                             &    Proto-stellar   &   1     &      107      &     22.70  &   0.03       &       0.28        &         22.67    &  22.14 &  23.92      \\
                             &                    &   2     &      185      &     22.67  &   0.02       &       0.27        &         22.64    &  22.00 &  23.71      \\
                             &    HII             &   1     &      72       &     22.76  &   0.04       &       0.33        &         22.74    &  22.12 &  23.53      \\
                             &                    &   2     &      163      &     22.77  &   0.03       &       0.40        &         22.69    &  21.90 &  23.93      \\
                             &    PDR             &   1     &      9        &     22.45  &   0.11       &       0.32        &         22.56    &  21.91 &  22.81      \\
                             &                    &   2     &      62       &     22.53  &   0.04       &       0.28        &         22.50    &  22.11 &  23.46      \\
$S_{int}$                    &    Pre-stellar     &   1     &      28       &     9.83   &   1.81       &       9.55        &         7.86     &  2.20  &  51.81      \\
($Jy$)                       &                    &   2     &      40       &     10.90  &   1.78       &       11.25       &         8.32     &  1.30  &  66.94      \\
                             &    Proto-stellar   &   1     &      107      &     15.54  &   3.24       &       33.54       &         9.01     &  1.84  &  343.64     \\
                             &                    &   2     &      185      &     16.48  &   1.49       &       20.25       &         9.75     &  1.15  &  139.92     \\
                             &    HII             &   1     &      72       &     21.62  &   2.09       &       17.76       &         15.19    &  1.56  &  89.14      \\
                             &                    &   2     &      163      &     35.40  &   3.91       &       49.90       &         16.99    &  1.75  &  339.23     \\
                             &    PDR             &   1     &      9        &     20.32  &   4.37       &       13.10       &         18.47    &  4.90  &  51.93      \\
                             &                    &   2     &      62       &     24.46  &   2.98       &       23.50       &         15.63    &  2.91  &  111.08     \\
Log($M_{clump}$)             &    All sources     &   1     &      213      &     3.02   &   0.04       &       0.65        &         2.99     &  1.00  &  4.67       \\
($M_{\odot}$)                &                    &   2     &      462      &     3.14   &   0.03       &       0.72        &         3.09     &  0.92  &  5.24       \\
                             &    Pre-stellar     &   1     &      20       &     2.66   &   0.16       &       0.73        &         2.70     &  1.36  &  4.27       \\
                             &                    &   2     &      35       &     2.76   &   0.12       &       0.73        &         2.77     &  0.92  &  4.23       \\
                             &    Proto-stellar   &   1     &      99       &     3.00   &   0.07       &       0.66        &         2.94     &  1.00  &  4.67       \\
                             &                    &   2     &      163      &     3.06   &   0.05       &       0.68        &         2.94     &  1.33  &  4.86       \\
                             &    HII             &   1     &      70       &     3.16   &   0.07       &       0.58        &         3.23     &  1.63  &  4.48       \\
                             &                    &   2     &      154      &     3.34   &   0.05       &       0.64        &         3.36     &  1.59  &  5.24       \\
                             &    PDR             &   1     &      9        &     3.07   &   0.16       &       0.49        &         3.03     &  2.10  &  3.71       \\
                             &                    &   2     &      62       &     3.20   &   0.10       &       0.79        &         3.11     &  1.47  &  4.89       \\
Log(Volume density)          &    All sources     &   1     &      206      &     4.12   &   0.03       &       0.50        &         4.10     &  3.00  &  5.47       \\
($cm^{-3}$)                  &                    &   2     &      444      &     3.94   &   0.03       &       0.54        &         3.92     &  2.48  &  5.74       \\
                             &    Pre-stellar     &   1     &      18       &     4.26   &   0.13       &       0.57        &         4.25     &  3.28  &  5.38       \\
                             &                    &   2     &      32       &     4.16   &   0.11       &       0.60        &         4.13     &  2.90  &  5.28       \\
                             &    Proto-stellar   &   1     &      95       &     4.26   &   0.05       &       0.48        &         4.23     &  3.11  &  5.39       \\
                             &                    &   2     &      154      &     4.07   &   0.05       &       0.58        &         4.07     &  2.48  &  5.74       \\
                             &    HII             &   1     &      69       &     3.96   &   0.05       &       0.40        &         3.91     &  3.30  &  5.47       \\
                             &                    &   2     &      150      &     3.80   &   0.03       &       0.42        &         3.81     &  2.78  &  4.91       \\
                             &    PDR             &   1     &      9        &     3.81   &   0.16       &       0.49        &         3.76     &  3.26  &  4.99       \\
                             &                    &   2     &      62       &     3.76   &   0.06       &       0.50        &         3.70     &  2.78  &  5.07       \\
Log($\dot{M}$)               &    All sources     &   1     &      206      &    -2.27   &   0.03       &       0.42        &        -2.22     & -3.59  & -1.29       \\
M$_{\odot}$ yr$^{-1}$        &    Pre-stellar     &   1     &      18       &    -2.50   &   0.11       &       0.45        &        -2.58     & -3.41  & -1.71       \\
                             &    Proto-stellar   &   1     &      95       &    -2.22   &   0.04       &       0.42        &        -2.19     & -3.24  & -1.29       \\
                             &    HII             &   1     &      69       &    -2.25   &   0.05       &       0.43        &        -2.19     & -3.59  & -1.31       \\
                             &    PDR             &   1     &      9        &    -2.44   &   0.16       &       0.49        &        -2.26     & -3.59  & -1.94       \\ 
      \hline
     \end{tabular}
     \medskip
     NOTE. Column 3 notes: (1) - infall candidates, (2) - non-infall clumps, (3) - infall candidates + non-infall clumps.\\
  \end{minipage}
\end{table*}
\normalsize

\section{Conclusions}
We performed a search for infall candidates and studied the properties of a sample of 732 high-mass clumps selected from the MALT90 survey. The sample included 405 sources from HYX15, and 327 new sources presented in this work. Among of them, 68 are pre-stellar, 292 are proto-stellar, 235 are HII, 71 are PDR, while the remaining 66 sources could not be classified into these categories in GAE15. Our main conclusions can be summarized as follows.

(i) We identified 100 new reliable infall candidates on clump scale in this work. Combined with the results in HYX15, we identified 231 infall candidates from a total of 732 sources.

(ii) Simulations based on 1-dimensional spherically symmetric RATRAN model suggest that HCO$^{+}$(1-0) and HNC(1-0) lines are sensitive to infall motions on spatial scales at radii $<$133000 and 5600 $-$ 28000 AU, respectively.

(iii) The detection rates of the infall candidates in the pre-stellar, proto-stellar, HII and PDR stages are 41.2\%, 36.6\%, 30.6\% and 12.7\%, respectively. The detection rate decreases as clumps evolve, which is consistent with the dynamical evolution of high-mass star-forming regions.

(iv) The average and median temperatures derived for the total infall candidates (19.6 and 18.9K) are smaller than those measured for the clumps where infall is not detected (21.6 and 21.0K).

(v) The average values of the aspect ratio of the infall candidates at pre-stellar, proto-stellar, HII and PDR stages are 1.76, 1.59, 1.51 and 1.77, respectively. The corresponding values are 1.85, 1.64, 1.64 and 1.85 for the clumps where infall is not detected. K-S test suggests that there is a systematic difference in aspect ratio between the infall candidates and clumps where infall is not detected, which suggests that clumps with observed infall are more spherical than those without.

(vi) Both infall candidates and clumps where infall is not detected show an obvious trend of increasing in mass from the pre-stellar to proto-stellar, and to the HII stages, which indicates clumps still accumulate material efficiently as they evolve.

(vii) We identified two new MPC candidates (G333.018+0.766 and G348.531-0.972) and confirmed four highly reliable MPC candidates (G008.691-0.401, G348.183+0.482, G348.759-0.946 and G351.774-0.537) in HYX15.

(viii) The power indices of the ClMF are 2.04$\pm$0.16 and 2.17$\pm$0.31 for infall candidates and clumps where infall is not detected, respectively, which are similar to the power index of the IMF (2.35) and the cold Planck sources (2.0).

(ix) The median values of the infall rates of the infall candidates at the pre-stellar, proto-stellar, HII and PDR stages are 2.6$\times$10$^{-3}$, 7.0$\times$10$^{-3}$, 6.5$\times$10$^{-3}$ and 5.5$\times$10$^{-3}$ M$_\odot$ yr$^{-1}$, respectively. This also supports that infall candidates at later evolutionary stages still efficiently accumulate material.

(x) The power-law correlation between the infall velocity and the total clump mass ($M_{clump}>100M_{\odot}$) is $V_{in}=(0.62\pm0.12)M_{clump}^{0.03\pm0.03}$, our power index (0.03) is significantly less than the power index of high mass cores of 0.36 presented by \citet{2013ApJ...776...29L}. This suggests that infall rate of high-mass cores is significantly greater than that of high-mass clumps.

These infall candidates provide a very important sample for studying physical properties of regions where massive stars are forming. In subsequent papers, we will investigate the clump fragmentation and star cluster formation through high spatial resolution observations.

\section*{Acknowledgments}
This research has made use of the data products from the Millimetre Astronomy Legacy Team 90 GHz (MALT90) survey, the SIMBAD database, operated at CDS, Strasbourg, France, the data from \emph{Herschel}, a European Space Agency space observatory with science instruments provided by European led
consortia, the \emph{APEX} Telescope Large Area Survey of the Galaxy (ATLASGAL) survey, which is a collaboration between the Max-Planck-Gesellschaft, the European Southern Observatory (ESO) and the Universidad de Chile, and also used NASA/IPAC Infrared Science Archive, which is operated by the Jet Propulsion Laboratory, California Institute of Technology, under contract with the National Aeronautics and Space Administration.

This work was funded by The National Natural Science foundation of China under grant 11433008 and partly supported by National Basic Research Program of
China(973 program, 2012CB821802) and the National Natural Science foundation of China under grant 11373062 and 11303081, and The Program of the
Light in China¡¯s Western Region (LCRW) under grant Nos. RCPY201202 and XBBS-2014-24.

\appendix
\section[]{Dust temperature}
Approximately 92\% of 732 clumps of our sample were classified into one of the three preceding evolutionary stages using the criterion presented in HYX15: pre-stellar, proto-stellar and UCHII. UCHII clumps containing objects obeying the criteria defined in \citet{1989ApJS...69..831W}. Proto-stellar clumps containing objects obeying the \textbf{criteria:} a point-source should have [4.5] - [5.8] $>$ 1.0 and be detected at 8 $\mu$m; or a point-source should have [4.5] - [5.8] $>$ 0.7, [3.6] - [4.5] $>$ 0.7 and be detected at 8$\mu$m; or 24$\mu$m point point-source. Clumps associated with saturated 24$\mu$m sources or extended 8$\mu$m emission (e.g. photo-dissociation region) were indicated as ``Non". We refer to the remaining clumps as pre-stellar. Columns 5 and 10 of Table A1 list the evolutionary stages of all the clumps. Note that 63 clumps with ``Non" represent undetermined evolutionary stages using the classification criterion presented in HYX15.

In GAE15, they first tested the model of \citet{2011A&A...532A..43O} based on 14 sources using data at frequency $<$600 GHz, and found that the spectral index is in agreement with the absorption coefficient law of silicate-graphite grains, with $3\times10^4$ years of coagulation, and without ice coatings. They used this model of dust for the SED fitting, and derived dust temperature of their sample. Also, they excluded the 70 $\mu$m data of \emph{Herschel} survey and included the 870 $\mu$m data of ATLASGAL survey. We set $\beta$ as a free parameter, and derived it with dust temperature and critical frequency simultaneously using the modified blackbody model \citep{2012MNRAS.426..402F}. \emph{Herschel Hi-GAL} survey data at 70, 160, 250, 350 and 500 $\mu$m were used for SED fitting for proto-stellar and UCHII stage clumps, though 70 $\mu$m emission likely contains some contribution from a warmer dust component. As sources in the pre-stellar stage do not have 70$\mu$m emission, we fitted those just using 160, 250, 350 and 500 $\mu$m. There are 86 clumps whose Hi-GAL data are affected by saturation. \textbf{Overall}, we derived dust temperature for 646 clumps (see Columns 2 and 7 of Table A1).

In Figure A1, we show the dust temperature distributions derived from GAE15 (gray filled histogram) and in this paper (red histogram). The median dust temperatures are 20.4 and 23.3K, respectively. The K-S test (probability of $\ll$0.01\%) indicates that the difference between them is significant. Figure A2 shows the dust temperature distributions of infall candidates (blue histogram) separated by the pre-stellar, proto-stellar and UCHII clumps, where the median values are indicated by dashed black vertical lines, which are 19.9, 29.9 and 26.0K, respectively. For all clumps at different stages (gray filled histogram in Figure A2), the corresponding values are 20.3, 23.3 and 27.1K.

The dust opacity spectral index $\beta$ is sensitive to the maximum size of dust grains \citep{2014prpl.conf..339T}.  For dust grains with different chemical composition and porosity, the spectral index $\beta$ is larger than 2.0 at lower values of the maximum grain size \citep[see Figures 4 of][]{2014prpl.conf..339T}. Moreover, in the theoretical dust opacity models, the possibility of having $\beta>$2.0 is due to the effect of disordered charge distribution \citep{2007A&A...468..171M, 2011ApJ...728..143S}. Uncertainties in observed fluxes may also lead to incorrect spectral index (e.g. $\beta>$2.0) estimates from SED fitting \citep{2009ApJ...696.2234S}. Also, we should note that the T-$\beta$ degeneracy \citep{2012ApJ...752...55K,2013A&A...556A..63J,2015A&A...584A..94J} and strong temperature gradients can induce incorrect results in blackbody determinations \citep{2009A&A...504..415S,2009ApJ...696.2234S,2012A&A...542A..21Y}. We plotted the distribution of $\beta$ (upper panel) and T-$\beta$ correlation (bottom panel) in Figure A3. In fact, about twelve percent sources have $\beta>$2.0, while only fourteen sources have $\beta>$2.5. From T-$\beta$ correlation, we can see that only sources with $\beta>$2.5 show a relationship with temperature. These clumps with controversial $\beta$ represent only about ten per cent of our final sample and are therefore unlikely to affect the results significantly. Uncertainties in temperature are small (often $<$5\%), and errors in $\beta$ are generally 10\%.

\begin{figure}
  \begin{center}
  \includegraphics[width=0.45\textwidth]{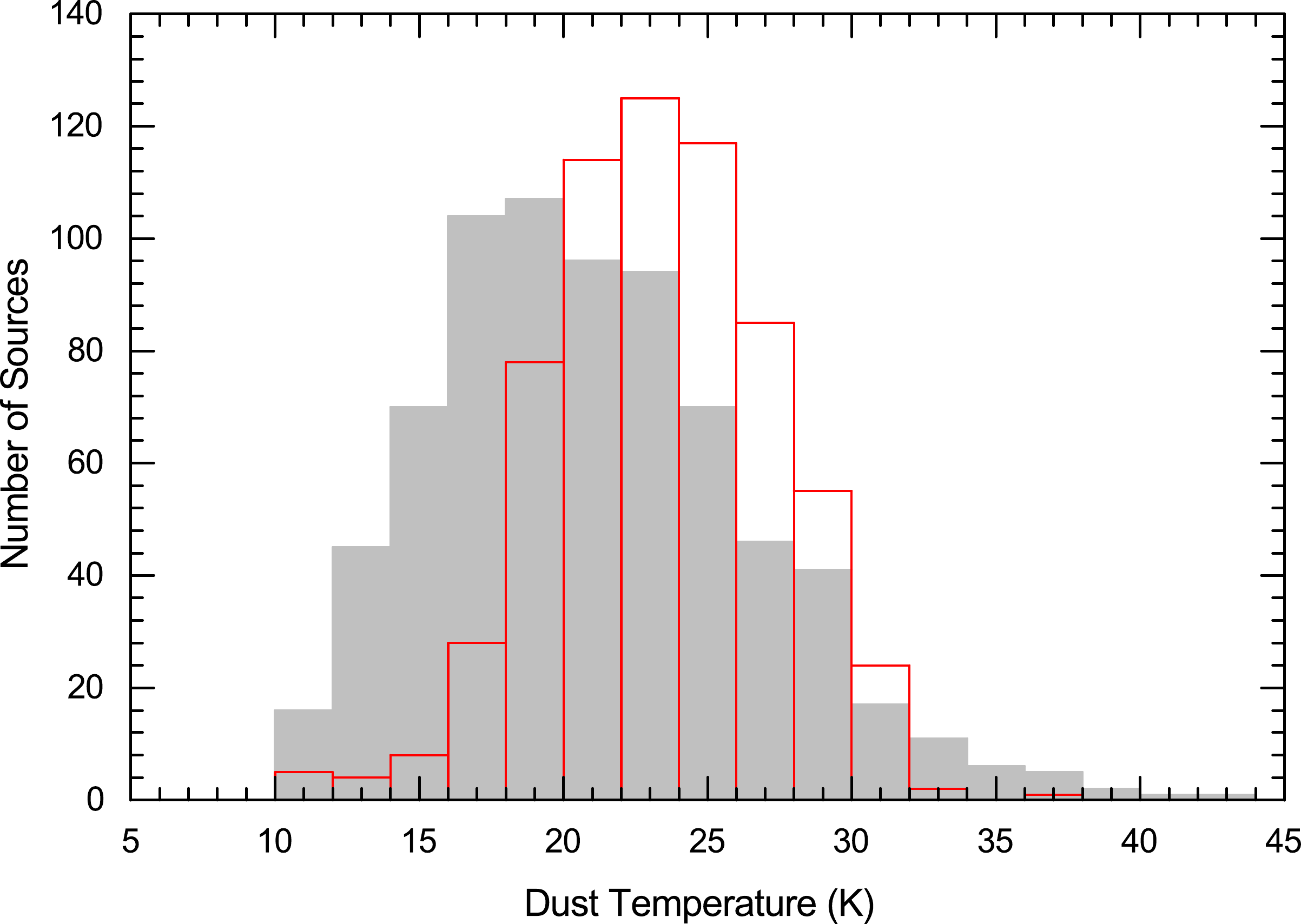}
  \end{center}
  \caption[dum]{Distributions of the dust temperatures derived from GAE15 (gray filled histogram) and this paper (red histogram).}
\end{figure}

\begin{figure}
  \begin{center}
  \includegraphics[width=0.45\textwidth]{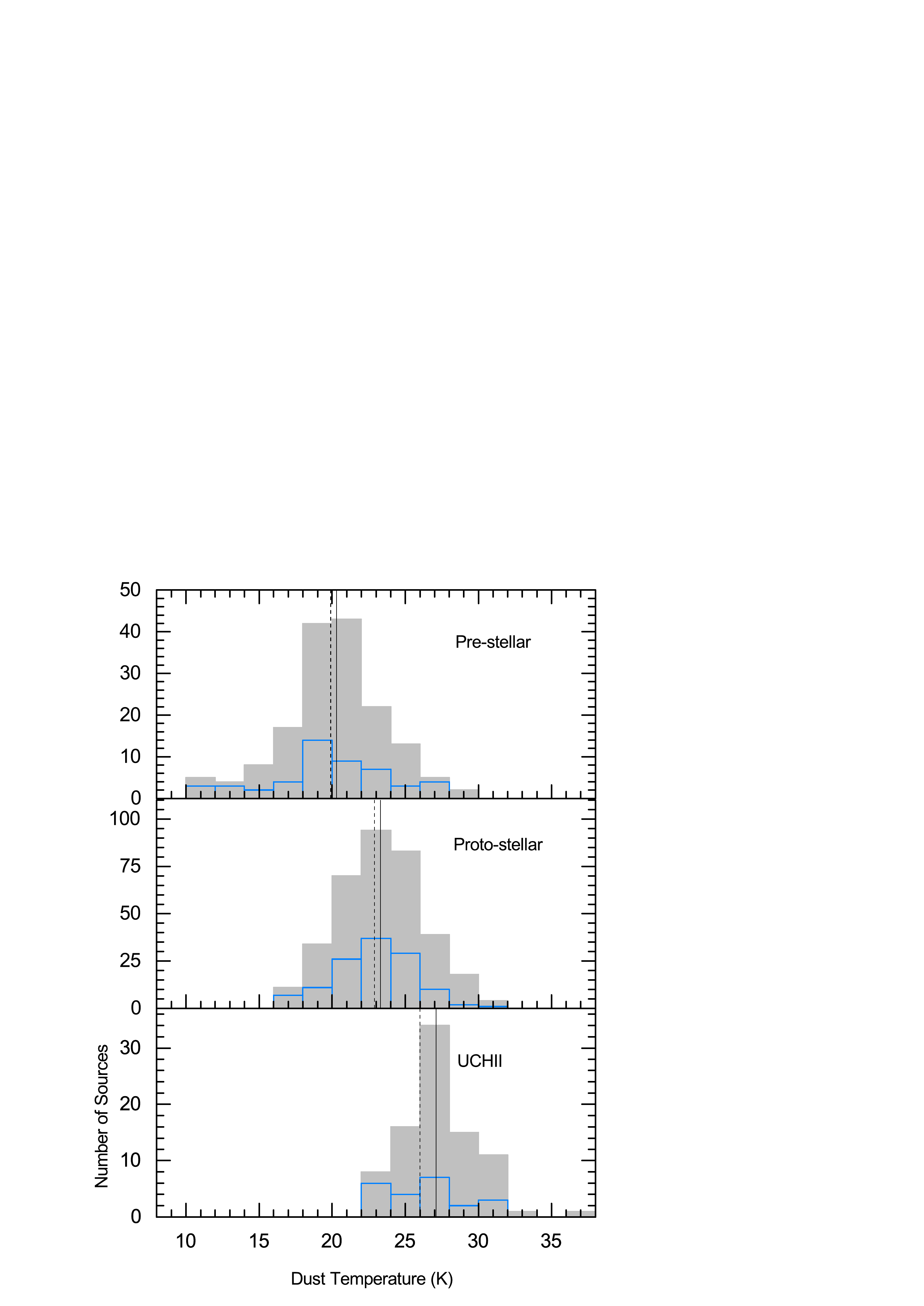}
  \end{center}
  \caption[dum]{Dust temperature distributions derived from \emph{Herschel}/Hi-GAL data separated by evolutionary stages. The corresponding stage are given on the top right of each panel. The median temperature for infall candidates (blue histogram) and the whole clumps (gray filled histogram) in each stage are indicated by the dashed and solid vertical black line.}
\end{figure}

\begin{figure}
  \begin{center}
  \includegraphics[width=0.45\textwidth]{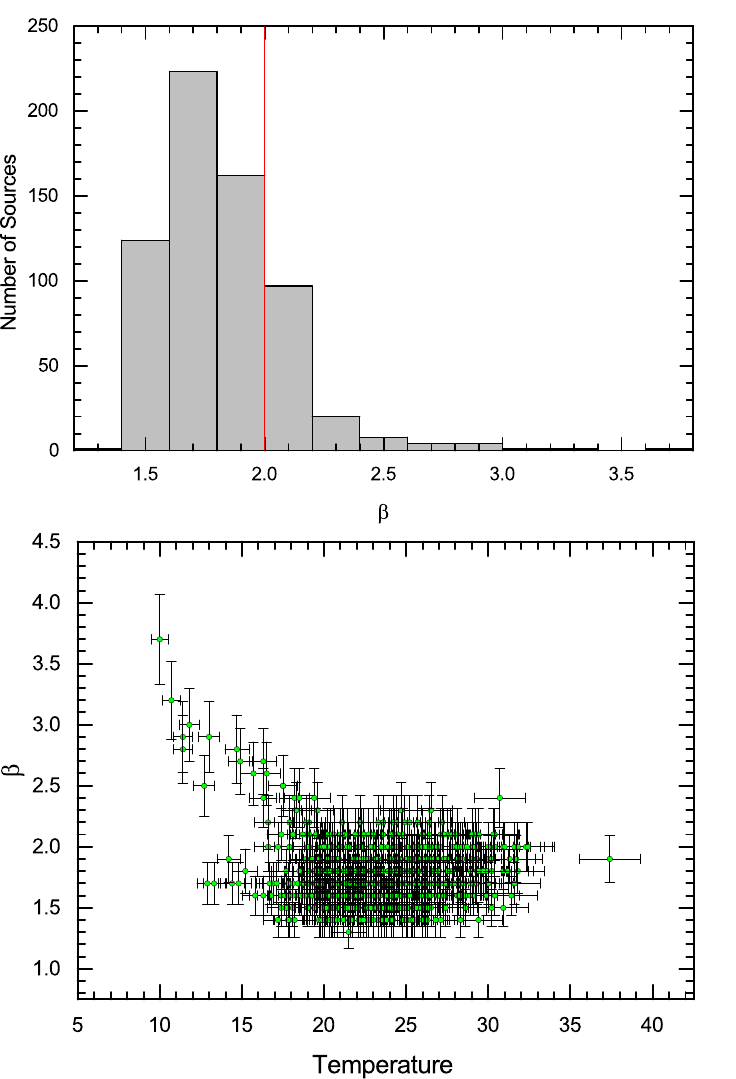}
  \end{center}
  \caption[dum]{Upper panel: The dust emissivity index $\beta$ distribution. The red solid line indicates $\beta$ = 2. Lower panel: The relationship between the fitted temperature and $\beta$.}
\end{figure}

\onecolumn
\clearpage
 \scriptsize
  \setlength{\LTcapwidth}{100mm}
    \begin{flushleft}
     Table A1. Dust properties derived from SEDs. The columns are as follows: (1) Clump names; (2) dust temperatures; (3) dust emissivity index; (4) critical frequency;
     (5) \emph{Spitzer} classification; (6) Clump names; (7) dust temperatures; (8) dust emissivity index; (9) critical frequency; (10) \emph{Spitzer} classification;. \\
    \end{flushleft}
\begin{longtable}{cccccccccc}

   \hline
    \multicolumn{1}{c}{Clump name} & \multicolumn{1}{c}{T$_{d}$} & \multicolumn{1}{c}{$\beta$} &  \multicolumn{1}{c}{$\nu_{c}$}  & \multicolumn{1}{c}{Classification} & \multicolumn{1}{c}{Clump name} & \multicolumn{1}{c}{T$_{d}$} & \multicolumn{1}{c}{$\beta$} &  \multicolumn{1}{c}{$\nu_{c}$}  & \multicolumn{1}{c}{Classification} \\

                                   & \multicolumn{1}{c}{T}       &                             &   \multicolumn{1}{c}{$\times10^{4}$GHz}       &                                    &                                &     K                       &                             &   \multicolumn{1}{c}{$\times10^{4}$GHz}       &                                     \\

    \multicolumn{1}{c}{(1)}        & \multicolumn{1}{c}{(2)}     & \multicolumn{1}{c}{(3)}     & \multicolumn{1}{c}{(4)}         & \multicolumn{1}{c}{(5)}            & \multicolumn{1}{c}{(6)}        & \multicolumn{1}{c}{(7)}     & \multicolumn{1}{c}{(8)}     & \multicolumn{1}{c}{(9)}         & \multicolumn{1}{c}{(10)}             \\
  \hline
  \endfirsthead

  Table A1 $-$ Continued\\
   \hline
    \multicolumn{1}{c}{Clump name} & \multicolumn{1}{c}{T$_{d}$} & \multicolumn{1}{c}{$\beta$} &  \multicolumn{1}{c}{$\nu_{c}$}  & \multicolumn{1}{c}{Classification} & \multicolumn{1}{c}{Clump name} & \multicolumn{1}{c}{T$_{d}$} & \multicolumn{1}{c}{$\beta$} &  \multicolumn{1}{c}{$\nu_{c}$}  & \multicolumn{1}{c}{Classification} \\

                                   & \multicolumn{1}{c}{T}       &                             &   \multicolumn{1}{c}{$\times10^{4}$GHz}       &                                    &                                &     K                       &                             &   \multicolumn{1}{c}{$\times10^{4}$GHz}       &                                     \\

    \multicolumn{1}{c}{(1)}        & \multicolumn{1}{c}{(2)}     & \multicolumn{1}{c}{(3)}     & \multicolumn{1}{c}{(4)}         & \multicolumn{1}{c}{(5)}            & \multicolumn{1}{c}{(6)}        & \multicolumn{1}{c}{(7)}     & \multicolumn{1}{c}{(8)}     & \multicolumn{1}{c}{(9)}         & \multicolumn{1}{c}{(10)}             \\
  \hline
  \endhead

  \hline
  \endfoot

  $*$G000.156$-$0.452    &    20.7  &   1.9   &  3.85    &    Pre-stellar       &     G301.814+0.781        &   31.7  &   1.9   &  5.49    &    Proto-stellar \\
$*$G000.534$-$0.644    &    24.3  &   1.7   &  5.06    &    Pre-stellar       &     G304.556+0.327        &   27.8  &   1.8   &  6.41    &    Proto-stellar \\
$*$G000.591$-$0.851    &    22.3  &   2.1   &  3.06    &    Pre-stellar       &     G305.137+0.069        &   24.6  &   1.8   &  4.24    &    Proto-stellar \\
$*$G003.416$-$0.354    &    11.4  &   2.8   &  0.61    &    Pre-stellar       &     G305.226+0.274        &   22.9  &   1.7   &  3.69    &    Proto-stellar \\
$*$G003.434$-$0.572    &    14.2  &   1.9   &  3.04    &    Pre-stellar       &     G305.242$-$0.041      &   24.3  &   1.5   &  10.20   &    Proto-stellar \\
$*$G005.917$-$0.311    &    20.7  &   1.8   &  4.87    &    Pre-stellar       &     G308.057$-$0.397      &   25.9  &   2.0   &  2.59    &    Proto-stellar \\
$*$G007.234$-$0.101    &    17.4  &   1.6   &  5.97    &    Pre-stellar       &     G308.754+0.547        &   28.1  &   1.9   &  4.70    &    Proto-stellar \\
$*$G010.199$-$0.316    &    20.2  &   2.1   &  1.53    &    Pre-stellar       &     G311.044+0.687        &   21.0  &   1.7   &  6.97    &    Proto-stellar \\
\end{longtable}
  NOTE. An $\ast$ indicates infall candidates.\\

\begin{thebibliography}{99}
\bibitem[\protect\citeauthoryear{Allen}{1973}]{1973asqu.book.....A} Allen C.~W., 1973, asqu.book,
\bibitem[\protect\citeauthoryear{Anderson et al.}{2014}]{2014ApJS..212....1A} Anderson L.~D., Bania T.~M., Balser D.~S., Cunningham V., Wenger T.~V., Johnstone B.~M., Armentrout W.~P., 2014, ApJS, 212, 1
\bibitem[\protect\citeauthoryear{Behrend \& Maeder}{2001}]{2001A&A...373..190B} Behrend R., Maeder A., 2001, A\&A, 373, 190
\bibitem[\protect\citeauthoryear{Benjamin et al.}{2003}]{2003PASP..115..953B} Benjamin R.~A., et al., 2003, PASP, 115, 953
\bibitem[\protect\citeauthoryear{Bergin \& Tafalla}{2007}]{2007ARA&A..45..339B} Bergin E.~A., Tafalla M., 2007, ARA\&A, 45, 339
\bibitem[\protect\citeauthoryear{Bernasconi \& Maeder}{1996}]{1996A&A...307..829B} Bernasconi P.~A., Maeder A., 1996, A\&A, 307, 829
\bibitem[\protect\citeauthoryear{Bonnell et al.}{2001}]{2001MNRAS.324..573B} Bonnell I.~A., Clarke C.~J., Bate M.~R., Pringle J.~E., 2001, MNRAS, 324, 573
\bibitem[\protect\citeauthoryear{Bonnell, Vine, \& Bate}{2004}]{2004MNRAS.349..735B} Bonnell I.~A., Vine S.~G., Bate M.~R., 2004, MNRAS, 349, 735
\bibitem[\protect\citeauthoryear{Bonnell \& Bate}{2006}]{2006MNRAS.370..488B} Bonnell I.~A., Bate M.~R., 2006, MNRAS, 370, 488
\bibitem[\protect\citeauthoryear{Bressert et al.}{2012}]{2012ApJ...758L..28B} Bressert E., Ginsburg A., Bally J., Battersby C., Longmore S., Testi L., 2012, ApJ, 758, L28
\bibitem[\protect\citeauthoryear{Busfield et al.}{2006}]{2006MNRAS.366.1096B} Busfield A.~L., Purcell C.~R., Hoare M.~G., Lumsden S.~L., Moore T.~J.~T., Oudmaijer R.~D., 2006, MNRAS, 366, 1096
\bibitem[\protect\citeauthoryear{Carey et al.}{2009}]{2009PASP..121...76C} Carey S.~J., et al., 2009, PASP, 121, 76
\bibitem[\protect\citeauthoryear{Chen et al.}{2013}]{2013ApJS..206....9C} Chen X., Gan C.-G., Ellingsen S.~P., He J.-H., Shen Z.-Q., Titmarsh A., 2013, ApJS, 206, 9
\bibitem[\protect\citeauthoryear{Churchwell et al.}{2009}]{2009PASP..121..213C} Churchwell E., et al., 2009, PASP, 121, 213
\bibitem[\protect\citeauthoryear{Dunham et al.}{2011}]{2011ApJ...731...90D} Dunham M.~K., Robitaille T.~P., Evans N.~J., II, Schlingman W.~M., Cyganowski C.~J., Urquhart J., 2011, ApJ, 731, 90
\bibitem[\protect\citeauthoryear{Ellsworth-Bowers et al.}{2013}]{2013ApJ...770...39E} Ellsworth-Bowers T.~P., et al., 2013, ApJ, 770, 39
\bibitem[\protect\citeauthoryear{Faimali et al.}{2012}]{2012MNRAS.426..402F} Faimali A., et al., 2012, MNRAS, 426, 402
\bibitem[\protect\citeauthoryear{Fa{\'u}ndez et al.}{2004}]{2004A&A...426...97F} Fa{\'u}ndez S., Bronfman L., Garay G., Chini R., Nyman L.-{\AA}., May J., 2004, A\&A, 426, 97
\bibitem[\protect\citeauthoryear{Giannetti et al.}{2014}]{2014A&A...570A..65G} Giannetti A., et al., 2014, A\&A, 570, A65
\bibitem[\protect\citeauthoryear{Gregersen et al.}{2000}]{2000ApJ...533..440G} Gregersen E.~M., Evans N.~J., II, Mardones D., Myers P.~C., 2000, ApJ, 533, 440
\bibitem[\protect\citeauthoryear{Griffin et al.}{2010}]{2010A&A...518L...3G} Griffin M.~J., et al., 2010, A\&A, 518, L3
\bibitem[\protect\citeauthoryear{Gutermuth \& Heyer}{2015}]{2015AJ....149...64G} Gutermuth R.~A., Heyer M., 2015, AJ, 149, 64
\bibitem[\protect\citeauthoryear{Guzm{\'a}n et al.}{2015}]{2015arXiv151100762G} Guzm{\'a}n A.~E., Sanhueza P., Contreras Y., Smith H.~A., Jackson J.~M., Hoq S., Rathborne J.~M., 2015, arXiv, arXiv:1511.00762
\bibitem[\protect\citeauthoryear{He et al.}{2015}]{2015MNRAS.450.1926H} He Y.-X., et al., 2015, MNRAS, 450, 1926
\bibitem[\protect\citeauthoryear{Hill et al.}{2005}]{2005MNRAS.363..405H} Hill T., Burton M.~G., Minier V., Thompson M.~A., Walsh A.~J., Hunt-Cunningham M., Garay G., 2005, MNRAS, 363, 405
\bibitem[\protect\citeauthoryear{Hogerheijde \& van der Tak}{2000}]{2000A&A...362..697H} Hogerheijde M.~R., van der Tak F.~F.~S., 2000, A\&A, 362, 697
\bibitem[\protect\citeauthoryear{Hoq et al.}{2013}]{2013ApJ...777..157H} Hoq S., et al., 2013, ApJ, 777, 157
\bibitem[\protect\citeauthoryear{Hosokawa, Yorke, \& Omukai}{2010}]{2010ApJ...721..478H} Hosokawa T., Yorke H.~W., Omukai K., 2010, ApJ, 721, 478
\bibitem[\protect\citeauthoryear{Ikeda \& Kitamura}{2009}]{2009ApJ...705L..95I} Ikeda N., Kitamura Y., 2009, ApJ, 705, L95
\bibitem[\protect\citeauthoryear{Jackson et al.}{2013}]{2013PASA...30...57J} Jackson J.~M., et al., 2013, PASA, 30, e057
\bibitem[\protect\citeauthoryear{Johnstone et al.}{2001}]{2001ApJ...559..307J} Johnstone D., Fich M., Mitchell G.~F., Moriarty-Schieven G., 2001, ApJ, 559, 307
\bibitem[\protect\citeauthoryear{Juvela et al.}{2013}]{2013A&A...556A..63J} Juvela M., Montillaud J., Ysard N., Lunttila T., 2013, A\&A, 556, A63
\bibitem[\protect\citeauthoryear{Juvela et al.}{2015}]{2015A&A...584A..94J} Juvela M., et al., 2015, A\&A, 584, A94
\bibitem[\protect\citeauthoryear{Kauffmann \& Pillai}{2010}]{2010ApJ...723L...7K} Kauffmann J., Pillai T., 2010, ApJ, 723, L7
\bibitem[\protect\citeauthoryear{Kauffmann et al.}{2010}]{2010ApJ...716..433K} Kauffmann J., Pillai T., Shetty R., Myers P.~C., Goodman A.~A., 2010, ApJ, 716, 433
\bibitem[\protect\citeauthoryear{Kelly et al.}{2012}]{2012ApJ...752...55K} Kelly B.~C., Shetty R., Stutz A.~M., Kauffmann J., Goodman A.~A., Launhardt R., 2012, ApJ, 752, 55
\bibitem[\protect\citeauthoryear{Klaassen \& Wilson}{2007}]{2007ApJ...663.1092K} Klaassen P.~D., Wilson C.~D., 2007, ApJ, 663, 1092
\bibitem[\protect\citeauthoryear{L{\'o}pez-Sepulcre, Cesaroni,\& Walmsley}{2010}]{2010A&A...517A..66L} L{\'o}pez-Sepulcre A., Cesaroni R., Walmsley C.~M., 2010, A\&A, 517, A66
\bibitem[\protect\citeauthoryear{Li et al.}{2015}]{2015AJ....150...60L} Li Y., Xu Y., Yang J., Du X.-Y., Lu D.-R., Li F.-C., 2015, AJ, 150, 60
\bibitem[\protect\citeauthoryear{Liu, Wu, \& Zhang}{2013}]{2013ApJ...776...29L} Liu T., Wu Y., Zhang H., 2013, ApJ, 776, 29
\bibitem[\protect\citeauthoryear{Lumsden et al.}{2013}]{2013ApJS..208...11L} Lumsden S.~L., Hoare M.~G., Urquhart J.~S., Oudmaijer R.~D., Davies B., Mottram J.~C., Cooper H.~D.~B., Moore T.~J.~T., 2013, ApJS, 208, 11
\bibitem[\protect\citeauthoryear{Mardones et al.}{1997}]{1997ApJ...489..719M} Mardones D., Myers P.~C., Tafalla M., Wilner D.~J., Bachiller R., Garay G., 1997, ApJ, 489, 719
\bibitem[\protect\citeauthoryear{McClure-Griffiths et al.}{2005}]{2005ApJS..158..178M} McClure-Griffiths N.~M., Dickey J.~M., Gaensler B.~M., Green A.~J., Haverkorn M., Strasser S., 2005, ApJS, 158, 178
\bibitem[\protect\citeauthoryear{McKee \& Tan}{2002}]{2002Natur.416...59M} McKee C.~F., Tan J.~C., 2002, Natur, 416, 59
\bibitem[\protect\citeauthoryear{McKee \& Tan}{2003}]{2003ApJ...585..850M} McKee C.~F., Tan J.~C., 2003, ApJ, 585, 850
\bibitem[\protect\citeauthoryear{Meny et al.}{2007}]{2007A&A...468..171M} Meny C., Gromov V., Boudet N., Bernard J.-P., Paradis D., Nayral C., 2007, A\&A, 468, 171
\bibitem[\protect\citeauthoryear{Miettinen}{2014}]{2014A&A...562A...3M} Miettinen O., 2014, A\&A, 562, A3
\bibitem[\protect\citeauthoryear{Molinari et al.}{2010}]{2010PASP..122..314M} Molinari S., et al., 2010, PASP, 122, 314
\bibitem[\protect\citeauthoryear{Montillaud et al.}{2015}]{2015A&A...584A..92M} Montillaud J., et al., 2015, A\&A, 584, A92
\bibitem[\protect\citeauthoryear{Morales et al.}{2013}]{2013A&A...560A..76M} Morales E.~F.~E., Wyrowski F., Schuller F., Menten K.~M., 2013, A\&A, 560, A76
\bibitem[\protect\citeauthoryear{Motte, Andre, \& Neri}{1998}]{1998A&A...336..150M} Motte F., Andre P., Neri R., 1998, A\&A, 336, 150
\bibitem[\protect\citeauthoryear{Mottram et al.}{2013}]{2013A&A...558A.126M} Mottram J.~C., van Dishoeck E.~F., Schmalzl M., Kristensen L.~E., Visser R., Hogerheijde M.~R., Bruderer S., 2013, A\&A, 558, A126
\bibitem[\protect\citeauthoryear{Ormel et al.}{2011}]{2011A&A...532A..43O} Ormel C.~W., Min M., Tielens A.~G.~G.~M., Dominik C., Paszun D., 2011, A\&A, 532, A43
\bibitem[\protect\citeauthoryear{Ossenkopf \& Henning}{1994}]{1994A&A...291..943O} Ossenkopf V., Henning T., 1994, A\&A, 291, 943
\bibitem[\protect\citeauthoryear{Ott}{2010}]{2010ASPC..434..139O} Ott S., 2010, ASPC, 434, 139
\bibitem[\protect\citeauthoryear{Peretto \& Fuller}{2009}]{2009A&A...505..405P} Peretto N., Fuller G.~A., 2009, A\&A, 505, 405
\bibitem[\protect\citeauthoryear{Pilbratt et al.}{2010}]{2010A&A...518L...1P} Pilbratt G.~L., et al., 2010, A\&A, 518, L1
\bibitem[\protect\citeauthoryear{Poglitsch et al.}{2010}]{2010A&A...518L...2P} Poglitsch A., et al., 2010, A\&A, 518, L2
\bibitem[\protect\citeauthoryear{Peretto et al.}{2013}]{2013A&A...555A.112P} Peretto N., et al., 2013, A\&A, 555, A112
\bibitem[\protect\citeauthoryear{Ragan, Henning, \& Beuther}{2013}]{2013A&A...559A..79R} Ragan S.~E., Henning T., Beuther H., 2013, A\&A, 559, A79
\bibitem[\protect\citeauthoryear{Reid et al.}{2009}]{2009ApJ...700..137R} Reid M.~J., et al., 2009, ApJ, 700, 137
\bibitem[\protect\citeauthoryear{S{\'a}nchez-Monge et al.}{2013}]{2013A&A...550A..21S} S{\'a}nchez-Monge {\'A}., Beltr{\'a}n M.~T., Cesaroni R., Fontani F., Brand J., Molinari S., Testi L., Burton M., 2013, A\&A, 550, A21
\bibitem[\protect\citeauthoryear{Salpeter}{1955}]{1955ApJ...121..161S} Salpeter E.~E., 1955, ApJ, 121, 161
\bibitem[\protect\citeauthoryear{Sanhueza et al.}{2012}]{2012ApJ...756...60S} Sanhueza P., Jackson J.~M., Foster J.~B., Garay G., Silva A., Finn S.~C., 2012, ApJ, 756, 60
\bibitem[\protect\citeauthoryear{Schuller et al.}{2009}]{2009A&A...504..415S} Schuller F., et al., 2009, A\&A, 504, 415
\bibitem[\protect\citeauthoryear{Shetty et al.}{2009a}]{2009ApJ...696..676S} Shetty R., Kauffmann J., Schnee S., Goodman A.~A., 2009a, ApJ, 696, 676
\bibitem[\protect\citeauthoryear{Shetty et al.}{2009b}]{2009ApJ...696.2234S} Shetty R., Kauffmann J., Schnee S., Goodman A.~A., Ercolano B., 2009b, ApJ, 696, 2234
\bibitem[\protect\citeauthoryear{Shirley et al.}{2011}]{2011ApJ...728..143S} Shirley Y.~L., Huard T.~L., Pontoppidan K.~M., Wilner D.~J., Stutz A.~M., Bieging J.~H., Evans N.~J., II, 2011, ApJ, 728, 143
\bibitem[\protect\citeauthoryear{Siringo et al.}{2009}]{2009A&A...497..945S} Siringo G., et al., 2009, A\&A, 497, 945
\bibitem[\protect\citeauthoryear{Sun \& Gao}{2009}]{2009MNRAS.392..170S} Sun Y., Gao Y., 2009, MNRAS, 392, 170
\bibitem[\protect\citeauthoryear{Tan et al.}{2014}]{2014prpl.conf..149T} Tan J.~C., Beltr{\'a}n M.~T., Caselli P., Fontani F., Fuente A., Krumholz M.~R., McKee C.~F., Stolte A., 2014, prpl.conf, 149
\bibitem[\protect\citeauthoryear{Testi et al.}{2014}]{2014prpl.conf..339T} Testi L., et al., 2014, prpl.conf, 339
\bibitem[\protect\citeauthoryear{Traficante et al.}{2015}]{2015MNRAS.451.3089T} Traficante A., Fuller G.~A., Peretto N., Pineda J.~E., Molinari S., 2015, MNRAS, 451, 3089
\bibitem[\protect\citeauthoryear{Urquhart et al.}{2012}]{2012MNRAS.420.1656U} Urquhart J.~S., et al., 2012, MNRAS, 420, 1656
\bibitem[\protect\citeauthoryear{Urquhart et al.}{2013a}]{2013MNRAS.431.1752U} Urquhart J.~S., et al., 2013a, MNRAS, 431, 1752
\bibitem[\protect\citeauthoryear{Urquhart et al.}{2013b}]{2013MNRAS.435..400U} Urquhart J.~S., et al., 2013b, MNRAS, 435, 400
\bibitem[\protect\citeauthoryear{Urquhart et al.}{2014a}]{2014MNRAS.437.1791U} Urquhart J.~S., Figura C.~C., Moore T.~J.~T., Hoare M.~G., Lumsden S.~L., Mottram J.~C., Thompson M.~A., Oudmaijer R.~D., 2014a, MNRAS, 437, 1791
\bibitem[\protect\citeauthoryear{Urquhart et al.}{2014b}]{2014MNRAS.443.1555U} Urquhart J.~S., et al., 2014b, MNRAS, 443, 1555
\bibitem[\protect\citeauthoryear{Urquhart et al.}{2014c}]{2014A&A...568A..41U} Urquhart J.~S., et al., 2014c, A\&A, 568, A41
\bibitem[\protect\citeauthoryear{Vasyunina et al.}{2014}]{2014ApJ...780...85V} Vasyunina T., Vasyunin A.~I., Herbst E., Linz H., Voronkov M., Britton T., Zinchenko I., Schuller F., 2014, ApJ, 780, 85
\bibitem[\protect\citeauthoryear{Ward-Thompson \& Robson}{1990}]{1990MNRAS.244..458W} Ward-Thompson D., Robson E.~I., 1990, MNRAS, 244, 458
\bibitem[\protect\citeauthoryear{Williams, Blitz, \& McKee}{2000}]{2000prpl.conf...97W} Williams J.~P., Blitz L., McKee C.~F., 2000, prpl.conf, 97
\bibitem[\protect\citeauthoryear{Wood \& Churchwell}{1989}]{1989ApJS...69..831W} Wood D.~O.~S., Churchwell E., 1989, ApJS, 69, 831
\bibitem[\protect\citeauthoryear{Wu \& Evans}{2003}]{2003ApJ...592L..79W} Wu J., Evans N.~J., II, 2003, ApJ, 592, L79
\bibitem[\protect\citeauthoryear{Ysard et al.}{2012}]{2012A&A...542A..21Y} Ysard N., et al., 2012, A\&A, 542, A21
\end{thebibliography}
\end{document}